\documentclass[preprint,10pt]{elsarticle}
\usepackage{amsmath}
\usepackage{amssymb}
\usepackage{verbatim}
\usepackage[colorlinks,linkcolor=blue,citecolor=blue,urlcolor=blue]{hyperref}
\usepackage{booktabs}
\usepackage[dvipsnames]{xcolor}
\usepackage{listings}

\usepackage[nomain,toc,acronym,automake, nonumberlist, style=superheaderborder,nogroupskip]{glossaries} 	
\makeglossaries										

\newcommand{\libcntr}{{\tt libcntr}}
\newcommand{\nessi}{{\tt NESSi}}

\newcommand{\mat}{\mathrm{M}}
\newcommand{\ret}{\mathrm{R}}

\newcommand{\beq}[1]{\begin{align} #1 \end{align}}

\newcommand{\I}{i}

\newcommand{\ha}{\hat{a}}
\newcommand{\hc}{\hat{c}}

\newcommand{\hn}{\hat{n}}
\newcommand{\hX}{\hat{X}}
\newcommand{\hP}{\hat{P}}

\newcounter{bla}

\journal{Computer Physics Communications}

\begin{document}

\newacronym{KB}{KB}{Kadanoff-Baym}
\newacronym{GF}{GF}{Green's function}
\newacronym{GFs}{GFs}{Green's functions}
\newacronym{NEGF}{NEGF}{nonequilibrium Green's function}
\newacronym{NEGFs}{NEGFs}{nonequilibrium Green's functions}
\newacronym{dmft}{DMFT}{dynamical mean-field theory}
\newacronym{ppsc}{PPSC}{Pseudo-Particle Strong Coupling}
\newacronym{nca}{NCA}{Non-Crossing Approximation}
\newacronym{oca}{OCA}{One-Crossing Approximation}
\newacronym{KMS}{KMS}{Kubo-Martin-Schwinger}
\newacronym{BDF}{BDF}{backward differentiation formula}
\newacronym{vie}{VIE}{Volterra integral equation}
\newacronym{vide}{VIDE}{Volterra integro-differential equation}
\newacronym{vies}{VIEs}{Volterra integral equations}
\newacronym{vides}{VIDEs}{Volterra integro-differential equations}
\newacronym{mpi}{MPI}{Message passing interface}
\newacronym{BZ}{BZ}{Brillouin zone}
\newacronym{2B}{2B}{second-Born}
\newacronym{HF}{HF}{Hartree-Fock}
\newacronym{hdf5}{HDF5}{Hierarchical Data Format version 5}
\newacronym{1D}{1D}{one-dimensional}
\newacronym{sMig}{sMig}{self-consistent Migdal approximation}
\newacronym{uMig}{uMig}{unrenormalized Migdal approximation}
\newacronym{SC}{SC}{superconducting}

\renewcommand*{\acronymname}{List of Abbreviations}

\begin{frontmatter}



\title{\textbf{NESSi}: The \textbf{N}on-\textbf{E}quilibrium \textbf{S}ystems \textbf{Si}mulation package}


\author[a,b]{Michael Sch\"uler}
\author[a,c]{Denis Gole\v{z}}
\author[a,d]{Yuta Murakami}
\author[a]{Nikolaj Bittner}
\author[a]{Andreas Hermann}
\author[c,e]{Hugo~U.~R.~Strand}
\author[a]{Philipp Werner}
\author[f]{Martin Eckstein\corref{author}}

\cortext[author] {Corresponding author.\\\textit{E-mail address:} martin.eckstein@fau.de}
\address[a]{Department of Physics, University of Fribourg, 1700 Fribourg, Switzerland}
\address[b]{Stanford Institute for Materials and Energy Sciences, SLAC \& Stanford University, Stanford, California 94025, USA}
\address[c]{Center for Computational Quantum Physics, Flatiron Institute, 162 Fifth avenue, New York, NY 10010, USA}
\address[d]{Department of Physics, Tokyo Institute of Technology, Meguro, Tokyo 152-8551, Japan}
\address[e]{Department of Physics, Chalmers University of Technology, SE-412 96 Gothenburg, Sweden}
\address[f]{Department of Physics, University of Erlangen-N\"urnberg, 91058 Erlangen, Germany }

\begin{abstract}
  The nonequilibrium dynamics of correlated many-particle systems is of interest in 
  connection with pump-probe experiments on molecular systems and solids, 
  as well as theoretical investigations of transport properties and relaxation processes. 
  Nonequilibrium Green's functions are a powerful tool to study interaction effects 
  in quantum many-particle systems out of equilibrium,
  and to extract physically relevant information
  for the interpretation of experiments.
  We present the open-source software package \nessi{} (The \textbf{N}on-\textbf{E}quilibrium \textbf{S}ystems \textbf{Si}mulation package)
 which allows to perform  many-body dynamics simulations based on
   Green's functions on the L-shaped Kadanoff-Baym 
  contour. \nessi{} contains  
the library \libcntr{} which implements
  tools for basic operations on these nonequilibrium
  Green's functions, for constructing Feynman diagrams, and for the solution of
  integral and integro-differential equations involving contour Green's functions.
     The library employs a discretization of the Kadanoff-Baym contour into time $N$ points 
  and a high-order implementation of integration routines. The total
  integrated error scales 
  up to $\mathcal{O}(N^{-7})$, 
  which is important since the  
  numerical effort increases at least cubically with the simulation time. 
    A distributed-memory
  parallelization over reciprocal space allows large-scale simulations of lattice systems. 
 We provide a collection of example programs ranging from dynamics in simple two-level systems to problems relevant in contemporary condensed matter physics, including Hubbard clusters and Hubbard or Holstein lattice models.  
The \libcntr{} library
  is the basis of a follow-up software package for nonequilibrium
  dynamical mean-field theory calculations based on strong-coupling perturbative impurity solvers. 
\end{abstract}


\end{frontmatter}



{\bf PROGRAM SUMMARY}

\begin{small}
\noindent
{\em Manuscript Title:} \textbf{NESSi}: The \textbf{N}on-\textbf{E}quilibrium \textbf{S}ystems \textbf{Si}mulation package 		\\
{\em Authors:}    Michael Sch\"uler, Denis Gole\v{z}, Yuta Murakami, Nikolaj Bittner, Hugo U.~R. Strand, Andreas Hermann, Philipp Werner, Martin Eckstein                                       \\
{\em Program Title:} NESSi		\\
{\em Journal Reference:}                                      \\
{\em Catalogue identifier:}                                   \\
{\em Licensing provisions:}          MPL v2.0                         \\
{\em Programming language:}  C++, python      \\
{\em Computer:} Any architecture with suitable compilers including PCs and clusters.             \\
{\em Operating system:}    Unix, Linux, OSX       \\
{\em RAM:} Highly problem dependent  \\
{\em Classification:}                                         \\
{\em External routines/libraries:} cmake, eigen3, hdf5 (optional), mpi (optional), omp (optional)   \\
{\em Nature of problem:} Solves equations of motion of time-dependent Green's functions on the Kadanoff-Baym contour.\\
   {\em Solution method:} Higher-order solution methods of integral and integro-differential equations on the Kadanoff-Baym contour.
   \\

\end{small}

\tableofcontents

\printglossaries

\part{Core functionalities and usage of the library}
\label{partI}

\section{Introduction}



Calculating the time evolution of an interacting quantum many-body system poses significant computational challenges. 
For instance, in wave-function based methods such as exact diagonalization or the density matrix renormalization group \cite{Daley2004, White2004},  one has to solve the time-dependent Schr\"odinger equation.
The main obstacles here are the exponential scaling of the Hilbert space with system size, or the rapid entanglement growth. Variational methods avoid this problem \cite{Ido2015}, but their accuracy depends on the ansatz for the wave function.
The Green's function formalism \cite{MahanBook} provides
a versatile framework to derive systematic approximations or to develop numerical techniques (e.g.~Quantum Monte Carlo \cite{GubernatisBook}) that circumvent the exponential scaling of the Hilbert space. Moreover, the Green's functions contain useful physical information that can be directly related to measurable quantities such as the photoemission spectrum.  

The \gls*{NEGF} approach, as pioneered by Keldysh, Kadanoff and Baym \cite{KadanoffBaym_1962,Keldysh1964}, is an extension of the equilibrium (Matsubara) Green's function technique \cite{AbrikosovGorkovDzyaloshinskiBook, stefanucci_nonequilibrium_2013}. It defines the analytical foundation for important concepts in nonequilibrium theory, such as the quantum Boltzmann equation \cite{KamenevBook}, but it also serves as a basis for numerical simulations. The direct numerical solution of the equations of motion for the real-time \gls*{NEGFs} has been successfully applied to the study of open and closed systems ranging from molecules to condensed matter, with various types of interactions including electron-electron, electron-phonon, or electron-photon couplings \cite{stefanucci_nonequilibrium_2013}. At the heart of these simulations lies the solution of integro-differential equations which constitute non-Markovian equations of motion for the \gls*{NEGFs}, the so-called Kadanoff-Baym equations. Even in combination with simple perturbative approximations, their solution remains a formidable numerical task.

\begin{figure}[tbp]
  \centering
  \includegraphics[width=0.5\textwidth]{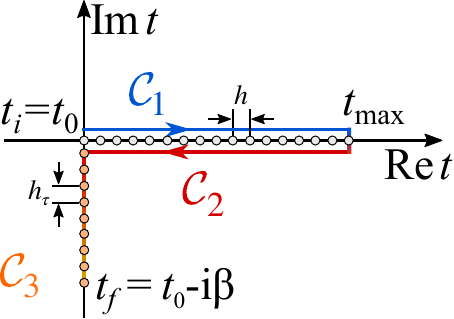}
  \caption{L-shaped Kadanoff-Baym contour $\mathcal{C}$ in the complex time
  plane, containing the forward branch $\mathcal{C}_1$,
  backward branch $\mathcal{C}_2$, and the imaginary (Matsubara) branch
  $\mathcal{C}_3$. Gray dots indicate the discretization on
  the real-time branches, while orange dots represent the
  discretization of the Matsubara branch. The arrows indicate the contour ordering.
  \label{fig:contour1}}
\end{figure}

A general \gls*{NEGF} calculation is based on the L-shaped \gls*{KB} contour $\mathcal{C}=\mathcal{C}_1\cup\mathcal{C}_2\cup\mathcal{C}_3$ in the complex time plane, which is sketched in Fig.~\ref{fig:contour1}. Here, the vertical (imaginary-time) branch $\mathcal{C}_3$ of the contour represents the initial equilibrium state of the system ($\beta=1/k_BT$ is the inverse temperature), while the horizontal branches $\mathcal{C}_{1,2}$ represent the time evolution of the system starting from this equilibrium state. Correlation functions with time arguments on this contour, and a time ordering defined by the arrows on the contour, are a direct generalization of the corresponding imaginary-time quantities. Hence, diagrammatic techniques and concepts which have been established for equilibrium many-body problems can be directly extended to the nonequilibrium domain.

In this paper, we introduce the 
\textbf{N}on\textbf{E}quilibrium \textbf{S}ystems \textbf{Si}mulation package (\nessi) as a state-of-the art tool for solving nonequilibrium many-body problems. \nessi{}
provides an efficient framework for representing various types of \gls*{GFs} on the discretized \gls*{KB} contour, implements the basic operations on these functions and allows to solve the
corresponding equations of motion. The library is aimed at the study of transient dynamics from an initial equilibrium state, induced by parameter modulations or electric field excitations. It is not designed for the direct study of nonequilibrium steady states or time-periodic Floquet states, where the memory of the initial state is lost and thus the branch $\mathcal{C}_3$ is not needed. While steady states can be reached in open systems in a relatively short time, depending on parameters and driving conditions, such simulations may be more efficiently implemented with a dedicated steady-state or Floquet code.  

The two-fold purpose of this paper is to explain both the numerical details underlying the solution of the integro-differential equations, and the usage and core functionalities of the library. The paper is therefore structured such that a reader who is mainly interested in using the library may consider only part I of the text (Sections~\ref{sec:overview}--\ref{sec:mpi}), while part II (Sections~\ref{sec:polynum}--\ref{sec:impl_green}) contains an explanation of the numerical methods. We remark that the usage of the library is also explained in a detailed an independent online manual on the webpage {\tt www.nessi.tuxfamily.org}.

This  paper is  organized as  follows. In  Section~\ref{sec:overview} we
present an overview of the basic structure of the \nessi{} 
software package, its core ingredients, and the main functionalities. This
overview is kept brief to serve as a reference for readers who are
familiar with the formalism. 
A detailed description of the general formalism is provided in
Section~\ref{sec:basic}, while Section~\ref{sec:integ_on_c} introduces the
fundamental equations of motion on the \gls*{KB} contour and discusses their solution, as 
implemented within \nessi.
Section~\ref{sec:compile_use} explains how to compile the \nessi{}
software and how to use its functionalities in custom projects. Several illustrative examples
are presented and discussed in Section~\ref{sec:example_programs}, and Section~\ref{sec:mpi} explains how to
use the code package to solve a large number of coupled integral equations with a distributed memory parallelization. 
Finally, the numerical details are presented in
Sections~\ref{sec:polynum}--\ref{sec:num_convolution}: starting from 
highly accurate methods for quadrature and integration
(Section~\ref{sec:polynum}), we explain the numerical procedures underlying the
solution of the equations on the \gls*{KB}
contour. In Section~\ref{sec:impl_convolution}--\ref{sec:impl_green} we
discuss the implementation of the main functions.

\section{Overview of the program package \label{sec:overview}}

\begin{figure}[ht]
  \centering
  \includegraphics[width=0.95\textwidth]{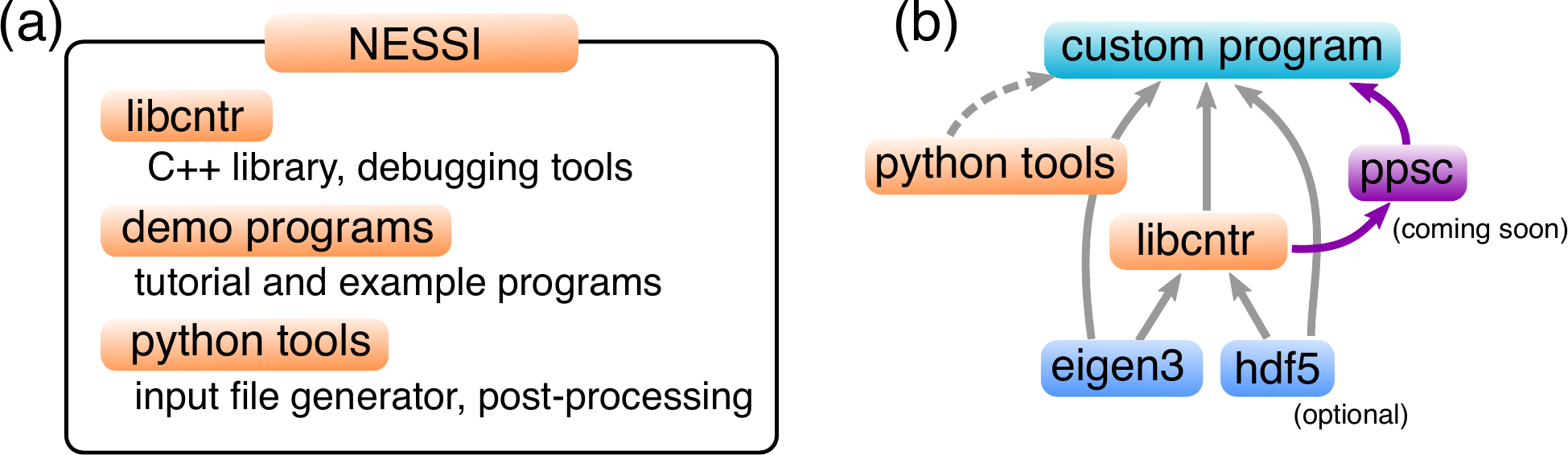}
  \caption{(a) Basic structure of the \nessi{}
  software package. The
    core ingredient is the shared library \libcntr{}, which 
    contains basic classes and routines for storing and manipulating 
    nonequilibrium Green functions.
    Furthermore, we provide a tutorial and
    demonstration programs which illustrate the usage and
    functionalities of \libcntr{}. (b) Custom programs based on
    \libcntr{} should be linked against the \libcntr{} library and
    the dependencies {\tt eigen3} and {\tt hdf5} (optional). An extension of the \libcntr{} library for dynamical mean field theory calculations is in preparation and will be published separately (PPSC library). 
    \label{fig:nessy}}
\end{figure}

\subsection{Structure of the software}

Figure~\ref{fig:nessy} summarizes the content of the \nessi{}
package. 
The core constituent of \nessi{}
is the shared library \libcntr{}. The \libcntr{} library is written in
C++ and provides the essential functionalities to treat \gls*{GFs} on the \gls*{KB}
contour. To solve a particular problem within the \gls*{NEGF} formalism, the
user can write a custom C++ program based on the extensive and
easy-to-use \libcntr{} library (see Fig.~\ref{fig:nessy}(b)).  The \nessi{}
package also contains a number of simple example programs, which
demonstrate the usage of \libcntr{}. All 
important
callable routines perform
various sanity checks in debugging mode, which enables an efficient
debugging of \libcntr{}-based programs.  Furthermore, we provide a
number of python tools for pre- and post-processing to assist the use
of programs based on \libcntr{}. More details can be found in
Section~\ref{sec:example_programs}, where we present a number of example
programs demonstrating the usage of \libcntr{} and the python tools.
The \libcntr{} library and the example programs depend on the {\tt
  eigen3} library 
  which implements efficient matrix operations.
  Furthermore, the {\tt hdf5} library and file format can be
used for creating binary, machine-independent output such as \gls*{GFs}, for instance. We further provide python tools for reading and
post-processing \gls*{GFs} from {\tt hdf5} format via the {\tt h5py} python
package. The usage of the {\tt hdf5} library in the \nessi{} package is, however, optional.


\subsection{Core functionalities}

The central task within the \gls*{NEGF} framework is calculating the
single-particle \gls*{GF}, from which all single-particle
observables such as the density or the current can be evaluated. Let
us consider
the generic many-body Hamiltonian
\begin{align}
\label{gggge772}
  \hat{H}(t) = \sum_{a,b} \epsilon_{a,b}(t) \hat{c}^\dagger_{a}
  \hat{c}_b + \hat{V}(t) \ ,
\end{align}
where $\hat{c}^\dagger_a$ ($\hat{c}_a$) denotes the fermionic or
bosonic creation (annihilation) operator with respect to some basis
labelled by $a$ and $\epsilon_{a,b}(t)$ the corresponding single-particle
Hamiltonian, while $\hat{V}(t)$ represents 
an arbitrary interaction term or the coupling to a bath.
In typical problems, the \gls*{GF} is obtained by solving
the Dyson equation 
\begin{align}
  \label{eq:dyson_1}
  \left[i \partial_t -\epsilon(t)\right]G(t,t^\prime) -
  \int_{\mathcal{C}} d\bar{t} \, \Sigma(t,\bar{t}) G(\bar{t},t^\prime)
  = \delta_\mathcal{C}(t,t^\prime) \,
\end{align}
or related integral equations. Here all objects are matrices in orbital indices. The symbols 
$\int_\mathcal{C}$ and $\delta_\mathcal{C}(t,t^\prime)$ 
denote an integral and the Dirac delta function defined on the \gls*{KB}
contour, respectively, while $\Sigma(t,t^\prime)$ is the 
self-energy, which captures all interaction effects originating from
$\hat{V}(t)$. 
Details are discussed in
Section~\ref{sec:basic}. 

\begin{figure}[ht]
  \centering
  \includegraphics[width=0.7\textwidth]{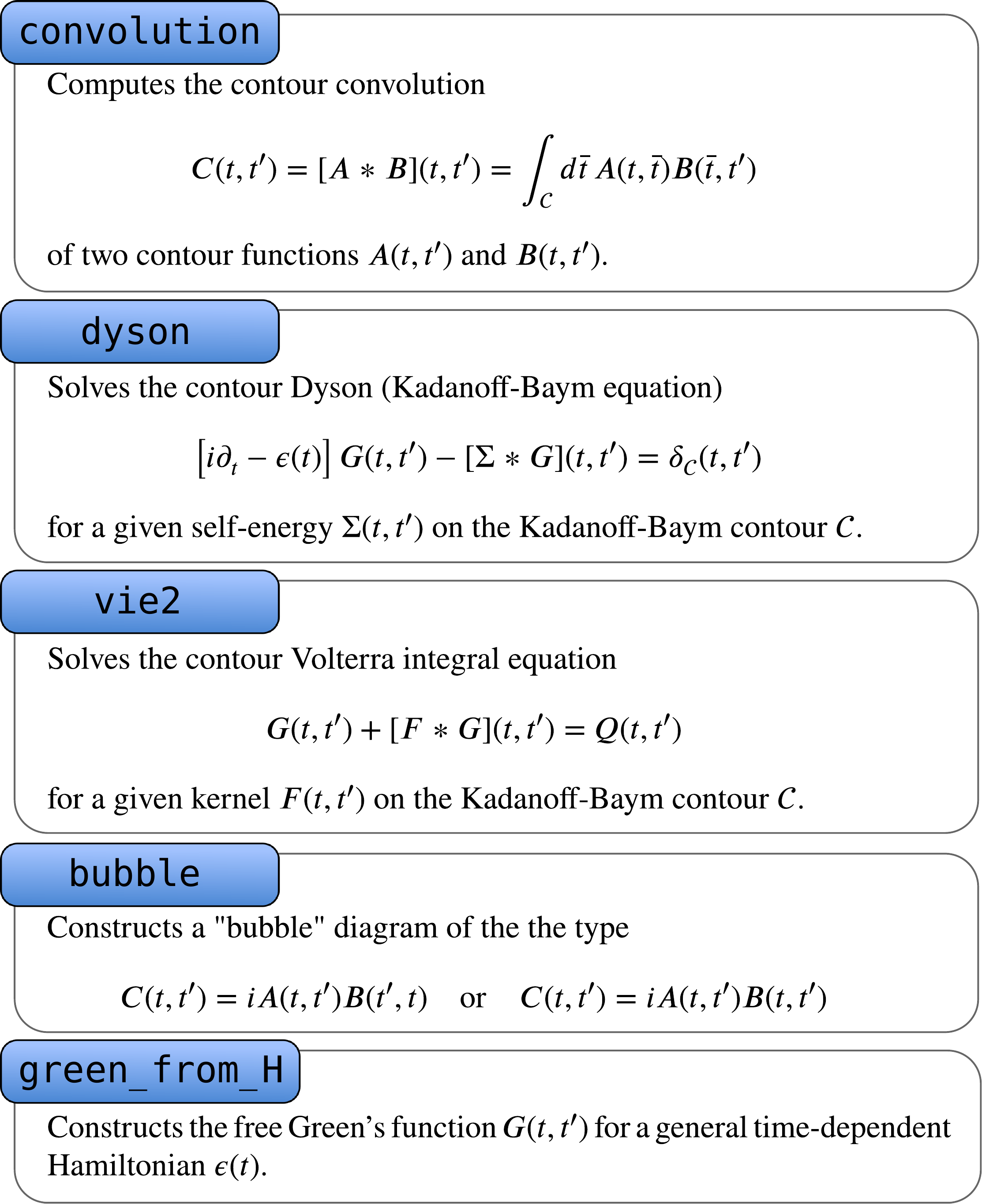}
  \caption{Main routines for constructing and manipulating objects
    on the \gls*{KB} contour and solving the corresponding equations of motion.
    \label{fig:main_routines}}
\end{figure}

The \libcntr{} library provides accurate methods for solving
the Dyson equation~\eqref{eq:dyson_1} and related problems. A brief overview of
the core routines is presented in Fig.~\ref{fig:main_routines}. It
includes routines for computing the convolution
$[A\ast B](t,t^\prime) = \int_{\mathcal{C}} d\bar{t} \, A(t,\bar{t})
B(\bar{t},t^\prime)$ ({\tt convolution}), which is an essential part
of solving Eq.~\eqref{eq:dyson_1}. It furthermore provides a
high-order solver for Eq.~\eqref{eq:dyson_1} ({\tt dyson}) along the
full \gls*{KB} contour. In particular, the initial thermal equilibrium state and the time evolution
are treated on equal footing. Moreover, contour integral equations of
the type
\begin{align}
  \label{eq:vie2_1}
  G(t,t^\prime) + [F\ast G](t,t^\prime) = Q(t,t^\prime)
\end{align}
can be solved efficiently via {\tt vie2}. A typical example is the self-consistent
$GW$ approximation~\cite{hedin_correlation_1999}: the screened interaction obeys the Dyson equation
$W = V + V\ast \Pi \ast W$, where $V$ denotes the bare Coulomb
interaction, while $\Pi$ stands for the irreducible
polarization, see Section~\ref{sec:example_programs}.
Free \gls*{GF} with respect to a given single particle Hamiltonian $\epsilon(t)$ are computed by {\tt
  green\_from\_H}. 
Finally, in many-body theories, the self-energy $\Sigma(t,t^\prime)$ can be
expressed in terms of Feynman diagrams using the \gls*{GF}
themselves. The most common Feynman diagrams consist of products of
two two-time functions, implemented as {\tt bubble} in \libcntr{}. All routines work both for fermions and bosons.

\subsection{Perspective: dynamical mean-field theory}

While the \nessi{} package provides a general framework for the manipulation of real-time \gls*{GFs} and can be used in different types of applications, one particularly
fruitful application has been the nonequilibrium extension of \gls*{dmft}~\cite{aoki2014_rev}.
In order to perform \gls*{dmft} calculations the library needs to be supplemented 
with a solver for the \gls*{dmft} effective impurity problem. Typical approximate
approaches are weak coupling expansions \cite{Tsuji2013} (in particular Iterated Perturbation Theory, IPT)
and strong coupling methods.
The weak coupling expansions can be directly implemented with the help of the routines provided by \libcntr{}.
The strong coupling based methods \cite{Eckstein:2010fk} are based on pseudo-particle \gls*{GFs},
which are defined for each state in the local Hilbert space and have properties different from the normal \gls*{GFs} introduced below.
 This formulation solves the atomic problem
exactly and treats the hybridization with the environment perturbatively.
The first and second order dressed expansion of this method is commonly
known as the \gls*{nca} and the \gls*{oca}~\cite{keiter_diagrammatic_1971,pruschke_anderson_1989}.
Currently we are working on a library implementing these methods
called the \gls*{ppsc} library -- based on
\libcntr{} -- and plan to 
release it as a future extension of \nessi{}.


\section{Basic formalism: NEGFs on the contour \label{sec:basic}}


\gls*{GFs} are objects depending on position, orbital and spin arguments (or an equivalent basis representation), as well as on two time arguments. The dependence on
multiple time arguments does not only account for the explicit
time dependence of observables, it also allows to store information on the
characteristic energy scales of the system and its thermal
equilibrium state. All these ingredients can be incorporated on equal
footing by choosing the time arguments on the \gls*{KB} contour $\mathcal{C}$ illustrated in Fig.~\ref{fig:contour1}: $t \in \mathcal{C}=\mathcal{C}_1\cup\mathcal{C}_2\cup\mathcal{C}_3$. The directions of the arrows in Fig.~\ref{fig:contour1} define the
induced ordering of time arguments $t_1,t_2 \in \mathcal{C}$: we call
$t_2$ later than $t_1$ (denoted by $t_2\succ t_1$) if $t_2$ can be
reached by progressing along $\mathcal{C}$ as indicated by the arrows. 
Thus, contour arguments on the backward branch
$t_2\in\mathcal{C}_2$ are always later than $t_1\in \mathcal{C}_1$.

Let us furthermore define the many-body Hamiltonian
$\hat{H}_\mathcal{C}(t)$ on the contour
$\mathcal{C}$ by $\hat{H}_\mathcal{C}(t) = \hat{H}(t)-\mu \hat{N}$ for
$t\in\mathcal{C}_{1,2}$ and $\hat{H}_\mathcal{C}(t) =
\hat{H}_\mathrm{eq}-\mu \hat{N}$ for $t\in\mathcal{C}_3$. Here,
$\hat{H}(t)$ denotes the real-time Hamiltonian, while
$\hat{H}_\mathrm{eq}$ describes the system in the thermal equilibrium of a
grand-canonical ensemble with chemical potential $\mu$ and particle
number operator $\hat{N}$. The thermal equilibrium is described by
the many-body density matrix
\begin{align}
  \hat{\rho} = \frac{1}{Z} e^{-\beta (\hat{H}_\mathrm{eq}-\mu
  \hat{N})} =  \frac{1}{Z} e^{-\beta \hat{H}_\mathcal{C}(-i \beta)} 
\end{align}
and the partition function by
$Z=\mathrm{Tr}[e^{-\beta (\hat{H}_\mathrm{eq}-\mu
  \hat{N})}]=\mathrm{Tr}[e^{-\beta \hat{H}_\mathcal{C}(-i\beta)}]$.
The time evolution of any observable is governed by the
time-evolution operator
\begin{subequations}
  \label{eq:timevol}
  \begin{equation}
  \hat{U}(t_1,t_2) = T \,\exp\left[-i
    \int^{t_1}_{t_2}\!d t\, \hat{H}_\mathcal{C}(t) \right] \ ,\hspace{2mm}
  t_1> t_2 \ , 
\end{equation}
\begin{equation}
  \hat{U}(t_1,t_2) = \bar{T} \,\exp\left[i
    \int^{t_2}_{t_1}\!d t\, \hat{H}_\mathcal{C}(t) \right] \ ,\hspace{2mm}
  t_2 >  t_1 \ ,
\end{equation}
\end{subequations}
where $T$ ($\bar{T}$) denotes the chronological (anti-chronological)
time ordering symbol.  
Time-dependent ensemble averages of an operator $\hat{A}(t)$ (here we
refer to an explicit time dependence in the Schr\"odinger picture) are
given by time evolving the density matrix according to
\begin{align}
 \langle \hat{A}(t) \rangle = \mathrm{Tr}\left[
  \hat{U}(t,0) \hat{\rho} \hat{U}(0,t) \hat{A}(t)  \right] \ ,
\end{align}
which we can formally rewrite as
\begin{align}
  \label{eq:expectA}
  \langle \hat{A}(t) \rangle = \frac{1}{Z} \mathrm{Tr}\left[
  \hat{U}(-i \beta,0) \hat{U}(0,t) \hat{A}(t) \hat{U}(t,0) \right] \ .
\end{align}
This is where the contour $\mathcal{C}$ comes into play: the time arguments
(from right to left) in
Eq.~\eqref{eq:expectA} follow the \gls*{KB} contour, passing through
$\mathcal{C}_1$, $\mathcal{C}_2$ and finally through
$\mathcal{C}_3$. If we now introduce the 
contour ordering symbol $T_\mathcal{C}$
which orders the operators along the \gls*{KB} contour,
any expectation value can be written as
\begin{align}
  \langle \hat{A}(t) \rangle = \frac{\mathrm{Tr}\left[T_\mathcal{C}
  \exp\left(-i \int_\mathcal{C}\!d\bar{t}\,
  \hat{H}_\mathcal{C}(\bar{t}) \right) \hat{A}(t) \right]}{\mathrm{Tr}\left[T_\mathcal{C}
  \exp\left(-i \int_\mathcal{C}\!d\bar{t}\,
  \hat{H}_\mathcal{C}(\bar{t}) \right) \right]}  .
\end{align}
Note that the integrals over $\mathcal{C}_1$ and $\mathcal{C}_2$ in the
denominator cancel, such that it becomes equivalent to the partition
function $Z$, while the matrix exponential in the numerator is
equivalent to the time evolution in Eq.~\eqref{eq:expectA}.

General correlators with respect to operators $\hat{A}(t)$ and
$\hat{B}(t^\prime)$ 
are similarly defined on $\mathcal{C}$ as
\begin{align}
  \label{eq:corrgen}
  C_{AB}(t,t^\prime) = \frac{\mathrm{Tr}\left[T_\mathcal{C}
  \exp\left(-i \int_\mathcal{C}\!d\bar{t}\,
  \hat{H}_\mathcal{C}(\bar{t}) \right) \hat{A}(t) \hat{B}(t^\prime)\right]}{\mathrm{Tr}\left[T_\mathcal{C}
  \exp\left(-i \int_\mathcal{C}\!d\bar{t}\,
  \hat{H}_\mathcal{C}(\bar{t}) \right) \right]} \equiv \langle
  T_\mathcal{C} \hat{A}(t) \hat{B}(t^\prime)  \rangle\ . 
\end{align}
Fermionic and bosonic particles are characterized by different commutation relations.
Throughout this paper, we
associate fermions (bosons) with the negative (positive) sign $\xi=-1$
($\xi =1$). Assuming $\hat{A}$, $\hat{B}$ to be pure fermionic or
bosonic operators, the contour ordering in the
definition~\eqref{eq:corrgen} is defined by
\begin{align}
  T_\mathcal{C} \big\{\hat{A}(t_1) \hat{B}(t_2)  \big\}
  = \begin{cases}
    \hat{A}(t_1) \hat{B}(t_2) & : t_1 \succ t_2 \\
    \xi\hat{B}(t_2) \hat{A}(t_1) & : t_2 \succ t_1 \ .
  \end{cases}
\end{align}
All two-time correlators like the \gls*{GF}~\eqref{eq:def_green} fulfill the
\gls*{KMS} boundary conditions
\begin{align}
  \label{eq:kmsbound}
  G(0,t^\prime)  = \xi G(-i\beta,t^\prime) \ , \ G(t,0) = \xi G(t,-i\beta)
  \ .
\end{align}
One of the most important correlators on the \gls*{KB} contour is the
single-particle \gls*{GF}, defined by
\begin{align}
  \label{eq:def_green}
  G_{ab}(t,t^\prime) = -i\langle T_\mathcal{C} \hat{c}_a(t)
  \hat{c}^\dagger_b(t^\prime) \rangle \ ,
\end{align}
where $\hat{c}^\dagger_a$ ($\hat{c}_a$) denotes the fermionic or
bosonic creation (annihilation) operator with respect to the
single-particle state $a$. 
Higher-order correlators with multiple
contour arguments, like two-particle
\gls*{GFs}, are defined in an analogous way.

Generalizing Eq.~\eqref{eq:corrgen}, contour correlators can also be defined with respect to any action $\hat{S}$, 
\begin{align}
  C_{AB}(t,t^\prime) = \frac{\mathrm{Tr}\left[T_\mathcal{C}
  e^{\hat{S}}\hat{A}(t) \hat{B}(t^\prime)\right]}{\mathrm{Tr}\left[T_\mathcal{C}
  e^{\hat{S}} \right]} \equiv \langle
  T_\mathcal{C} \hat{A}(t) \hat{B}(t^\prime)  \rangle_\mathcal{S}\ . 
\end{align}
For instance, Eq.~\eqref{eq:corrgen} corresponds to the action $\hat{S}=-i \int_\mathcal{C}\!d \bar{t}\,
\hat{H}_\mathcal{C}(\bar{t})$, but  all the properties and procedures discussed in this work remain valid if the action $\hat{S}$
is extended to a more general form. A typical example is the
action encountered in the framework of \gls*{dmft}~\cite{aoki2014_rev},
\begin{align}
  \hat{S}=-i \int_\mathcal{C}\!d t\, \hat{H}_\mathcal{C}(t) - i
  \int_\mathcal{C}\!d t\!\int_\mathcal{C}\!d t&^\prime\,
                                                \hat{c}^\dagger(t)\Delta(t,t^\prime)
                                                \hat{c}(t^\prime) \ ,
\end{align}
where $\Delta(t,t^\prime)$ is the so-called hybridization function.

\subsection{Contour decomposition}
\label{sSec:ProjObsTime}

While the real-time \gls*{GFs} is defined for any pair of arguments ($t$, $t^\prime$) on the L-shaped \gls*{KB} contour $\mathcal{C}$, it can be decomposed in a number of components 
where each of the two time arguments are constrained to a specific branch $\mathcal{C}_j$ of the KB contour. These, which will generally be called Keldysh components in the following,\footnote{In the literature (Ref.~\cite{aoki2014_rev}, for
  instance), often only the combination $G^\mathrm{K}(t,t^\prime) = G^<(t,t^\prime)  + G^>(t,t^\prime)$ is  referred to as the Keldysh component or  Keldysh \gls*{GF}.}
  are summarized in Table~\ref{tab:keldcomp}.
  We note that time arguments $t,t^\prime$ are used both to represent
contour arguments as well as real times, and whenever a correlator $C(t,t^\prime)$ occurs
without a superscript specifying the Keldysh components, the time
arguments $t, t^\prime$ are to be understood as contour arguments.

\begin{table}[ht]
  \caption{Keldysh components of a function $C(t_1,t_2)$ with arguments on $\mathcal{C}$. \label{tab:keldcomp}}
         \vspace*{2mm}
  \centerline{
  \renewcommand{\arraystretch}{1.2}
  \begin{tabular}{|c|c|c|c|}
    \hline
    $t_1 \in$ & $t_2 \in$ & notation & name \\
    \hline
    \hline
    $\mathcal{C}_1$ & $\mathcal{C}_1$ & $C^\mathrm{T}(t_1,t_2)$ &
                                                                  causal (time-ordered) \\
    \hline 
    $\mathcal{C}_1$ & $\mathcal{C}_2$ & $C^<(t_1,t_2)$ &
                                                         lesser \\
    \hline
    $\mathcal{C}_1$ & $\mathcal{C}_3 (t_2=-i\tau_2)$ & $C^\rceil(t_1,\tau_2)$ &
                                                                                left-mixing \\
    \hline
    $\mathcal{C}_2$ & $\mathcal{C}_1 $ & $C^>(t_1,t_2)$ &
                                                          greater \\
    \hline 
    $\mathcal{C}_2$ & $\mathcal{C}_2$ & $C^{\bar{\mathrm{T}}}(t_1,t_2)$ &
                                                                          anti-causal \\
    \hline 
    $\mathcal{C}_2$ & $\mathcal{C}_3 (t_2=-i\tau_2)$ & $C^\rceil(t_1,\tau_2)$ &
                                                                                left-mixing \\
    \hline
    $\mathcal{C}_3 (t_1=-i\tau_1)$ & $\mathcal{C}_1 $ & $C^\lceil(\tau_1,t_2)$ &
                                                                                 right-mixing \\
    \hline 
    $\mathcal{C}_3 (t_1=-i\tau_1)$ & $\mathcal{C}_2 $ & $C^\lceil(\tau_1,t_2)$ &
                                                                                 right-mixing \\
    \hline 
    $\mathcal{C}_3 (t_1=-i\tau_1)$ & $\mathcal{C}_3 (t_2=-i\tau_2) $ & $C(-i\tau_1,-i\tau_2)$ &
                                                                                imaginary time-ordered
    \\
      \hline                                                           
  \end{tabular}
  \renewcommand{\arraystretch}{1.0}
  }
\end{table}

In addition to the Keldysh components defined in Table~\ref{tab:keldcomp}, 
one defines the retarded (advanced) component
$C^\mathrm{R}(t,t^\prime)$ ($C^\mathrm{A}(t,t^\prime)$) by
\begin{align}
\label{eq:defretarded}
  C^\mathrm{R}(t,t^\prime) = \theta(t-t^\prime) \left[C^>(t,t^\prime)
  - C^<(t,t^\prime) \right] \ ,
\end{align}
\begin{align}
  C^\mathrm{A}(t,t^\prime) = \theta(t^\prime-t) \left[C^<(t,t^\prime)
  - C^>(t,t^\prime) \right] \ .
\end{align}
Here, $\theta(t)$ denotes the Heaviside step function.

For the component with imaginary time arguments only (last entry in
Table~\ref{tab:keldcomp}), we employ the convention to represent it by the Matsubara component
\begin{align}
  \label{eq:matsconv}
  C^\mat(\tau_1-\tau_2) = -i C(-i\tau_1, -i\tau_2) \ .
\end{align}
As the Matsubara function is defined by the thermal equilibrium
state, it depends on the difference of the imaginary time arguments
only. For the single-particle \gls*{GF}~\eqref{eq:def_green}, the
corresponding Matsubara \gls*{GF} 
$G^\mathrm{M}_{ab}(\tau)$ corresponds to a hermitian matrix,
$G^\mathrm{M}_{ab}(\tau) = [G^\mathrm{M}_{ba}(\tau)]^*$.

Extending the concept of the hermitian conjugate to the real-time and
mixed components will prove very useful for the numerical implementation as
detailed below. Thus, we formally define the \emph{hermitian}
conjugate
%
%
$[C^\ddagger](t,t^\prime)$ of a general correlator $C(t,t^\prime)$ by
\begin{subequations}
  \label{eq:hermconjg}
  \begin{align}
    \label{eq:hermconjg_gtrless}
    C^\gtrless(t,t^\prime) &= -\left( [C^\ddagger]^\gtrless(t^\prime,t) \right)^\dagger \ ,\\
    \label{eq:hermconjg_ret}
    C^\mathrm{R}(t,t^\prime) &= \left( [C^\ddagger]^\mathrm{A}(t^\prime,t)\right)^\dagger \ , \\
    \label{eq:hermconjg_tv}
    C^{\rceil}(t,\tau) &= - \xi \left( [C^\ddagger]^\lceil(\beta-\tau,t) \right)^\dagger \ ,\\
    \label{eq:hermconjg_vt}
    C^{\lceil}(\tau,t) &= - \xi \left( [C^\ddagger]^\rceil(t,\beta-\tau) \right)^\dagger \ , \\
    \label{eq:hermconjg_mat}
    C^{\mathrm{M}}(\tau) & =\left( [C^\ddagger]^{\mathrm{M}}(\tau) \right)^\dagger \ .
  \end{align}
\end{subequations}
Here the superscript $\dagger$ refers to the usual hermitian conjugate of a complex matrix.
The definition is reciprocal, $[C^\ddagger]^\ddagger(t,t^\prime) = C(t,t^\prime)$. 
A contour  function $C$  is called hermitian symmetric if $C=C^\ddagger$ (which does not mean that $C(t,t')$ is a hermitian matrix, see definition above). In particular, the \gls*{GF} defined by Eq.~\eqref{eq:def_green} possesses
hermitian symmetry. 
In contrast, more general objects, such as
convolutions (see Section~\ref{subsec:contourconv}), do not possess
a hermitian symmetry, and hence $C(t,t^\prime)$ and
$[C^\ddagger](t,t^\prime)$ are independent.

Note that $C^\ret(t,t^\prime) = 0$ if $t^\prime > t$, which expresses
the causality of the retarded component. However, 
for the implementation of numerical algorithms, it can be convenient to drop the Heavyside function in Eq.~\eqref{eq:defretarded}.
Therefore, we define a modified retarded
component by
\begin{align}
  \label{eq:Cretmod}
  \tilde{C}^{\ret}(t,t^\prime) = C^>(t,t^\prime) - C^<(t,t^\prime)
  \ .
\end{align}
The modified retarded component of the hermitian conjugate
$[C^\ddagger](t,t^\prime)$ then assumes a similar form as the greater
and lesser components:
\begin{align}
  \tilde{C}^{\ret}(t,t^\prime) = -
  \left([\tilde{C}^\ddagger]^{\ret}(t^\prime,t) \right)^\dagger \ .
\end{align}

Assuming the hermitian symmetry $C=C^\ddagger$, the number of independent
Keldysh components is limited to four. From $C^>(t,t^\prime)-C^<(t,t^\prime)=C^\ret(t,t^\prime)-C^\mathrm{A}(t,t^\prime)$
and Eq.~\eqref{eq:hermconjg_ret} one finds
that the pair $\{C^>,C^<\}$ or $\{C^\ret,C^<\}$ determines the other
real-time components. Furthermore, the hermitian symmetry for the left-mixing
component (Eq.~\eqref{eq:hermconjg_tv}) renders the $C^\lceil(\tau,t)$
redundant if $C^\rceil(t,\tau)$ is known. Hence, we use in \libcntr{} $\{C^<, C^\mathrm{R},
C^\rceil,C^\mathrm{M}\}$ as the minimal set of independent Keldysh components.

The \gls*{KMS} boundary conditions~\eqref{eq:kmsbound} establish further
relations between the Keldysh components. For the minimal set used
here, the corresponding relations are
\begin{subequations}
  \label{eq:kmsbound_keld}
  \begin{align}
    C^\mat(\tau+\beta) &= \xi C^\mat(\tau) \ ,\\
    C^\rceil(0,\tau) &= i C^\mat(-\tau) \ ,\\
    C^<(t,0) &= C^\rceil(t,0^+) \ .
  \end{align}
\end{subequations}
For the \gls*{GFs} $G(t,t^\prime)$, the anti-commutation (commutation) relations
for fermions (bosons) determine the retarded component at equal times
by
\begin{align}
  \label{eq:gretdiag}
  G^\ret_{ab}(t,t) = -i \delta_{a,b} \ . 
\end{align}
These conditions are used to numerically solve the Dyson equation, see below.

\subsection{Numerical representation of NEGFs\label{subsec:numrep}}

In the solvers used for computing the \gls*{GF} numerically, the
contour arguments are discretized according to the sketch in
Fig.~\ref{fig:contour1}. The contour $\mathcal{C}$ is divided into
$(N_t+1)$ equidistant points $t_n=n h$, $n=0,\dots,N_t$ on the real
axis (the points correspond to both real time branches $\mathcal{C}_{1,2}$), while $\tau_m= m h_\tau$,
$m=0, \dots,N_\tau$ with $\tau_0=0^+$, $\tau_{N_\tau}=\beta^-$ samples
the Matsubara branch.  The corresponding discretized contour is
denoted by $\mathcal{C}[h,N_t,h_\tau,N_\tau]$. 

\begin{figure}[t]
  \centering
  \includegraphics[width=\textwidth]{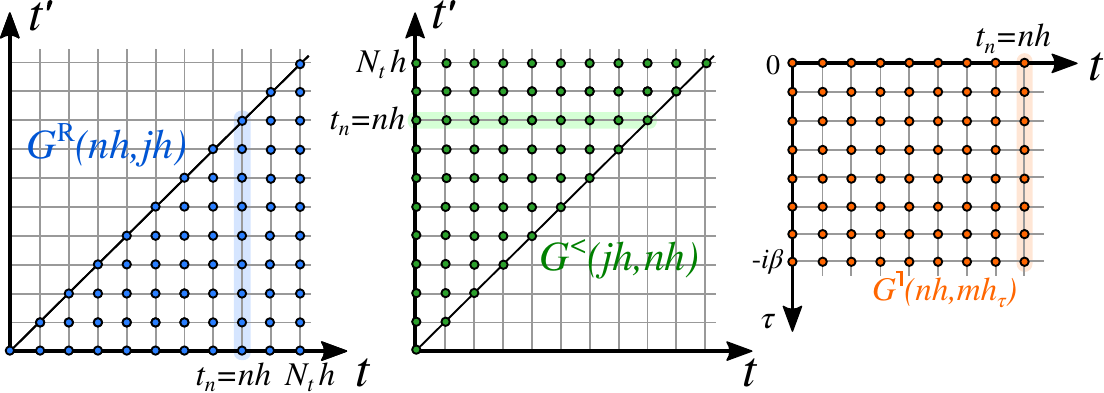}
  \caption{Storage scheme of the {\tt herm\_matrix} class: for
    $0\le n \le N_t$ time steps, the class saves $G^\ret(n h, j h)$
    and $G^<(j h, n h)$ for $0 \le j \le n$ along with the left-mixing
    component $G^\rceil(n h,m h_\tau)$ for $m=0,\dots,N_\tau$. The
    shaded background represents the storage scheme of the time slice
    $\mathcal{T}[G]_n$, represented by the class
    {\tt herm\_matrix\_timestep}.  \label{fig:storage_herm}}
\end{figure}

As discussed in Section~\ref{sSec:ProjObsTime}, the contour correlators $C(t,t^\prime)$ with hermitian
symmetry are represented in \libcntr{} by the minimal set of Keldysh components
$\{C^<,C^\ret,C^\rceil,C^\mat\}$ on
$\mathcal{C}[h,N_t,h_\tau,N_\tau]$. The hermitian
symmetry~\eqref{eq:hermconjg_gtrless} allows to further reduce the
number of points to be stored. We gather this representation of
$C(t,t^\prime)$ in the class {\tt herm\_matrix}, which stores
\begin{subequations}
  \label{eq:storage_herm}
  \begin{align}
    C^\mat_m &= C^\mat(m h_\tau) \ , \ m=0,\dots,N_\tau \ , \\
    C^<_{jn} &= C^<(j h, n h) \ , \ n=0,\dots,N_t, j=0,\dots,n \ , \\
    C^\ret_{nj} &= C^\ret(n h, j h) \ , \ n=0,\dots,N_t, j=0,\dots,n \ , \\
    C^\rceil_{nm} &= C^\rceil(n h, m h_\tau) \ , \ n=0,\dots,N_t, m=0,\dots,N_\tau \ .
  \end{align}
\end{subequations}
Hence, the retarded component is only stored on the lower triangle in the two-time plane,
while only the upper triangle is required to represent the lesser
component (see Fig.~\ref{fig:storage_herm}). For fixed time arguments,
the contour function $C$ represents a $d\times d$ square matrix. Note that general two-time functions $C$ (without hermitian symmetry) are also stored in the form of Eq.~\eqref{eq:storage_herm}. Hence, to recover the full two-time dependence $C(t,t^\prime)$, $C^\ddagger(t,t^\prime)$ is required.

For some of the algorithms described below, not the full two-time
correlator but only a slice with one fixed contour argument is
required. To this end, we define a time step $\mathcal{T}[C]_n$ represented by the class
{\tt herm\_matrix\_timestep}, which stores the Keldysh components
\begin{subequations}
  \begin{align}
    (\mathcal{T}[C]_n)^\mat_m & = C^\mat(m h_\tau) \ , \ m=0,\dots,N_\tau \ , \\
    (\mathcal{T}[C]_n)^<_{j} &= C^<(j h, n h) \ ,  j=0,\dots,n \ , \\
    (\mathcal{T}[C]_n)^\ret_{j} &= C^\ret(n h, j h) \ ,  j=0,\dots,n \ , \\
    (\mathcal{T}[C]_n)^\rceil_{m} &= C^\rceil(n h, m h_\tau) \ ,  m=0,\dots,N_\tau \ .
  \end{align}
\end{subequations}
For later convenience we define $\mathcal{T}[C]_{-1}$, which refers to the
Matsubara component only. The stored points in the two-time plane are
indicated by the shaded background in
Fig.~\ref{fig:storage_herm}. Note that $\mathcal{T}[C]_n$ is a
$d\times d$ square matrix for fixed contour argument.

Finally, we introduce contour functions with a single contour argument $f(t)$. While $f(t)$ corresponds to the real times for $t\in \mathcal{C}_1\cup\mathcal{C}_2$, the function value on the imaginary branch is defined by $f(-i\tau) = f(0^-)$.
Single-time contour functions are represented by the {\tt function} class, storing
\begin{align}
  f_n = \begin{cases} f(0^-) & : n = -1 \\ f(n h) & :
    n=0,\dots,N_t 
    \end{cases} \ .
\end{align}
For fixed $n$, $f_n$ can be matrix valued ($d\times d$ square matrix).

\begin{table}[ht]
  \caption{Constructor of the classes  {\tt herm\_matrix}, {\tt
      herm\_matrix\_timestep} and {\tt function}.
      The arguments in this table correspond to the number of
points and the storage scheme discussed above: {\tt nt} is the number
of real-time points $N_t$, {\tt ntau} stands for the number of points $N_\tau$
on the Matsubara branch, {\tt tstp} marks the current timestep $t_n$,  whereas {\tt size1} denotes the
number of basis functions (orbitals) $d$. The last argument {\tt sig} for the
{\tt herm\_matrix} and {\tt herm\_matrix\_timestep} 
specifies the fermionic ({\tt sig = -1}) or bosonic ({\tt sig =
  +1}) 
statistics. 
       \label{tab:herm_init}}
       \vspace*{2mm}
  \begin{small}
  \begin{tabular}{|l||l|}
    \hline
    class & constructor \\
    \hline
    \hline
  {\tt herm\_matrix} & {\tt herm\_matrix(int nt, int ntau, int size1,}\\
     &                            {\tt  int sig) }\\
    \hline 
    {\tt herm\_matrix\_timestep} &{\tt herm\_matrix\_timestep(int tstp, int ntau,}\\&{\tt int size1, int sig)} \\
    \hline 
     {\tt function } & {\tt function(int nt, int size1)} \\
    \hline 
  \end{tabular}
  \end{small}
\end{table}

The initialization of the above contour functions as C++ classes in \libcntr{} is summarized
in Table~\ref{tab:herm_init}.

\subsection{Contour multiplication and convolution\label{subsec:contourconv}}

The basic Feynman diagrams can be constructed from products and convolutions of \gls*{GFs}. In this subsection we summarize how such operations can be expressed in terms of Keldysh components.

\paragraph{Product $C(t,t^\prime) = i A(t,t^\prime) B(t^\prime,t)$}
--- This type of product is often encountered in diagrammatic calculations. 
For
instance, the polarization entering the $GW$ approximation is of this
form~\cite{Aryasetiawan_1998}.
Its representation in terms of the
Keldysh components follows from the Langreth
rules~\cite{stefanucci_nonequilibrium_2013}:
\begin{subequations}
  \label{eq:def_bubble1}
  \begin{align}
    C^\gtrless(t,t^\prime) &= i A^\gtrless(t,t^\prime)
    B^\lessgtr(t^\prime,t) \ , \\
    C^\ret(t,t^\prime) &= i A^\ret(t,t^\prime) B^<(t^\prime,t) +
    i A^<(t^\prime,t) B^\mathrm{A}(t^\prime,t) \ , \\
    C^\rceil(t,\tau) &= i A^\rceil(t,\tau)
    B^\lceil(\tau,t) \ , \\
    C^\mat(\tau) &= A^\mat(\tau)
    B^\mat(-\tau) \ .
  \end{align}
\end{subequations}
In \libcntr{}, we refer to this contour product as {\tt Bubble1}.

\paragraph{Product $C(t,t^\prime) = i A(t,t^\prime) B(t,t^\prime)$} ---
The direct product of this form also represents a bubble. It is used, for instance, in the calculation of the $GW$ self-energy diagram (for additional examples of usage see Section~\ref{sec:example_programs}). The corresponding representation in terms of the Keldysh components is
analogous to the above:
\begin{subequations}
  \label{eq:def_bubble2}
  \begin{align}
    C^\gtrless(t,t^\prime) &= i A^\gtrless(t,t^\prime)
    B^\gtrless(t,t^\prime) \ , \\
    C^\ret(t,t^\prime) & = i A^\ret(t,t^\prime) B^\ret(t,t^\prime) +
    i A^<(t,t^\prime) B^\ret(t,t^\prime) + A^\ret(t,t^\prime)
    i B^<(t,t^\prime) \ , \\
    C^\rceil(t,\tau) & = i A^\rceil(t,\tau)
    B^\rceil(t,\tau) \ , \\
    C^\mat(\tau) & = A^\mat(\tau)
    B^\mat(\tau) \ .
  \end{align}
\end{subequations}
In \libcntr{}, we refer to this contour product as {\tt Bubble2}.

\paragraph{Convolution $C = A\ast B$} --- The convolution of the
correlators
\begin{align}
   [A\ast B](t,t^\prime) = \int_{\mathcal{C}} d \bar{t}\, A(t,\bar{t}) B(\bar{t},t^\prime)
\end{align}
is one of the most basic operations on the
contour.
Using the Langreth rules for the convolution, one obtains
\begin{align}
    C^\gtrless(t,t^\prime) &=\int^t_0\!d \bar{t}\, A^\ret(t,\bar t)
    B^\gtrless(\bar t, t^\prime) + \int^{t^\prime}_0\! d \bar{t}\,
    A^\gtrless(t,\bar{t}) B^\mathrm{A}(\bar{t},t^\prime) \nonumber \\
                           &\quad
                             -i \int^\beta_0\! d\bar{\tau}\,
                             A^\rceil(t,\bar{\tau}) B^\lceil(\bar{\tau},t^\prime)  ,\\
    C^\ret(t,t^\prime) &=\int^t_{t^\prime}\!d \bar{t}\, A^\ret(t,\bar t)
    B^\ret(\bar t, t^\prime) ,\\
    C^\rceil(t,\tau) &=\int^t_{0}\!d \bar{t}\, A^\ret(t,\bar t)
    B^\rceil(\bar t, \tau) + \int^\beta_0\! d\tau^\prime\,
                       A^\rceil(t,\tau^\prime)
                       B^\mat(\tau^\prime-\tau) \ , \\
    C^\mat(\tau) &=\int^\beta_{0}\!d \bar{\tau}\, A^\mat(\tau-\bar{\tau})
    B^\mat(\bar\tau) \ .
\end{align}
For the hermitian conjugate one finds $[C^\ddagger](t,t^\prime) =
[B^\ddagger\ast A^\ddagger](t,t^\prime)$.

\subsection{Free Green's functions}
\label{sec:impl_green_from_H_part_I}

Free \gls*{GFs} $G_0(t,t^\prime)$ are often required when solving the Dyson
equation in integral form. A free Green's function for a time-dependent Hamiltonian 
$\epsilon(t)$ [Eq.~\eqref{gggge772}] is obtained from the solution of the equation
\begin{align}
\label{freefreefree}
  \left[i \partial_t - \epsilon(t)\right] G_0(t,t^\prime) = \delta_\mathcal{C}(t,t^\prime)
\end{align}
with \gls*{KMS} boundary conditions. This defines a regular differential equation, which can be solved by various standard algortithms. 
In \libcntr{}, free Green's functions can be obtained by the call to a function {\tt green\_from\_H}. There are several rather obvious interfaces to this function, and we refer to the examples  (and the online manual) for more details. The numerical implementation is described in Section.~\ref{sec:impl_green}.

\section{Integral equations on $\mathcal{C}$: Overview\label{sec:integ_on_c}}


\subsection{Equations with causal time-dependence}

In applications of the Keldysh formalism to problems involving real-time dynamics, one needs to solve various types of differential and integral equations on the contour $\mathcal{C}$.  \libcntr{} provides algorithms to solve the three most common tasks (convolution of two \gls*{GFs}, solution of a Dyson equation in both integro-differential and 
integral form) on an equidistant contour mesh $\mathcal{C}[h,N_t,h_\tau,N_\tau]$ with a global error that scales like $\mathcal{O}(h^k,h_\tau^{k})$ with $k$ up to $k=5$. 

The precise equations are summarized in Secs.~\ref{secygxeqw01} to \ref{secygxeqw03} below. A common property of all equations is their 
{\em causal} structure, i.\,e., the 
solution for the time slice $\mathcal{T}[G]_n$ of the unknown $G$ does
not depend on the time slices $m>n$. 
This causality allows to
transform the $k$th order accurate solution of all integral equations
on $\mathcal{C}$ into a time-stepping procedure with the following three steps, which are executed consecutively:
\begin{itemize}
\item[1)] {\bf Matsubara:} Solve the equation for the Matsubara
  time slice $\mathcal{T}[G]_{-1}$, using the input at time slice
  $m=-1$.
\item[2)] {\bf Start-up:} Solve the equation for $\mathcal{T}[G]_{j}$,
  $j=0,...,k$, using $\mathcal{T}[G]_{-1}$ and the input at
  time slices $j=-1,...,k$. The start-up procedure is essential to
  keep the $\mathcal{O}(h^k)$ accuracy of the  algorithm, as
  explained in the numerical details (Section~\ref{sec:num_vie}).
\item[3)] {\bf Time-stepping:} For time slices $n>k$, successively
  solve the equation for $\mathcal{T}[G]_{n}$, using
  $\mathcal{T}[G]_{j}$ for $j=-1,...,n-1$ and the input at time slices
  $j=-1,...,n$.
\end{itemize}
The causality is preserved {\em exactly} by these algorithms
for all time slices $n=-1$ and $n\ge k$. Only for the starting
time slices $n=0,...,k$, the numerical error 
$\mathcal{O}(h^k,h_\tau^{k})$ can also depend on the input at later
time slices $j=n+1,...,k$.

In Sections \ref{secygxeqw01}--\ref{secygxeqw03} we
specify the integral equations implemented in
\libcntr{}, and present an overview over their input and dependencies
on the Matsubara, start-up, and time-stepping parts. The details of the 
numerical implementation of the $k^\mathrm{th}$-order accurate algorithm are
explained in Sections~\ref{sec:num_vie}--\ref{sec:impl_vie2}.

\subsection{{\tt dyson:} Dyson equation in integro-differential form}
\label{secygxeqw01}

The Dyson equation for the Green's function $G(t,t^\prime)$ can be written as 
\begin{subequations}
\label{dyson-all}
\begin{align}
\label{dyson}
i\partial_t G(t,t^\prime) - \epsilon(t) G(t,t^\prime) - 
\int_\mathcal{C} d\bar t\, \Sigma(t,\bar t) G(\bar t,t^\prime) = \delta_{\mathcal{C}}(t,t^\prime).
\end{align}
This equation is to be solved for $G(t,t^\prime)$ for given input
$\epsilon(t)$ and $\Sigma(t,t^\prime)$, and the \gls*{KMS} boundary conditions
\eqref{eq:kmsbound}. It is assumed that $\Sigma=\Sigma^\ddagger$ 
is hermitian
(according to Eq.~\eqref{eq:hermconjg}), and $\epsilon(t) =\epsilon(t)^\dagger$, which
implies that also the solution $G$ possesses hermitian symmetry. All
quantities $\Sigma(t,t^\prime)$, $G(t,t^\prime)$, and $\epsilon(t)$ can be square
matrices of dimension $d\ge 1$. Because of the hermitian symmetry, $G$
can also be determined from the equivalent conjugate equation
\begin{align}
\label{dyson_cc}
-i\partial_{t^\prime} G(t,t^\prime) -  G(t,t^\prime) \epsilon(t^\prime)- 
\int_\mathcal{C} d\bar t \,G(t,\bar t) \Sigma(\bar t,t^\prime) = \delta_{\mathcal{C}}(t,t^\prime).
\end{align}
\end{subequations}

In \libcntr{}, Eq.~\eqref{dyson-all} is referred to as {\tt dyson}
equation. The dependencies between the input and output for the Matsubara,
start-up, and time-stepping routines related to the solution of
Eqs.~\eqref{dyson-all} are summarized in Table \ref{tab:dyson}.

A typical application of Eqs.~\eqref{dyson} and \eqref{dyson_cc} is
the solution of the Dyson series in diagrammatic perturbation theory,
i.\,e., a differential formulation of the problem
\begin{align}
G 
&= G_0 + G_0\ast \Sigma \ast G_0 +  G_0\ast \Sigma \ast G_0\ast \Sigma \ast G_0 + \cdots\nonumber
\\
\label{.dm;camx}
&= G_0 + G_0\ast \Sigma \ast G,
\end{align}
where $G_0$ satisfies the differential equation 
\begin{align}
i\partial_t G_0(t,t^\prime) - \epsilon(t) G_0(t,t^\prime) = \delta_{\mathcal{C}}(t,t^\prime).
\end{align}
In this case $\epsilon(t)$ is a (possibly time-dependent) single-particle or mean-field Hamiltonian.

\begin{table}[tbp]
\centerline{
\begin{tabular}{|l||l|l|l|}
\hline
Routine(s)
&
Input
&
Output
\\
\hline
\hline
\tt{dyson\_mat}
&
\begin{minipage}{0.35\textwidth}
~\\
$\mathcal{T}[\Sigma]_{-1}$, $\epsilon_{-1}$ \\[-2mm]
\end{minipage}
&
$\mathcal{T}[G]_{-1}$
\\
\hline
\tt{dyson\_start}
&
\begin{minipage}{0.35\textwidth}
~\\
$\mathcal{T}[\Sigma]_{j}$ for $j=-1,...,k$,\\
$\epsilon_{j}$ for $j=-1,...,k$,\\
$\mathcal{T}[G]_{-1}$ \\[-2mm]
\end{minipage}
&
$\mathcal{T}[G]_{j}$, $j=0,...,k$
\\
\hline
\begin{minipage}{0.23\textwidth}
{\tt dyson\_timestep}(n)
\\
$n>k$
\end{minipage}
&
\begin{minipage}{0.35\textwidth}
~\\
$\mathcal{T}[\Sigma]_{j}$ for $j=-1,...,n$,\\
$\epsilon_{j}$ for $j=-1,...,n$,\\
$\mathcal{T}[G]_{j}$  for $j=-1,...,n-1$\\[-2mm]
\end{minipage}
&
$\mathcal{T}[G]_{n}$
\\
\hline
\end{tabular}
}
\caption{Dependencies between input and output for the Matsubara, start-up, and time-stepping routines associated with the solution of Eqs.~\eqref{dyson-all}.}
\label{tab:dyson}
\end{table}

\subsection{{\tt vie2}: Dyson equation in integral form}
\label{secygxeqw02}

The second important equation is an integral equation of the form
\begin{subequations}
\label{vie2-all}
\begin{align}
\label{vie2}
G(t,t^\prime) + \int_\mathcal{C} d\bar t\, F(t,\bar t) G(\bar t,t^\prime) = Q(t,t^\prime) &\quad \Leftrightarrow\quad (1+F)*G=Q,
\\
 G(t,t^\prime) + \int_\mathcal{C} d\bar t \,G(t,\bar t) F^\ddagger(\bar t,t^\prime) = Q(t,t^\prime) &\quad \Leftrightarrow\quad G*(1+F^\ddagger)=Q.
\label{vie2_cc}
\end{align}
\end{subequations}
This linear equation is to be solved for $G(t,t^\prime)$ for a given input  kernel $F(t,t^\prime)$, its hermitian conjugate $F^\ddagger(t,t^\prime)$, and a source term $Q(t,t^\prime)$, assuming the \gls*{KMS} boundary conditions~\eqref{eq:kmsbound_keld}. In the solution of this linear equation, we assume that both $Q$ and $G$ are  hermitian. In general, the  hermitian symmetry would not hold for an arbitrary input $F$ and $Q$. However, it does hold when $F$ and $Q$ satisfy the relation
\begin{align}
\label{,masbX,.}
F\ast Q=Q\ast F^\ddagger, \,\,\,Q=Q^\ddagger,
\end{align}
which is the case for the typical applications discussed below. In this case, Eqs.~\eqref{vie2} and \eqref{vie2_cc} are equivalent. 

In \libcntr{}, Eq.~\eqref{vie2-all} is referred to as {\tt vie2}. The
nomenclature refers to the fact that the equation can be reduced to a
\underline{V}olterra \underline{I}ntegral \underline{E}quation of
\underline{$2$}nd kind (see below). The dependencies between the input and
output for the Matsubara, start-up, and time-stepping routines associated
with the solution of the {\tt vie2} equation are summarized in Table
\ref{tab:vie2}.

A typical physical application of Eqs.~\eqref{vie2-all} is given by the summation of a random phase approximation (RPA) series for a susceptibility
\begin{align}
\label{eq:rpa_vie2}
\chi 
&= \chi_0 + \chi_0\ast V \ast \chi_0 +  \chi_0\ast V \ast \chi_0\ast V \ast \chi_0 + \cdots \nonumber\\
&= \chi_0 + \chi_0\ast V \ast \chi \ .
\end{align}
Here $\chi_0$ is a bare susceptibility in a given channel (charge,
spin, etc,...), and $V$ is a (possibly retarded) interaction in that
channel. Since $\chi_0$ and $V$ are \gls*{GFs} with hermitian symmetry, the 
equation~\eqref{eq:rpa_vie2} can be recast in the form \eqref{vie2} with
\begin{align}
F=-\chi_0\ast V,\,\,\,F^\ddagger =- V\ast \chi_0,\,\,\,Q=\chi_0.
\end{align}
One can easily verify Eq.~\eqref{,masbX,.}. Equivalently, one can also recast the Dyson series \eqref{.dm;camx} into the form of a {\tt vie2} equation, with $F=-G_0\ast \Sigma$, $F^\ddagger=- \Sigma\ast G_0$, and $Q=G_0$.

\begin{table}[tbp]
\centerline{
\begin{tabular}{|c||l|l|c|}
\hline
Routine(s)
&
Input
&
Output
\\
\hline
\hline
\tt{vie2\_mat}
&
\begin{minipage}{0.4\textwidth}
~\\
$\mathcal{T}[F]_{-1},\mathcal{T}[F^\ddagger]_{-1}$, $\mathcal{T}[Q]_{-1}$\\[-2mm]
\end{minipage}
&
$\mathcal{T}[G]_{-1}$
\\
\hline
\tt{vie2\_start}
&
\begin{minipage}{0.4\textwidth}
~\\
$\mathcal{T}[F]_{j},\mathcal{T}[F^\ddagger]_{j}$ for $j=-1,...,k$,\\
$\mathcal{T}[Q]_{j}$ for $j=-1,...,k$,\\
$\mathcal{T}[G]_{-1}$ \\[-2mm]
\end{minipage}
&
$\mathcal{T}[G]_{j}$, $j=0,...,k$
\\
\hline
\begin{minipage}{0.22\textwidth}
{\tt vie2\_timestep}(n)\\
$n>k$
\end{minipage}
&
\begin{minipage}{0.4\textwidth}
~\\
$\mathcal{T}[F]_{j},\mathcal{T}[F^\ddagger]_{j}$ for $j=-1,...,n$,\\
$\mathcal{T}[Q]_{n}$\\
$\mathcal{T}[G]_{j}$  for $j=-1,...,n-1$\\[-2mm]
\end{minipage}
&
$\mathcal{T}[G]_{n}$
\\
\hline
\end{tabular}
}
\caption{Dependencies between input and output for the Matsubara, Start-up, and time-stepping routines associated with the solution of Eqs.~\eqref{vie2-all}.}
\label{tab:vie2}
\end{table}

\subsection{{\tt convolution}}
\label{secygxeqw03}

The most general convolution of two contour Green's functions $A$ and $B$ and a time-dependent function $f$ is given by the integral
\begin{align}
\label{convolution}
C(t,t^\prime)
=
\int_\mathcal{C} d\bar t\, A(t,\bar t) f(\bar t) B(\bar t,t^\prime).
\end{align}
In \libcntr{} this integral is calculated by the {\tt convolution} routines. The dependencies between the input and output for the Matsubara, start-up, and time-stepping routines related to {\tt convolution} are summarized in Table \ref{tab:convolution}.  

In the evaluation of this integral we make in general no assumption on
the hermitian properties of $A$ and $B$. 
Since the input of the implemented routine is the class of the type {\tt herm\_matrix}, both $A$ and $B$ and
their hermitian conjugate $A^\ddagger$ and $B^\ddagger$ must be provided, so that
$A(t,t^\prime)$ and $B(t,t^\prime)$ can be restored for arbitrary $t,t^\prime$ on
$\mathcal{C}$ (see Section~\ref{sec:basic}).
Similarly, the implemented
routines calculate the convolution integral only for the components of $C$
corresponding to the domain of the {\tt herm\_matrix} type, i.\,e.,
the upper/lower triangle representation~\eqref{eq:storage_herm}. The
full two-time function
 $C(t,t^\prime)$ can be restored by calculating both $C$ and
$C^\ddagger$ on the domain of the {\tt herm\_matrix} type, where
$C^\ddagger$ is obtained from a second call to {\tt convolution},
\begin{align}
\label{khsabjx01}
C^\ddagger(t,t^\prime)
=
\int_\mathcal{C} d\bar t\, B^\ddagger(t,\bar t) f^\dagger(\bar t) A^\ddagger(\bar t,t^\prime).
\end{align}

\begin{table}[tbp]
\centerline{
\begin{tabular}{|l||l|l|l|}
\hline
Routine(s)
&
Input
&
Output
\\
\hline
\hline
\tt{convolution\_mat}
&
\begin{minipage}{0.35\textwidth}
~\\
$\mathcal{T}[A]_{-1},\mathcal{T}[A^\ddagger]_{-1}$, \\
$\mathcal{T}[B]_{-1},\mathcal{T}[B^\ddagger]_{-1}$, \\
$f_{-1}$\\[-2mm]
\end{minipage}
&
$\mathcal{T}[C]_{-1}$
\\
\hline
\begin{minipage}{0.33\textwidth}
  {\tt convolution\_timestep}(n)\\
  for $0 \le n \le k$ \end{minipage} & \begin{minipage}{0.35\textwidth}
  ~\\
  for $j=-1,...,k$:\\
  $\mathcal{T}[A]_{j},\mathcal{T}[A^\ddagger]_{j}$ ,\\
  $\mathcal{T}[B]_{j},\mathcal{T}[B^\ddagger]_{j}$ ,\\
  $f_{j}$\\[-2mm]
\end{minipage} & $\mathcal{T}[C]_{n}$
  \\
  \hline \begin{minipage}{0.33\textwidth} {\tt convolution\_timestep}(n)
    \\
    for $n>k$ \end{minipage} & \begin{minipage}{0.35\textwidth}
    ~\\
    for $j=-1,...,n$:\\
    $\mathcal{T}[A]_{j},\mathcal{T}[A^\ddagger]_{j}$ ,\\
    $\mathcal{T}[B]_{j},\mathcal{T}[B^\ddagger]_{j}$ ,\\
    $f_{j}$\\[-2mm]
  \end{minipage} & $\mathcal{T}[C]_{n}$
  \\
  \hline \end{tabular} } \caption{Dependencies between input and output for the Matsubara, start-up, and time-stepping routines associated with the solution of Eq.~\eqref{convolution}.}
\label{tab:convolution}
\end{table}

\section{Compiling and using \nessi
\label{sec:compile_use}}

\subsection{Main routines in \libcntr{}}

The main routines and classes in \libcntr{} are grouped under the
C++ name space {\tt cntr}. The important classes in {\tt cntr} are
summarized in Table~\ref{tab:cntr_classes}.
The main routines in the {\tt cntr} name space are presented in 
Table~\ref{tab:cntr_routines} along with a brief description. Most of
the routines have been introduced above; the remaining functions are
explained in the discussion of the example programs in
Section~\ref{sec:example_programs} and in \ref{app:utilities}.

\begin{table}[h]
  \caption{Classes grouped in the name space {\tt cntr}. \label{tab:cntr_classes}}
  \vspace*{2mm}
  \begin{small}
  \begin{tabular}{|l||l|}
    \hline
    class & purpose \\
    \hline
    \hline
  {\tt function} & \shortstack[l]{Class for representing single-time \\  functions $f(t)$ on the
                     \gls*{KB} contour.}\\
    \hline 
    {\tt herm\_matrix} & \shortstack[l]{Class for representing two-time functions
    $C(t,t^\prime)$ \\  with hermitian symmetry on the \gls*{KB} contour. }  \\
    \hline 
    {\tt herm\_matrix\_timestep} & \shortstack[l]{Class for representing a time
    slice $\mathcal{T}[G]_n$ of a \\  {\tt herm\_matrix}  at time step
    $n$.} \\
    \hline 
     {\tt herm\_matrix\_timestep\_view} & \shortstack[l]{Provides a pointer to a
    {\tt herm\_matrix\_timestep}  \\  or {\tt herm\_matrix} at a particular
                                          time step \\ without copying
                                          the data.} \\
    \hline 
    {\tt distributed\_array} & \shortstack[l]{Generic data structure for distributing
                               and \\ communicating a set of data blocks \\
                               by the
                               \gls*{mpi}.} \\
    \hline 
    {\tt distributed\_timestep\_array} & \shortstack[l]{Specialization of the {\tt
                                         distributed\_array}
                                         in \\ which data blocks are
                                         associated with \\the {\tt
                                         herm\_matrix\_timestep}
                                         objects.} \\
    \hline 
  \end{tabular}
  \end{small}
\end{table}

\begin{table}[ht!]
  \centering{}
  \caption{Functions available in the name space {\tt cntr}. \label{tab:cntr_routines}}
  \vspace*{2mm}
  \begin{small}
  \begin{tabular}{|l||l|l|}
    \hline
    class & purpose & reference section\\
    \hline
    \hline
    {\tt Bubble1} &  \shortstack[l]{Computes the bubble diagram
    \\  $C(t,t^\prime)=i A(t,t^\prime)B(t^\prime,t)$. }
          & \ref{subsec:contourconv}\\
    \hline 
   {\tt Bubble2} & \shortstack[l]{Computes the bubble diagram
    \\  $C(t,t^\prime)=i A(t,t^\prime)B(t,t^\prime)$. } & \ref{subsec:contourconv}\\
    \hline
      {\tt convolution} & \shortstack[l]{
                          Computes the convolution $C=A\ast B$ \\ in
    the full two-time plane.} & \ref{sec:impl_convolution}\\
    \hline
     {\tt convolution\_timestep} &\shortstack[l]{
                          Computes the time step $\mathcal{T}[C]_n$ of
                                   the \\ convolution $C=A\ast B$ 
    } & \ref{sec:impl_convolution}\\
    \hline
    {\tt convolution\_density\_matrix} & 
                          Computes the convolution $-i[A\ast
                                         B]^<(t,t)$.  & \ref{sec:impl_convolution}  \\
    \hline
     {\tt dyson} & \shortstack[l]{
                          Solves the Dyson equation for a given
                   \\ self-energy $\Sigma(t,t^\prime)$ in the full
                   two-time plane.}  &\ref{sec:impl_dyson}\\
    \hline
     {\tt dyson\_mat} & \shortstack[l]{Solves the Matsubara \\ Dyson
                        equation for $\mathcal{T}[G]_{-1}$.}  &\ref{subsec:impl_dyson_matsubara}\\
    \hline
    {\tt dyson\_start} &  \shortstack[l]{Solves the starting problem
                         of the \\ Dyson equation for
                         $\mathcal{T}[G]_{n}$, $n=0,\dots,k$.  }&    \ref{subsec:impl_dyson_start}\\
    \hline
     {\tt dyson\_timestep} & \shortstack[l]{Solves the Dyson equation
                         for \\ the time step
                         $\mathcal{T}[G]_{n}$. }  & \ref{subsec:impl_dyson_step} \\
    \hline
     {\tt green\_from\_H} & \shortstack[l]{Computes the free \gls*{GF}
                            $G_0(t,t^\prime)$ \\ for a given Hamiltonian
                            $\epsilon(t)$. } & \ref{sec:impl_green} \\
    \hline
     {\tt response\_convolution} &  \shortstack[l]{Computes the
                                   convolution $\int_\mathcal{C}\! d
                                   \bar{t}A(t,\bar{t}) f(\bar{t})$.  }
                    & \ref{sec:impl_convolution}\\
    \hline
     {\tt extrapolate\_timestep} & \shortstack[l]{Computes $\mathcal{T}[G]_{n+1}$ by
                                   polynomial \\ extrapolation.  }
                    & \ref{subsec:extrapolation} \\
    \hline
     {\tt correlation\_energy} & \shortstack[l]{Evaluates the
                                 Galitskii-Migdal formula
                                 \\ $E_\mathrm{corr} = 
                                 \frac{1}{2}\mathrm{Im}\mathrm{Tr}[\Sigma \ast G]^<(t,t)$.}
          &  \ref{sec:impl_convolution} \\
    \hline
    {\tt distance\_norm2} &\shortstack[l]{Computes the distance of
                            $A(t,t^\prime)$, $B(t,t^\prime)$\\ with
                            respect to the Euclidean norm \\ on the \gls*{KB}
    contour. }  & \ref{subsec:distance}  \\
    \hline
      {\tt vie2} &  \shortstack[l]{
                          Solves the VIE for given
                   \\ $F(t,t^\prime)$ and $Q(t,t^\prime)$  in the full
                   two-time plane.}& \ref{sec:impl_vie2} \\
    \hline
     {\tt vie2\_mat} & Solves the Matsubara VIE for
                       $\mathcal{T}[G]_{-1}$. & \ref{subsec:impl_vie_mat} \\
    \hline
     {\tt vie2\_start} &\shortstack[l]{Solves the starting problem of
                         \\ the VIE for $\mathcal{T}[G]_{n}$,
    $n=0,\dots,k$.} & \ref{subsec:impl_vie_start}  \\
    \hline
     {\tt vie2\_timestep} &  \shortstack[l]{Solves the VIE
                         for the time step
                         $\mathcal{T}[G]_{n}$. } & \ref{subsec:impl_vie_step} \\
    \hline
  \end{tabular}
  \end{small}
\end{table}

Furthermore, the name space {\tt integration} contains the {\tt  integrator} class, which contains all the coefficients for numerical
differentiation, interpolation and quadrature as explained in Section~\ref{sec:polynum}.

\subsection{Compilation of \libcntr{}\label{subsec:installation_libcntr}}

For compiling and installing the \libcntr{} library, we use the {\tt
  cmake} building environment \footnote{Version 2.8 or higher is required.}
to generate system specific {\tt make} files. {\tt cmake} can be
called directly from the terminal; however, it is more convenient to 
create a configure script with all compile options. We suggest the
following structure:
\begin{lstlisting}[language=bash]
CC=[C compiler] CXX=[C++ compiler] \
cmake \
    -DCMAKE_INSTALL_PREFIX=[install directory] \
    -DCMAKE_BUILD_TYPE=[Debug|Release] \
    -Domp=[ON|OFF] \
    -Dhdf5=[ON|OFF] \
    -Dmpi=[ON|OFF] \
    -DBUILD_DOC=[ON|OFF] \
    -DCMAKE_INCLUDE_PATH=[include directory] \
    -DCMAKE_LIBRARY_PATH=[library directory] \
    -DCMAKE_CXX_FLAGS="[compiling flags]" \
    ..
\end{lstlisting}
In the first line, the C and C++ compiler are set. The install
directory (for instance {\tt /home/opt}) is defined by the {\tt cmake}
variable {\tt CMAKE\_INSTALL\_PREFIX}. Debugging tools are switched on
by setting {\tt CMAKE\_BUILD\_TYPE} to {\tt Debug}; otherwise, all
assertions and sanity checks are turned off. The code is significantly
faster in {\tt Release} mode, which is recommended for production
runs. The {\tt Debug} mode, on the other hand, turns on assertions
(implemented as C++ standard assertions) of the consistency of the input
for all major routines.

The following three lines trigger optional (but recommended)
functionalities: Setting {\tt omp} to {\tt ON} turns on the
compilation of routines parallelized with {\tt openMP}, while setting
{\tt mpi} to {\tt ON} is required for compiling distributed-memory
routines based on {\tt MPI}. In this case, {\tt MPI} compilers have to
be specified in the first line. Finally, {\tt hdf5=ON} activates the
usage of the {\tt hdf5} library.

The path to the libraries that \libcntr{} depends upon ({\tt eigen3} and,
optionally, {\tt hdf5}) are provided by specifying the include
directory {\tt CMAKE\_INCLUDE\_PATH} and the library path {\tt
  CMAKE\_LIBRARY\_PATH}. Finally, the compilation flags are specified by
{\tt CMAKE\_CXX\_FLAGS}. To compile \libcntr{}, the flags should include
\begin{lstlisting}[language=bash]
-std=c++11
\end{lstlisting}

As the next step, create a build directory (for instance {\tt
  cbuild}). Navigate to this directory and run the configure script:
\begin{lstlisting}[language=bash]
sh ../configure.sh
\end{lstlisting}
After successful configuration (which generates the {\tt make} files),
compile the \mbox{library} by typing 
\begin{lstlisting}[language=bash]
make
\end{lstlisting}
and install it to the install directory by
\begin{lstlisting}[language=bash]
make install
\end{lstlisting}
After the compilation, the user can check the build by running
\begin{lstlisting}[language=bash]
make test
\end{lstlisting}
which runs a set of tests based on the {\tt catch} testing environment~\cite{noauthor_modern_2019}, checking every
functionality of \libcntr{}. After completing all test, the message
\begin{lstlisting}[language=bash]
All tests passed
\end{lstlisting}
indicates that the compiled version of \libcntr{} is fully functional.

The C++ code is 
documented using the automatic documentation tool
{\tt doxygen}. For generating the documentation, set the CMake
variable {\tt BUILD\_DOC} to {\tt ON} in the configure script. 
Running {\tt make} will then also generate an html description of many functions and classes in the {\tt doc/} directory.
A detailed and user-friendly manual is provided on the webpage {\tt www.nessi.tuxfamily.org}.

\subsection{Using \libcntr{} in custom programs}

In order to include the \libcntr{} routines in custom C++ programs,
the user needs to:
\begin{enumerate}
\item Include the declaration header by
\begin{lstlisting}[language=C++]
#include "cntr/cntr.hpp"
\end{lstlisting}
This makes available all main routines and classes in the 
C++ name space {\tt cntr}, as summarized in Table~\ref{tab:cntr_classes}.
We also offer tools for reading variables from an input file. The
respective routines can be used in a program by including
 \begin{lstlisting}[language=C++]
#include "cntr/utils/read_inputfile.hpp"
\end{lstlisting}
\item Compile the programs linking the \libcntr{} library with the
  flag {\tt -lcntr}.
\end{enumerate}
The example programs presented below in
Section~\ref{sec:example_programs} demonstrate how to integrate
\libcntr{} in custom programs. 

\subsection{HDF5 in/output}

In addition to simple input and output from and to text files (which is described in the manual on  {\tt www.nessi.tuxfamily.org}), \libcntr{} allows to use the \gls*{hdf5} to store basic data types for contour functions to disk. \gls*{hdf5} is an open source library and file format for numerical data which is widely used in the field of scientific computing. The format has two building blocks: (i) \emph{data sets}, that are general multi-dimensional arrays of a single type, and (ii) \emph{groups}, that are containers which can hold data sets and other groups. By nesting groups, it is possible to store arbitrarily complicated structured data, and to create a file-system-like hierarchy where groups can be indexed using standard POSIX format, e.g. \verb|/path/to/data|.

The \libcntr{} library comes with helper functions to store the basic contour response function data types in \gls*{hdf5} with a predefined structure of groups and data sets, defined in the header \verb|cntr/hdf5/hdf5_interface.hpp|. In particular, a \verb|herm_matrix| response function is stored as a group with a data set for each contour component \verb|mat| ($g^{\mat}(\tau)$), \verb|ret| ($g^\ret(t, t')$), \verb|les| ($g^<(t, t')$), and \verb|tv| ($g^\rceil(t, \tau)$), respectively, see Section~\ref{subsec:numrep}. The retarded and lesser components are stored in upper and lower triangular contiguous time order respectively. In the \libcntr{} \gls*{hdf5} format each component is stored as a rank 3 array where the first index is time, imaginary time, or triangular contiguous two-time, and the remaining two indices are orbital indices.

To store a contour \gls*{GF} of type \verb|cntr::herm_matrix|, one writes its components into a group of a \gls*{hdf5} file using the member function  \verb|write_to_hdf5|. 
In C++ this takes the form,
\begin{lstlisting}[language=C++]
#include <cntr/cntr.hpp>
..
// Create a contour Green's function
int nt = 200, ntau = 400, norb = 1;
GREEN A(nt, ntau, norb, FERMION);

// Open HDF5 file and write components of the Green's function A into a group g.
std::string filename = "data.h5";
A.write_to_hdf5(filename.c_str(), "g");
\end{lstlisting}

%
%
For another example of writing contour objects to file see the Holstein example program in Section~\ref{subsec:holstein}.
To understand the structure of the resulting \gls*{hdf5} file one can inspect it with the \verb|h5ls| command line program that can be used to list all groups and data sets in a \gls*{hdf5} file:
\begin{lstlisting}[language=bash]
$ h5ls -r data.h5
...
/g                       Group
/g/element_size          Dataset {1}
/g/les                   Dataset {20301, 1, 1}
/g/mat                   Dataset {401, 1, 1}
/g/nt                    Dataset {1}
/g/ntau                  Dataset {1}
/g/ret                   Dataset {20301, 1, 1}
/g/sig                   Dataset {1}
/g/size1                 Dataset {1}
/g/size2                 Dataset {1}
/g/tv                    Dataset {80601, 1, 1}
\end{lstlisting}
One can see that apart from the contour components the Green's function group \verb|g| contains additional information about the dimensions and the Fermi/Bose statistics (\verb|sig|$ = \mp 1$), for details see the API documentation of \verb|herm_matrix| and Section~\ref{subsec:numrep}. 
To understand the dimensions of the contour components we can look at the number of imaginary time steps \verb|ntau| and number of real time steps \verb|nt| using the \verb|h5dump| command line utility,
\begin{lstlisting}[language=bash]
$ h5dump -d /g/ntau data.h5
HDF5 "data.h5" {
DATASET "/g/ntau" {
   DATATYPE  H5T_STD_I32LE
   DATASPACE  SIMPLE { ( 1 ) / ( 1 ) }
   DATA {
   (0): 400
   }
}
}
$ h5dump -d /g/nt data.h5
HDF5 "data.h5" {
DATASET "/g/nt" {
   DATATYPE  H5T_STD_I32LE
   DATASPACE  SIMPLE { ( 1 ) / ( 1 ) }
   DATA {
   (0): 200
   }
}
}
\end{lstlisting}
which shows that the dimensions are $n_\tau = 400$ and $n_t=200$. The size of the \verb|/g/mat| component reveals that this corresponds to $n_\tau + 1 = 401$ imaginary time points. The mixed \verb|/g/tv| component has a slow time index and a fast imaginary time index and is of size $(n_t + 1)(n_\tau + 1) = 80601$ while the two time triangular storage of the \verb|/g/ret| and \verb|/g/les| components contains $(n_t + 1)(n_t + 2)/2 = 20301$ elements.

\begin{figure}[b]
  \centering
  \includegraphics[width=0.9\textwidth]{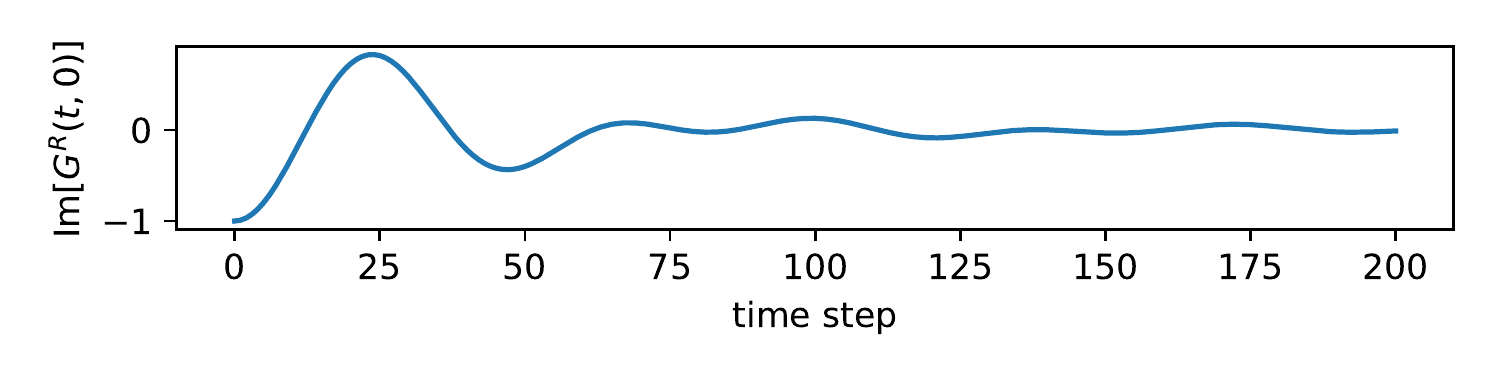}
  \caption{Result of the python example for reading and plotting $\textrm{Im}[G^{\ret}(t, t^\prime=0)]$ from a HDF5 file. Using the C++ example for generating the HDF5 file would give only zero values. Here, the result from a nontrivial time-dependent calculation is shown instead.
  \label{fig:g_ret_python}}
\end{figure}

To simplify postprocessing of contour \gls*{GFs}, \nessi{} also provides the python module  \verb|ReadCNTRhdf5.py| for reading the \gls*{hdf5} format (using the python modules \verb|numpy| and \verb|h5py|) producing python objects with the contour components as members.
The python module unrolls the triangular storage of the \verb|ret| and \verb|les| components making it simple to plot time slices. To plot the imaginary part of the retarded Green's function  $\textrm{Im}[G^{\ret}(t, t^\prime=0)]$ as a function of $t$ we may use the commands
\begin{lstlisting}[language=bash]
import h5py
from ReadCNTRhdf5 import read_group

with h5py.File('data.h5', 'r') as fd:
    g = read_group(fd).g
    
import matplotlib.pyplot as plt
plt.figure(figsize=(6., 1.5))

plt.plot(g.ret[:, 0, 0, 0].imag)

plt.xlabel(r'time step')
plt.ylabel(r'Im$[G^{R}(t, 0)]$')

plt.tight_layout(); plt.savefig('figure_g_ret.pdf')
\end{lstlisting}
which produce the plot shown in Fig.\ \ref{fig:g_ret_python}.
More advanced usage of the \gls*{hdf5} interface is exemplified in the example programs, and in the online manual.

\section{Example programs\label{sec:example_programs}}

In this section, we present a number of examples of how the described
routines can be used to solve typical nonequilibrium problems.
 It is assumed that the \libcntr{} library has been compiled and
installed. Furthermore, we assume that the collection of demonstration programs {\tt nessi\_demo} has been installed to some directory {\tt nessi\_demo/}
and compiled in, for instance, {\tt nessi\_demo/build/}. Detailed building
instructions can be found in \ref{app:install_nessidemo}.
Further examples can be found in the online manual, where one can also find a more detailed description of all member functions and simple helper routines (such as, e.g., adding up Green's functions, scalar multiplication, etc.).

\subsection{Test of accuracy and scaling analysis}
\label{sec:example-scaling}
\paragraph{Overview} ---
The first example both serves as a minimal application of the vie2 equation (without much physical significance), and at the same time it demonstrates  the convergence  of the methods described in Section~\ref{sec:num_vie} with the time discretization. We consider
a $2\times 2$ matrix-valued time-independent Hamiltonian
\begin{align}
  \label{eq:ham2x2}
  \epsilon = \begin{pmatrix} \epsilon_1 & i \lambda \\ -i\lambda &
    \epsilon_2 \end{pmatrix}  \ .
\end{align}
The corresponding numerically exact \gls*{GF} $G(t,t^\prime)$ (assuming fermions) is computed using the
routine {\tt green\_from\_H} mentioned in
Section~\ref{sec:impl_green_from_H_part_I}. 
Alternatively, one can compute the (1,1) component of the \gls*{GF}
by \emph{downfolding}: To this end, we solve
\begin{align}
  \label{eq:downfold1}
  \left(i \partial_t - \epsilon_1 \right) g_1(t,t^\prime)  =
  \delta_\mathcal{C}(t,t^\prime)
  + \int_{\mathcal{C}}d \bar{t}\, \Sigma(t,\bar t) g_1(\bar t,t^\prime)
\end{align}
with the embedding self-energy $\Sigma(t,t^\prime) = |\lambda|^2
g_2(t,t^\prime)$. Here, $g_2(t,t^\prime)$ is the free \gls*{GF} with respect
to $\epsilon_2$, 
\begin{align}
   \left(i \partial_t - \epsilon_2 \right) g_2(t,t^\prime)  =
  \delta_\mathcal{C}(t,t^\prime) \ .
\end{align}
The solution of the
Dyson equation~\eqref{eq:downfold1} then must be
identical to the $(1,1)$ matrix element of $G$:
$G_{1,1}(t,t^\prime)=g_1(t,t^\prime)$. The test programs {\tt
  test\_equilibrium.x} and {\tt test\_nonequilibrium.x} solve this problem in
equilibrium and nonequilibrium, respectively, and compare the error.
In the equilibrium case, we define
\begin{align}
  \label{eq:test_eq_err}
  \mathrm{err.} = \frac{1}{\beta} \int^\beta_0\! d\tau\,
  |G_{1,1}(\tau)-g_1(\tau)| \ ,
\end{align}
whereas
\begin{align}
  \label{eq:test_noneq_err}
  \mathrm{err.} &= \frac{1}{T^2} \int^T_0\! d t \!\int^t_0\! d t^\prime
                  |G^<_{1,1}(t^\prime,t)-g^<_1(t^\prime,t)| \nonumber
  \\ &\quad +
      \frac{1}{T^2} \int^T_0\! d t \!\int^t_0\! d t^\prime
       |G^\ret_{1,1}(t,t^\prime)-g^\ret_1(t,t^\prime)| \\ &\quad
            \frac{1}{T \beta} \int^T_0\! d t \!\int^\beta_0\! d \tau
       |G^\rceil_{1,1}(t,\tau)-g^\rceil_1(t,\tau)|                                                
\end{align}
for the nonequilibrium case. 

\paragraph{Implementation: Equilibrium} --- The implementation of the equilibrium solution of the example is found in \textcolor{blue}{\tt programs/test\_equilibrium.cpp}. 
In the following we summarize and explain the main parts:

In \libcntr{}, we define the following short-hand types
\begin{lstlisting}[language=C++]
  #define GREEN cntr::herm_matrix<double>
  #define GREEN_TSTP cntr::herm_matrix<double>
  #define CFUNC cntr::function<double>
\end{lstlisting}
for double-precision objects. They are available to any program including {\tt cntr.hpp}. In the main part of the C++ program,
the parameters of the Hamiltonian are defined as constants. In
particular, we fix $\epsilon_1=-1$, $\epsilon_2=1$, $\lambda=0.5$. The
chemical potential is set to $\mu=0$ and the inverse temperature fixed
to $\beta=20$. The input variables read from file are {\tt Ntau}
($N_\tau$) and {\tt SolveOrder} ($k=1,\dots,5$). After reading these
variables from file via
\begin{lstlisting}[language=C++]
  find_param(argv[1],"__Ntau=",Ntau);
  find_param(argv[1],"__SolveOrder=",SolveOrder);
\end{lstlisting}
we can define all quantities. 
First we define the Hamiltonian~\eqref{eq:ham2x2}
as an {\tt eigen3} complex matrix:
\begin{lstlisting}[language=C++]
  cdmatrix eps_2x2(2,2);
  eps_2x2(0,0) = eps1;
  eps_2x2(1,1) = eps2;
  eps_2x2(0,1) = I*lam;
  eps_2x2(1,0) = -I*lam;
\end{lstlisting}
The $1\times 1$ Hamiltonian representing $\epsilon_1$ is constructed as
\begin{lstlisting}[language=C++]
  CFUNC eps_11_func(-1,1);
  eps_11_func.set_constant(eps1*MatrixXcd::Identity(1,1));
\end{lstlisting}
Here, {\tt eps\_11\_func} is a contour function entering the solvers
below. Note the first argument in the constructor of {\tt CFUNC}: the number of real-time
points $N_t$ is set to $-1$. In this case, only the Matsubara part is
addressed.  Its value is fixed to the constant $1\times 1$ matrix by
the last line. With the Hamiltonians defined, we can initialize and
construct the free $2\times 2$ exact \gls*{GF} by
\begin{lstlisting}[language=C++]
  GREEN G2x2(-1,Ntau,2,FERMION);
  cntr::green_from_H(G2x2,mu,eps_2x2,beta,h);
\end{lstlisting}
Including the \libcntr{} header provides a number of constants for
convenience; here, we have used {\tt FERMION=-1} (bosons would be
described by  {\tt BOSON=+1}).
The time step {\tt h} is a dummy argument here, as the real-time
components are not addressed. From the exact \gls*{GF}, we extract the
submatrix $G_{1,1}$ by
\begin{lstlisting}[language=C++]
  GREEN G_exact(-1,Ntau,1,FERMION);
  G_exact.set_matrixelement(-1,0,0,G2x2);
\end{lstlisting}
Finally, we define the embedding self-energy by
\begin{lstlisting}[language=C++]
  GREEN Sigma(-1,Ntau,1,FERMION);
  cdmatrix eps_22=eps2*MatrixXcd::Identity(1,1);
  cntr::green_from_H(Sigma, mu, eps_22, beta, h);
  Sigma.smul(-1,lam*lam);
\end{lstlisting}
The last line performs the multiplication of $\mathcal{T}[\Sigma]_{-1}$
with the scalar $\lambda^2$. After initializing the approximate \gls*{GF} {\tt
  G\_approx}, we can solve the Matsubara Dyson equation and compute
the average error:
\begin{lstlisting}[language=C++]
  cntr::dyson_mat(G_approx, Sigma, mu, eps_11_func, beta, SolveOrder, CNTR_MAT_FOURIER);
  err_fourier = cntr::distance_norm2(-1,G_exact,G_approx) / Ntau;

  cntr::dyson_mat(G_approx, Sigma, mu, eps_11_func, beta,  SolveOrder, CNTR_MAT_FIXPOINT);
  err_fixpoint = cntr::distance_norm2(-1,G_exact,G_approx) / Ntau;
\end{lstlisting}
The error is then written to file. The function {\tt distance\_norm2} measures the Euclidean distance of two contour functions, as explained in \ref{subsec:distance}.

\begin{figure}[ht]
  \centering
  \includegraphics[width=\textwidth]{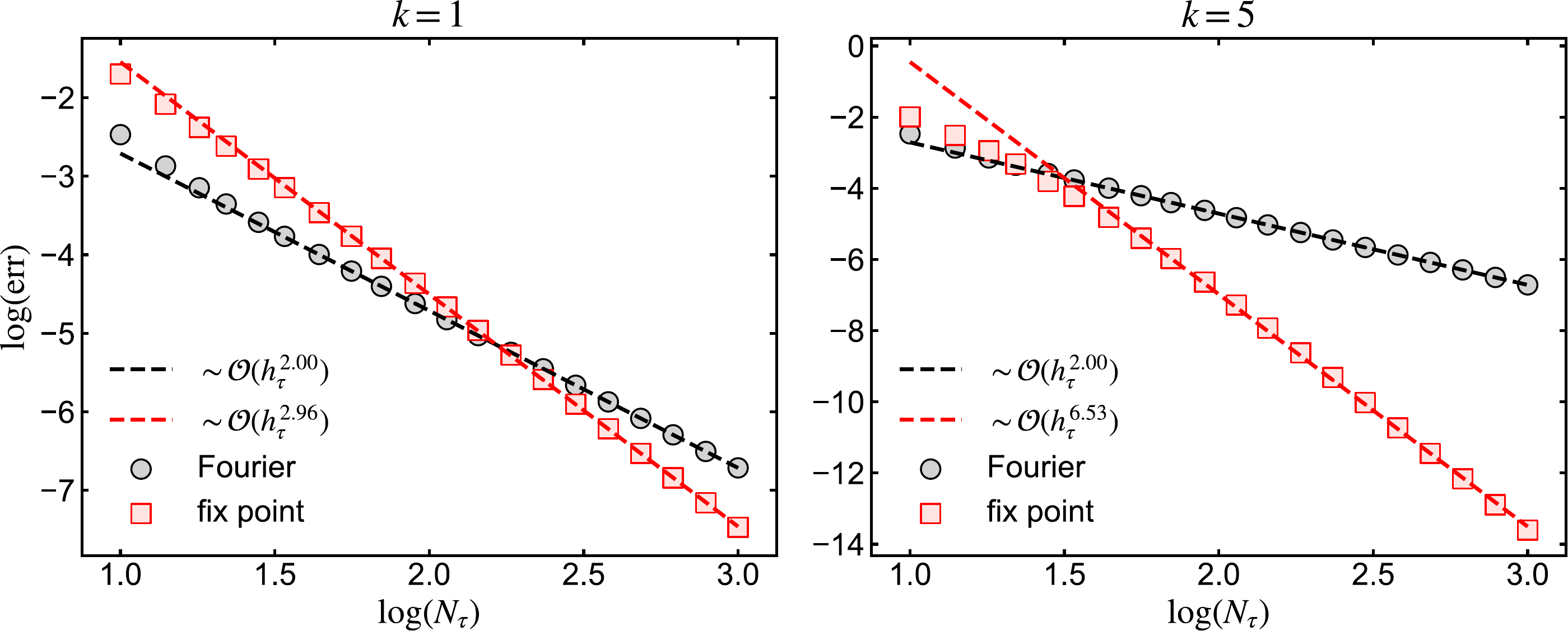}
  \caption{Average error according to Eq.~\eqref{eq:test_eq_err} for
    $\epsilon_1=-1$, $\epsilon_2=1$, $\lambda=0.5$, $\mu=0$,
    $\beta=20$ for $k=1$ and $k=5$. \label{fig:test_eq}}
\end{figure}

\paragraph{Running and output: Equilibrium} --- For convenience, we provide a
driver {\tt python3} script for creating the input file, running the
program and plotting the results. For running the equilibrium test, go
to {\tt nessi\_demo/} and run
\begin{lstlisting}[language=sh]
python3 utils/test_equilibrium.py k
\end{lstlisting}
where {\tt k=1,\dots,5} is the integration order. The test solves the
Matsubara Dyson equation for $N_\tau=10^x$ for 20 values of $x\in
[1,3]$. The results are plotted using {\tt
  matplotlib}. Figure~\ref{fig:test_eq} shows the corresponding plots
for $k=1$ and $k=5$. As Fig.~\ref{fig:test_eq} demonstrates, the
Fourier method described in subsection~\ref{subsec:impl_dyson_matsubara} scales
as $\mathcal{O}(h^{2}_\tau)$, while solving the Dyson equation in
integral form results approximately in a $\mathcal{O}(h^{k+2}_\tau)$ scaling of the
average error for small
enough $h_\tau$.

\paragraph{Implementation: Nonequilibrium} --- Testing the accuracy of the {\tt dyson} and {\tt vie2} solvers can be
done analogous to the equilibrium case above. The source code which is described below can be found in \textcolor{blue}{\tt programs/test\_nonequilibrium.cpp:}

We adopt the same parameters as for the
equilibrium case. To obtain the \gls*{NEGFs},  the Dyson
equation~\eqref{eq:downfold1} is propagated in time. Equivalently, one
can also solve the Dyson equation in integral form
\begin{subequations}
  \label{eq:downfold1_vie}
  \begin{equation}
  g_1(t,t^\prime) + [F\ast g_1](t,t^\prime) = g^{(0)}_1(t,t^\prime) \ ,
\end{equation}
  \begin{equation}
 g_1(t,t^\prime) +  [g_1\ast F^\ddagger](t,t^\prime) = g^{(0)}_1(t,t^\prime) \ ,
\end{equation}
\end{subequations}
where $F= -\Sigma\ast g_1^{(0)} $ and
$F^\ddagger=-g_1^{(0)}\ast \Sigma$, as explained in subsection~\ref{secygxeqw02}. The free \gls*{GF} $g_1^{(0)}(t,t^\prime)$ is
known analytically and computed by calling the routine {\tt
  green\_from\_H}.

The structure of the test program is analogous to the equilibrium
case. First, the input variables $N_t$, $N_\tau$, $T_\mathrm{max}$ and
$k$ are read from the input file:
\begin{lstlisting}[language=C++]
  find_param(flin,"__Nt=",Nt);
  find_param(flin,"__Ntau=",Ntau);
  find_param(flin,"__Tmax=",Tmax);
  find_param(flin,"__SolveOrder=",SolveOrder);
\end{lstlisting}
The time step is fixed by $h=T_\mathrm{max}/N_t$.
After initializing the Hamiltonian and the \gls*{GFs}, the embedding
self-energy is construced via
\begin{lstlisting}[language=C++]
  cntr::green_from_H(Sigma, mu, eps_22, beta, h);
  for(tstp=-1; tstp<=Nt; tstp++) {
    Sigma.smul(tstp,lam*lam);
  }
\end{lstlisting}
The generic procedure to solve a Dyson equation in the time domain
in \libcntr{} is
\begin{enumerate}
  \item Solve the equilibrium problem by solving the corresponding   Matsubara Dyson equation,
   \item Compute the \gls*{NEGFs} for time steps $n=0,\dots, k$
     by using the starting algorithm (bootstrapping), and
   \item Perform the time stepping for $n=k+1,\dots,N_t$.
\end{enumerate}
For Eq.~\eqref{eq:downfold1}, this task is accomplished by
\begin{lstlisting}[language=C++]
  GREEN G_approx(Nt, Ntau, 1, FERMION);

  // equilibrium
  cntr::dyson_mat(G_approx, mu, eps_11_func, Sigma, beta, SolveOrder);

  // start
  cntr::dyson_start(G_approx, mu, eps_11_func, Sigma, beta, h, SolveOrder);

  // time stepping
  for (tstp=SolverOrder+1; tstp<=Nt; tstp++) {
     cntr::dyson_timestep(tstp, G_approx, mu, eps_11_func, Sigma, beta, h, SolveOrder);
  }
\end{lstlisting}
The deviation of the nonequilbrium Keldysh components from the exact
solution is then calculated by
\begin{lstlisting}[language=C++]
  err_dyson=0.0;
  for(tstp=0; tstp<=Nt; tstp++){
    err_dyson += cntr::distance_norm2_les(tstp, G_exact, G_approx) / (Nt*Nt);
    err_dyson += cntr::distance_norm2_ret(tstp, G_exact, G_approx) / (Nt*Nt);
    err_dyson += cntr::distance_norm2_tv(tstp, G_exact, G_approx) / (Nt*Ntau);
  }
\end{lstlisting}

The solution of the corresponding integral
formulation~\eqref{eq:downfold1_vie} is peformed by the following
lines of source code:
\begin{lstlisting}[language=C++]
  // noninteracting 1x1 Greens function (Sigma=0)
  GREEN G0(Nt,Ntau,1,FERMION);
  cdmatrix eps_11=eps1*MatrixXcd::Identity(1,1);
  cntr::green_from_H(G0, mu, eps_11, beta, h);

  GREEN G_approx(Nt,Ntau,1,FERMION);
  GREEN F(Nt,Ntau,1,FERMION);
  GREEN Fcc(Nt,Ntau,1,FERMION);

  // equilibrium
  GenKernel(-1, G0, Sigma, F, Fcc, beta, h, SolverOrder);
  cntr::vie2_mat(G_approx, F, Fcc, G0, beta, SolverOrder);

  // start
  for(tstp=0; tstp <= SolveOrder; tstp++){
    GenKernel(tstp, G0, Sigma, F, Fcc, beta, h, SolverOrder);
  }
  cntr::vie2_start(G_approx, F, Fcc, G0, beta, h, SolveOrder);

  // time stepping
  for (tstp=SolveOrder+1; tstp<=Nt; tstp++) {
    GenKernel(tstp, G0, Sigma, F, Fcc, beta, h, SolverOrder);
    cntr::vie2_timestep(tstp, G_approx, F, Fcc, G0, beta, h, SolveOrder);
  }
\end{lstlisting}
For convenience, we have defined the routine {\tt GenKernel}, which
calculates the convolution kernels $F$ and $F^\ddagger$:
\begin{lstlisting}[language=C++]
  void GenKernel(int tstp, GREEN &G0, GREEN &Sigma, GREEN &F, GREEN &Fcc, const double beta, const double h, const int SolveOrder){
    cntr::convolution_timestep(tstp, F, G0, Sigma, beta, h, SolveOrder);
    cntr::convolution_timestep(tstp, Fcc, Sigma, G0, beta, h, SolveOrder);
    F.smul(tstp,-1);
    Fcc.smul(tstp,-1);
  }
\end{lstlisting}

\begin{figure}[ht]
  \centering
  \includegraphics[width=\textwidth]{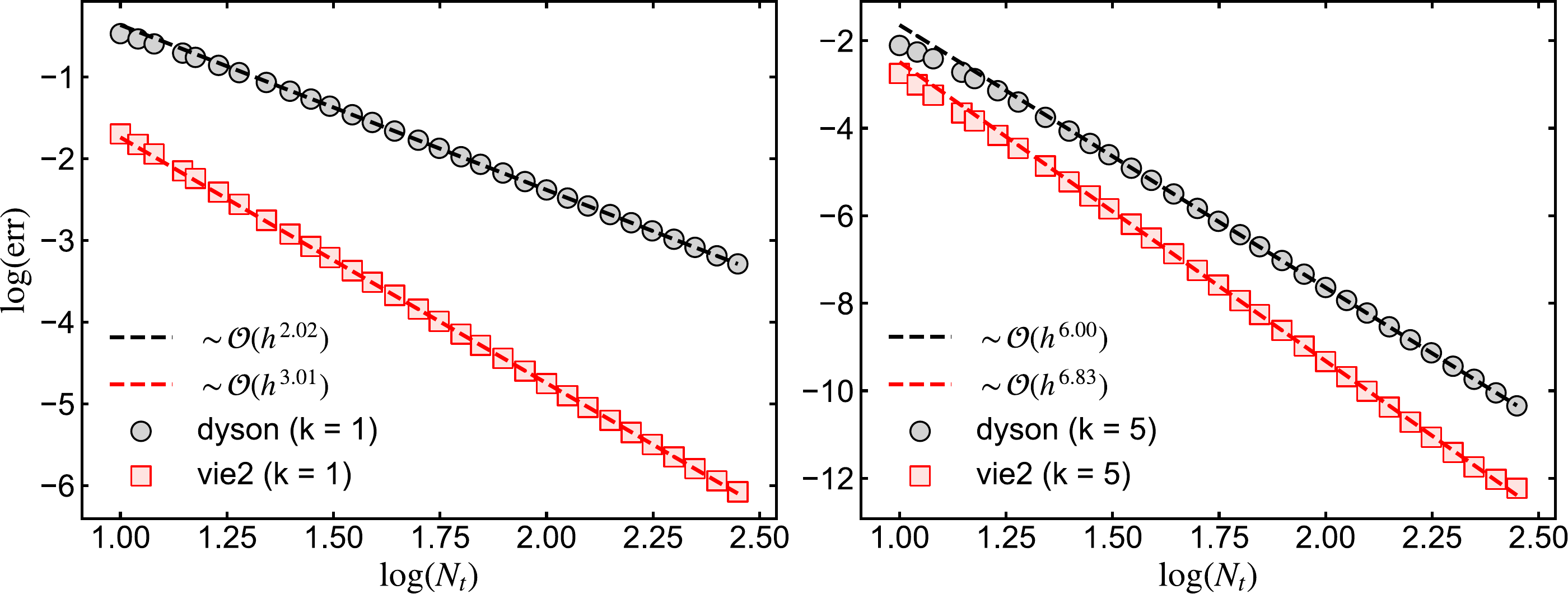}
  \caption{Average error according to Eq.~\eqref{eq:test_noneq_err} for
    $\epsilon_1=-1$, $\epsilon_2=1$, $\lambda=0.5$, $\mu=0$,
    $\beta=20$ for $k=1$ and $k=5$. We have fixed $N_\tau=800$ and
    $T_\mathrm{max}=5$.
    \label{fig:test_noneq}}
\end{figure}

\paragraph{Running and output: Nonequilibrium} --- The {\tt python3} driver script {\tt test\_nonequilibrium.py} provides 
an easy-to-use interface for running the accuracy test. In the  {\tt nessi\_demo/} directory, run
\begin{lstlisting}[language=sh]
python3 utils/test_nonequilbrium.py k
\end{lstlisting}
where {\tt k} is the solution order.
The average error of the numerical solution of
Eq.~\eqref{eq:downfold1_vie} is computed analogously to the Dyson
equation in integro-differential form. 
The output of {\tt test\_nonequilibrium.py} is shown in Fig.~\ref{fig:test_noneq}.
As this figure confirms, the average error of solving the
Dyson equation in the integro-differential form scales as
$\mathcal{O}(h^{k+1})$, while the corresponding integral form
yields a $\mathcal{O}(h^{k+2})$ scaling.

\subsection{Hubbard chain}\label{Sec:Chain}

\paragraph{Overview} --- The Hubbard model is one of the most basic models describing correlation
effects. It allows to demonstrate the performance, strengths and
also shortcomings of the \gls*{NEGF}
treatment~\cite{von_friesen_successes_2009,
  puig_von_friesen_kadanoff-baym_2010,schlunzen_nonequilibrium_2016}. Here,
we consider a \gls*{1D} 
configuration with the Hamiltonian
\begin{align}
  \label{eq:hubbmodel1}
  \hat{H}_0 = -J \sum_{\langle i,j \rangle, \sigma}
  \hat{c}^\dagger_{i\sigma} \hat{c}_{j \sigma} + U\sum_{i}
  (\hat{n}_{i\uparrow}-\bar{n}) (\hat{n}_{i\downarrow}-\bar{n}) \ ,
\end{align}
where $\langle i,j\rangle$ constrains the lattice sites $i,j$ to 
nearest neighbors, while $\bar\sigma=\uparrow,\downarrow$ for
$\sigma=\downarrow,\uparrow$. We consider $M$ lattice sites 
with open boundary conditions. Furthermore,
we restrict ourselves to the paramagnetic case with an equal number of
spin-up ($N_\uparrow$) and spin-down ($N_\downarrow$) particles. The number of particles determines the filling factor $\bar{n}=N_\uparrow/M$. Note that the Hamiltonian~\eqref{eq:hubbmodel1} contains a chemical potential shift, such that $\mu=0$ corresponds to filling $\bar{n}$.
In analogy to Ref.~\cite{puig_von_friesen_kadanoff-baym_2010}, the
system is excited with an instantaneous quench of the on-site potential
of the first lattice site to $w_0$:
\begin{align}
  \label{eq:hubbchain_quench}
  \hat{H}(t) = \hat{H}_0 + \theta(t) w_0 \sum_\sigma
  \hat{c}^\dagger_{1\sigma} \hat{c}_{1\sigma}  \ .
\end{align}
In this example, we treat the dynamics with respect to the
Hamiltonian~\eqref{eq:hubbchain_quench} within the \gls*{2B},
$GW$, and $T$-matrix (particle-particle ladder) approximations. 
A detailed description 
of these approximations can be found, for
instance, in Ref.~\cite{puig_von_friesen_kadanoff-baym_2010}. The
numerical representation of the respective self-energy expressions is
implemented in the C++ module {\tt
  hubbard\_chain\_selfen\_impl.cpp}. Below we explain the key
routines. 

\paragraph{Self-energy approximation: second-Born}---
The \gls*{2B} approximation corresponds to the second-order expansion in
terms of the Hubbard repulsion $U(t)$, which we treat here as time
dependent for generality. Defining the \gls*{GF} with respect to
the lattice basis, $G_{ij,\sigma}(t,t^\prime)=-i\langle
T_\mathcal{C}\hat{c}_{i\sigma}(t)
\hat{c}^\dagger_{j\sigma}(t^\prime)\rangle$,  the \gls*{2B} is defined by
\begin{align}
  \label{eq:sigma_hubb_2b}
  \Sigma_{ij,\sigma}(t,t^\prime) = U(t)U(t^\prime) G_{ij,\sigma}(t,t^\prime)
  G_{ij,\bar\sigma}(t,t^\prime)G_{ji,\bar\sigma}(t^\prime,t) \ .
\end{align}
The \gls*{2B} self-energy~\eqref{eq:sigma_hubb_2b} is implemented in two
steps. (i) The (spin-dependent) polarization $P_{ij,\sigma}(t,t^\prime)= -i
G_{ij,\sigma}(t,t^\prime)G_{ji,\sigma}(t^\prime,t) $ is computed using
the routine {\tt Bubble1} and subsequently multiplied by $-1$. (ii) The self-energy is then given by
$\Sigma_{ij,\sigma}(t,t^\prime)  = i U(t) U(t^\prime)
G_{ij,\sigma}(t,t^\prime) P_{ij,\bar\sigma}(t,t^\prime) $, which
corresponds to a bubble diagram computed by the routine {\tt Bubble2}. Inspecting
the Keldysh components of the \gls*{GFs}, one notices 
that the polarization $P_{ij,\sigma}(t,t^\prime)$ is needed on one time
slice only. As $G_{ij,\uparrow}(t,t^\prime)=G_{ij,\downarrow}(t,t^\prime)\equiv G_{ij}(t,t^\prime)$ (an analogous statement holds for other contour functions), the spin index can be dropped.
The \gls*{2B} self-energy is
computed by the routine {\tt Sigma\_2B} as follows:
\begin{lstlisting}[language=C++]
 void Sigma_2B(int tstp, GREEN &G, CFUNC &U, GREEN &Sigma){
    int nsites=G.size1();
    int ntau=G.ntau();
    GREEN_TSTP Pol(tstp,ntau,nsites,BOSON);

    Polarization(tstp, G, Pol);

    Pol.right_multiply(tstp, U);
    Pol.left_multiply(tstp, U);

    for(int i=0; i<nsites; i++){
      for(int j=0; j<nsites; j++){
      	cntr::Bubble2(tstp,Sigma,i,j,G,i,j,Pol,i,j);
      }
    }

 }
\end{lstlisting}
First, the polarization {\tt Pol}, which represents
$P_{ij}(t,t^\prime)$, is defined for the given
time step. After computing $P_{ij}(t,t^\prime)$ by the function
\begin{lstlisting}[language=C++]
  void Polarization(int tstp, GREEN &G, GREEN_TSTP &Pol){
    int nsites=G.size1();

    for(int i=0; i<nsites; i++){
      for(int j=0; j<nsites; j++){
        cntr::Bubble1(tstp,Pol,i,j,G,i,j,G,i,j);
      }
    }
    Pol.smul(-1.0);
  }
\end{lstlisting}
the lines
\begin{lstlisting}[language=C++]
Pol.right_multiply(tstp, U);
Pol.left_multiply(tstp, U);
\end{lstlisting}
perform the operation $P_{ij}(t,t^\prime) \rightarrow
P_{ij}(t,t^\prime) U(t^\prime)$ and $P_{ij}(t,t^\prime) \rightarrow
U(t)P_{ij}(t,t^\prime) $, respectively. Finally, {\tt Bubble2}
computes $\Sigma_{ij}(t,t^\prime)$. 

\paragraph{Self-energy approximation: $GW$} --- As the next
approximation to the self-energy, we consider the $GW$
approximation. We remark that we formally treat the Hubbard
interaction as spin-independent (as in
Ref.~\cite{puig_von_friesen_kadanoff-baym_2010}), while the
spin-summation in the polarization $P$ (which is forbidden by the Pauli
principle) is excluded by the corresponding prefactor. The analogous
approximation for the explicitly spin-dependent interaction (spin-$GW$)
is also discussed in Ref.~\cite{puig_von_friesen_kadanoff-baym_2010}.

Within the same setup as above, the $GW$ approximation is defined by
\begin{align}
\label{eq:SigmaGW}
  \Sigma_{ij}(t,t^\prime) = i G_{ij}(t,t^\prime) \delta
  W_{ij}(t,t^\prime) \ ,
\end{align}
where $\delta W_{ij}(t,t^\prime)$ denotes the dynamical part of the
screened interaction $W_{ij}(t,t^\prime) = U
\delta_{ij}\delta_\mathcal{C}(t,t^\prime) + \delta
W_{ij}(t,t^\prime)$. We compute $\delta W_{ij}(t,t^\prime)$ from the
charge susceptibility $\chi_{ij}(t,t^\prime)$  by $\delta
W_{ij}(t,t^\prime) = U(t) \chi_{ij}(t,t^\prime) U(t^\prime)$. 
This susceptibility obeys the Dyson equation 
\begin{align}
  \label{eq:dyson_chi}
  \chi = P + P\ast U \ast \chi \ ,
\end{align}
where $P$ stands for the irreducible polarization
$P_{ij}(t,t^\prime)=-i G_{ij}(t,t^\prime)G_{ji}(t^\prime,t)$. The
strategy to compute the $GW$ self-energy with \libcntr{} thus
consists of three steps:
\begin{enumerate}
  \item Computing the polarization $P_{ij}(t,t^\prime)$ by {\tt Bubble1}.
  \item Solving the Dyson equation~\eqref{eq:dyson_chi} as VIE. By defining the kernel $K_{ij}(t, t^\prime)=-P_{ij}(t,t^\prime)U(t^\prime)$
    and its hermitian conjugate, Eq.~\eqref{eq:dyson_chi} amounts to
    $[1+K]\ast \chi=P$, which is solved for $\chi$ using {\tt vie2}.
  \item Computing the self-energy~\eqref{eq:SigmaGW} by {\tt Bubble2}.
\end{enumerate}
The implementation of step 1 has been discussed above.
For step 2, we distinguish between the equilibrium (timestep {\tt tstp=-1}) and time stepping
on the one hand, and the starting phase on the other hand. For the
former, we have defined the routine
\begin{lstlisting}[language=C++]
  void GenChi(int tstp, double h, double beta, GREEN &Pol, 
  CFUNC &U, GREEN &PxU, GREEN &UxP, GREEN &Chi, int SolveOrder){

    PxU.set_timestep(tstp, Pol);
    UxP.set_timestep(tstp, Pol);
    PxU.right_multiply(tstp, U);
    UxP.left_multiply(tstp, U);
    PxU.smul(tstp,-1.0);
    UxP.smul(tstp,-1.0);

    if(tstp==-1){
      cntr::vie2_mat(Chi,PxU,UxP,Pol,beta,SolveOrder);
    } else{
      cntr::vie2_timestep(tstp,Chi,PxU,UxP,Pol,beta,h,SolveOrder);
    }
  }
\end{lstlisting}
Here, {\tt PxU} and {\tt UxP} correspond to the kernel $K_{ij}$ and
its hermitian conjugate, respectively. Analogously, the starting
routine is implemented as
\begin{lstlisting}[language=C++]
  void GenChi(double h, double beta, GREEN &Pol, CFUNC &U,
		  GREEN &PxU, GREEN &UxP, GREEN &Chi, int SolveOrder){

    for(int n = 0; n <= SolveOrder; n++){
      PxU.set_timestep(n, Pol);
      UxP.set_timestep(n, Pol);
      PxU.right_multiply(n, U);
      UxP.left_multiply(n, U);
      PxU.smul(n,-1.0);
      UxP.smul(n,-1.0);
    }

    cntr::vie2_start(Chi,PxU,UxP,Pol,beta,h,SolveOrder);

  }
\end{lstlisting}
Finally, the self-energy is computed by
\begin{lstlisting}[language=C++]
  void Sigma_GW(int tstp, GREEN &G, CFUNC &U, GREEN &Chi, GREEN &Sigma){
    int nsites=G.size1();
    int ntau=G.ntau();
    GREEN_TSTP deltaW(tstp,ntau,nsites,BOSON);

    Chi.get_timestep(tstp,deltaW);
    deltaW.left_multiply(tstp,U);
    deltaW.right_multiply(tstp,U);

    for(int i=0; i<nsites; i++){
      for(int j=0; j<nsites; j++){
	       cntr::Bubble2(tstp,Sigma,i,j,G,i,j,deltaW,i,j);
      }
    }
  }
\end{lstlisting}

\paragraph{Self-energy approximation: $T$-matrix} ---
The particle-particle ladder $T_{ij}(t,t^\prime)$ 
represents 
an effective particle-particle interaction, which defines the
corresponding self-energy 
\begin{align}
  \Sigma_{ij}(t,t^\prime) = i U(t) T_{ij}(t,t^\prime) U(t^\prime)
  G_{ji}(t^\prime,t) \ .
\end{align}
The $T$-matrix, in turn, is obtained by solving the Dyson equation
$T=\Phi - \Phi \ast U \ast T$, where $\Phi$ corresponds to the
particle-particle bubble $\Phi_{ij}(t,t^\prime) = -i G_{ij}(t,t^\prime)
G_{ij}(t,t^\prime) $. Hence, the procedure of numerically computing
the $\Sigma_{ij}(t,t^\prime)$ is analogous to the $GW$ approximation:
\begin{enumerate}
\item Compute $\Phi_{ij}(t,t^\prime) $ by {\tt Bubble2} and multiply
  by $-1$.
 \item Calculate the kernel $K_{ij}(t,t^\prime) =
   \Phi_{ij}(t,t^\prime)  U(t^\prime) $ and its hermitian conjugate
   and solve the VIE $[1+K]\ast T = \Phi$ for $T$ by
   {\tt vie2}.
 \item Perform the operation $T_{ij}(t,t^\prime) \rightarrow U(t)
   T_{ij}(t,t^\prime) U(t^\prime)$ and compute the self-energy by {\tt
   Bubble1}.
\end{enumerate}

\paragraph{Mean-field Hamiltonian and onsite quench} --- So far, we
have described how to compute the dynamical 
contribution to the self-energy. The total self-energy furthermore
includes the \gls*{HF} contribution, which we incorporate into the
mean-field Hamiltonian
$\epsilon^{\mathrm{MF}}_{ij}(t) = \epsilon^{(0)}_{ij}(t) + U (n_i-\bar{n}) $ with the occupation
(per spin) $n_i= \langle \hat{c}^\dagger_i \hat{c}_i \rangle$. The shift of chemical potential $-U \bar{n}$ is a convention to fix the chemical potential at half filling at $\mu=0$.
Note that 
the Fock term is zero because of the spin symmetry. In the
example program, the mean-field Hamiltonian is represented by the
contour function {\tt eps\_mf}. Updating {\tt eps\_mf} is accomplished by
computing the density matrix using the {\tt herm\_matrix} class routine
{\tt density\_matrix}. 

The general procedure to implement a quench of some parameter $\lambda$ at $t=0$
is to represent $\lambda$ by a contour function $\lambda_n$:
$\lambda_{-1}$ corresponds to the pre-quench value which determines the
thermal equilibrium, while $\lambda_n$ with $n\ge 0$ governs the
time evolution. In the example program, we simplify this procedure by
redefining $\epsilon^{(0)}_{ij} \rightarrow \epsilon^{(0)}_{ij} + w_0
\delta_{i,1}\delta_{j,1}$ after the Matsubara Dyson equation has been
solved.

\paragraph{Generic structure of the example program} ---

The source code 
for the \gls*{2B}, $GW$ and $T$-matrix approximation,
is found in \textcolor{blue}{\tt programs/hubbard\_chain\_2b.cpp,
}\textcolor{blue}{\tt programs/hubbard\_chain\_gw.cpp,}
\newline \textcolor{blue}{\tt programs/hubbard\_chain\_selfen\_impl.cpp}, respectively.
The programs are structured similarly as the previous examples. After
reading variables from file and initializing the variables and
classes, the Matsubara Dyson equation is solved in a self-consistent
fashion. The example below illustrates this procedure for the \gls*{2B} approximation.
\begin{lstlisting}[language=C++]
  tstp=-1;
  gtemp = GREEN(SolveOrder,Ntau,Nsites,FERMION); 
  gtemp.set_timestep(tstp,G);

  for(int iter=0;iter<=MatsMaxIter;iter++){
    // update mean field
    hubb::Ham_MF(tstp, G, Ut, eps0, eps_mf);

    // update self-energy
    hubb::Sigma_2B(tstp, G, Ut, Sigma);

    // solve Dyson equation
    cntr::dyson_mat(G, MuChem, eps_mf, Sigma, beta, SolveOrder);

    // self-consistency check
    err = cntr::distance_norm2(tstp,G,gtemp);

    if(err<MatsMaxErr){
       break;
    }
    gtemp.set_timestep(tstp,G);
  }
\end{lstlisting}
Updating the mean-field Hamiltonian ({\tt hubb::Ham\_MF}), the
self-energy ({\tt hubb::Sigma\_2B}) and solving the corresponding Dyson
equation ({\tt cntr::dyson\_mat}) is repeated until self-consistency
has been reached, which in practice means that the deviation between the previous and updated
\gls*{GF} is smaller than the given number {\tt MatsMaxErr}. For other self-energy
approximations, the steps described above (updating auxiliary
quantities) have to be performed before the self-energy can
be updated.

Once the Matsubara Dyson equation has been solved up to the required
convergence threshold, the starting algorithm for time steps
$n=0,\dots,k$ can be applied. To reach self-consistency for the first
few time steps, we employ the bootstrapping loop:
\begin{lstlisting}[language=C++]
  for (int iter = 0; iter <= BootstrapMaxIter; iter++) {
    // update mean field
    for(tstp=0; tstp<=SolveOrder; tstp++){
       hubb::Ham_MF(tstp, G, Ut, eps_0, eps_mf);
    }

    // update self-energy
    for(tstp=0; tstp<=SolveOrder; tstp++){
        hubb::Sigma_2B(tstp, G, Ut, Sigma);
    }

    // solve Dyson equation
    cntr::dyson_start(G, MuChem, eps_mf, Sigma, beta, h, SolveOrder);

    // self-consistency check
    err=0.0;
    for(tstp=0; tstp<=SolveOrder; tstp++) {
      err += cntr::distance_norm2(tstp,G,gtemp);
    }
  
    if(err<BootstrapMaxErr && iter>2){
      break;
    }
         
    for(tstp=0; tstp<=SolveOrder; tstp++) {
      gtemp.set_timestep(tstp,G);
    }
 }
\end{lstlisting}
Finally, after the bootstrapping iteration has converged, the time
propagation for time steps $n>k$ is launched. The self-consistency at
each time step is accomplished by iterating the update of the mean-field
Hamiltonian, \gls*{GF} and self-energy over a fixed number of {\tt
  CorrectorSteps}. As an initial guess, we employ a polynomial
extrapolation of the \gls*{GF} from time step $n-1$ to $n$, as implemented in
the routine {\tt extrapolate\_timestep} (see \ref{subsec:extrapolation}). Thus, the time
propagation loop takes the form
\begin{lstlisting}[language=C++]
  for(tstp = SolveOrder+1; tstp <= Nt; tstp++){
    // Predictor: extrapolation
    cntr::extrapolate_timestep(tstp-1, G ,SolveOrder);
    // Corrector
    for (int iter=0; iter < CorrectorSteps; iter++){
       // update mean field
       hubb::Ham_MF(tstp, G, Ut, eps0, eps_mf);

       // update self-energy
       hubb::Sigma_2B(tstp, G, Ut, Sigma);

       // solve Dyson equation
       cntr::dyson_timestep(tstp, G, MuChem, eps_mf, Sigma, beta, h, SolveOrder);
     }
 }
\end{lstlisting}
After the \gls*{GF} has been computed for all required time steps, we compute
the observables. In particular, the conservation of the total energy
provides a good criterion to assess the accuracy of the
calculation. The total energy per spin for the Hubbard
model~\eqref{eq:hubbmodel1} is given in terms of the Galitskii-Migdal
formula~\cite{stefanucci_nonequilibrium_2013}.
\begin{align}
  E = \frac{1}{2}\mathrm{Tr}\left[\rho(t)\left(\epsilon^{(0)}
  +\epsilon^{\mathrm{MF}}(t)\right)\right] + \frac{1}{2} \mathrm{Im}\mathrm{Tr}\left[\Sigma\ast
  G\right]^<(t,t) \ .\label{Eq:Total_ener}
\end{align}
The last term, known as the correlation energy, is most conveniently
computed by the routine 
\begin{lstlisting}[language=C++]
  Ecorr = cntr::correlation_energy(tstp, G, Sigma, beta, h, SolveOrder);
\end{lstlisting}

\paragraph{Running the example programs} ---
There are three programs for the \gls*{2B}, $GW$ and $T$-matrix approximation,
respectively: {\tt hubbard\_chain\_2b.x},  {\tt hubbard\_chain\_gw.x},
{\tt hubbard\_chain\_tpp.x}. The driver script {\tt
  demo\_hubbard\_chain.py} located in the {\tt utils/} directory
provides a simple interface to these programs. After defining the
parameters and convergence parameters, the script creates the
corresponding input file and launches all three programs in a
loop. The occupation of the first lattice
site $n_1(t)$ and the kinetic and total energy are then read from the
output files and plotted. The script  {\tt
  demo\_hubbard\_chain.py} also allows to pass reference data as
an optional argument, which can be used to compare, for instance, to
exact results.

\begin{figure}[ht]
  \centering
  \includegraphics[width=\textwidth]{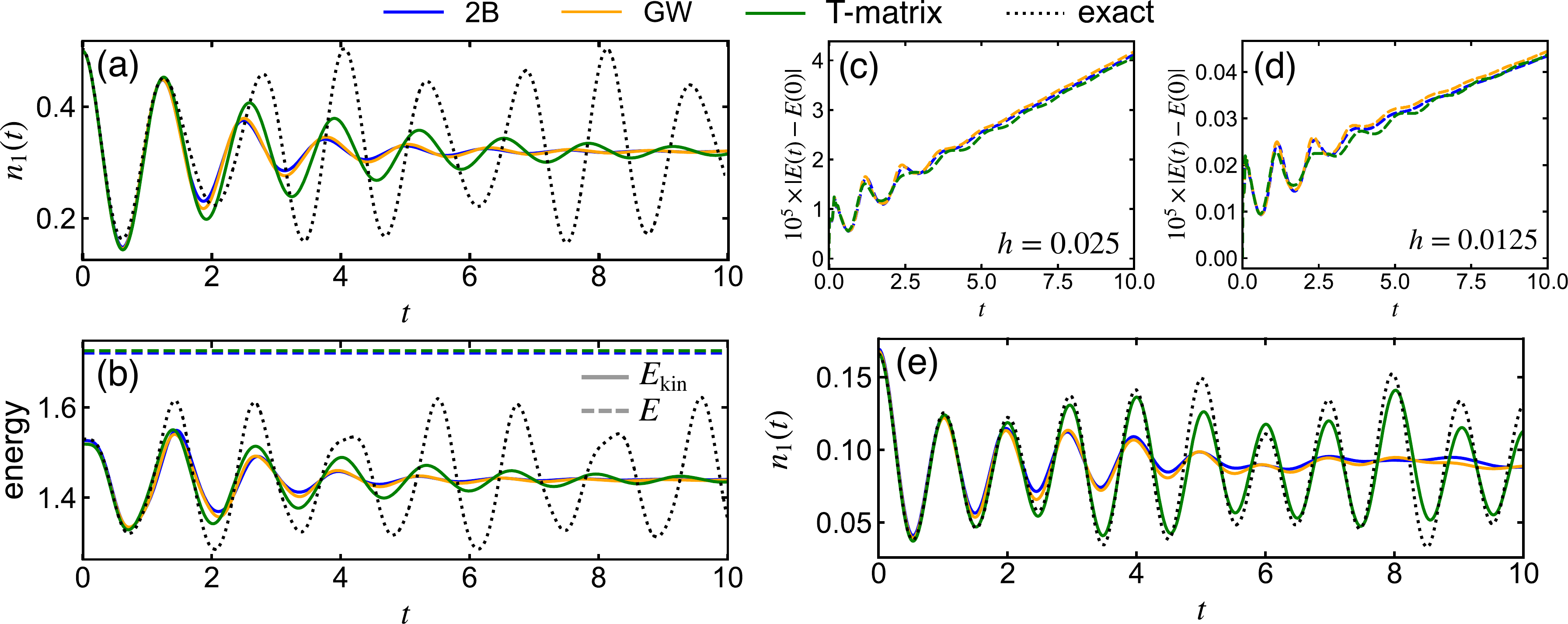}
  \caption{Dynamics in the Hubbard chain. (a) Occupation on the first
    site $n_1(t)$ for $M=2$, $U=1$, $n=1/2$ and $w_0=5$. (b)
    Corresponding kinetic (solid) and total (dashed lines) energy. (c) and (d): deviation of the total energy from the initial value,
    corresponding to (b), for time steps $h=0.025$ and $h=0.0125$,
    respectively. (e) Occupuation on the first site for $M=4$,
    $U=1.5$, $n=1/4$ and $w_0=5$.
    \label{fig:demo_hubbard}}
\end{figure}

\paragraph{Discussion} ---
Following Ref.~\cite{puig_von_friesen_kadanoff-baym_2010}, we have
selected two prominent examples illustrating the shortcomings of
weak-coupling diagrammatic treatments for finite systems and strong
excitations. The regimes where the discussed approximations work
well are systematically explored in Refs.~\cite{puig_von_friesen_kadanoff-baym_2010,
  schlunzen_nonequilibrium_2017}.

For the Hubbard dimer ($M=2$) at half filling ($\mu=0$, $\bar{n}=1/2$), a strong
excitation (here $w_0=5$) leads to the phenomenon of artificial
damping: although the density $n_1(t)$ exhibits an oscillatory behavior for all
times in an exact treatment, the approximate \gls*{NEGF} treatment -- with
either self-energy approximation considered here -- shows damping
to an unphysical steady state (see
Fig.~\ref{fig:demo_hubbard}(a)--(b)). It is instructive to look at the
total energy, shown as dashed lines in
Fig.~\ref{fig:demo_hubbard}(b). The conservation of total energy is
illustrated in Fig.~\ref{fig:demo_hubbard}(c)--(d). For the relatively
large time step $h=0.025$, the energy is conserved up to $4\times
10^{-5}$ in the considered time interval, while using a half as small
step $h=0.0125$ improves the accuracy of the energy conservation by
two orders of magnitude.

Fig.~\ref{fig:demo_hubbard}(e) shows the corresponding dynamics of the
occupation for $M=4$ and quarter filling. In the regime of small
filling, the $T$-matrix approximation is known to provide a good 
description for any strength of the interaction. This is confirmed by
Fig.~\ref{fig:demo_hubbard}(e), where the \gls*{2B} and $GW$ approximation
lead to artificial damping, while the $n_1(t)$ calculated by the $T$-matrix approximation
agrees well with the exact result.


\subsection{DMFT for the Holstein model}\label{subsec:holstein}

\paragraph{Overview} --- In this section, we study the dynamics of the Holstein model, which is a fundamental model for electron-phonon coupled systems.
This example demonstrates a minimal application of \libcntr{} within the non-equilibrium dynamical mean-field theory (DMFT) \cite{aoki2014_rev}, as well as the usage of the phonon (bosonic) Green's function.

The Hamiltonian of the single-band Holstein model is written as
\begin{align}
H(t)&=-J(t)\sum_{\langle  i,j\rangle,\sigma}\hc_{i,\sigma}^{\dagger}\hc_{j,\sigma}-\mu\sum_i \hn_i 
+\omega_0\sum_i \ha^{\dagger}_i \ha_i+g(t)\sum_i (\ha_i^{\dagger}+\ha_i)\hn_i.\label{eq:Holstein}
\end{align}
Here $J(t)$ is the hopping parameter of electrons, $\mu$ is the chemical potential, $\omega_0$ is
the phonon frequency, and $g(t)$ is the el-ph coupling. 
We consider modulation of the hopping parameter or the el-ph coupling as excitation protocols.
For simplicity, in the following we consider the Bethe lattice (with infinite coordination number). 
For this lattice, the free electrons has a semi-circular density of states, 
$\rho_0(\epsilon) = \frac{1}{2\pi J^{*2}}
\sqrt{4J^{*2}-\epsilon^2}$,  with $J^*$ a properly renormalized hopping amplitude \cite{Metzner1989}.  Here we take $J^*=1$.   
Assuming spin symmetry, the lattice \gls*{GFs} are introduced as 
\begin{subequations}
\begin{align}
G_{ij}(t,t') &= -i \langle  T_{\mathcal{C}}\hc_{i,\sigma}(t) \hc_{j,\sigma}^{\dagger}(t') \rangle, \\
D_{ij}(t,t') &= -i\langle T_{\mathcal{C}} \Delta\hX_i(t) \Delta\hX_j(t') \rangle.
\end{align}
\end{subequations}
Here $\hX_i = \ha^\dagger_i + \ha_i$ and $\Delta\hX_i(t) = \hX_i(t) - \langle \hX_i(t) \rangle $.

We treat the dynamics of the Holstein model using the \gls*{dmft} formalism \cite{Georges1996}.
In \gls*{dmft}, the lattice model is mapped to an effective impurity model with a properly adjusted free electron bath, which is characterized by the so-called hybridization function $\Delta(t,t)$, see Eq.~\eqref{eq:weisg} below.
The hybridization function is self-consistently determined, so that the impurity \gls*{GF} ($G_{\rm imp}(t,t')$) and the impurity self-energy ($\Sigma_{\rm imp}$) are identical 
to the local Green's function ($G_{\rm loc}=G_{ii}$) and the local self-energy of the lattice problem ($\Sigma_{\rm loc}$), respectively. 
In practice, the \gls*{dmft} implementation consists of (i), solving the impurity model for a given $\Delta(t,t')$ to obtain $G_{\rm imp}(t,t')$ and $\Sigma_{\rm imp}$, and (ii), the \gls*{dmft} lattice self-consistency part, where we update $G_{\rm loc}$ and $\Delta(t,t')$ assuming $\Sigma_{\bf k}=\Sigma_{\rm imp}$.
In the case of a Bethe lattice, the \gls*{dmft}  lattice self-consistency part is simplified and the hybridization function can be determined directly from the \gls*{GF}, 
 \begin{align}
 \Delta(t,t')=J^*(t) G_{\rm imp}(t,t') J^*(t').\label{eq:bethe_condition} 
\end{align}

The action of  the corresponding effective impurity model in the path integral formalism is \footnote{Here we denote the Grassmann fields by $c^\dagger$ and $c$ and the scalar field as $X$.} 
\begin{align}
\mathcal{S}_{\rm imp}&=i\sum_\sigma\int_{{\mathcal C}} dt dt'  c^{\dagger}_\sigma(t) \mathcal{G}^{-1}_{0} (t,t') c_\sigma(t')
+i\int_{{\mathcal C}} dt dt' X(t) \frac{D_0^{-1}(t,t')}{2}  X(t')\nonumber\\
&-ig\sum_\sigma \int_{{\mathcal C}}dt  X(t) c^{\dagger}_\sigma(t) c_\sigma(t), \label{eq:imp_Holstein}
\end{align}
where 
\begin{subequations}
\begin{align}
\mathcal{G}^{-1}_{0} (t,t') &= [i\partial_{t}+\mu]\delta_{\mathcal C}(t,t')-\Delta(t,t'),\label{eq:weisg}\\
D_0^{-1}(t,t') &= \frac{-\partial_t^2 - \omega_0^2}{2\omega_0} \delta_c(t,t').
\end{align}
\end{subequations}
The electron and phonon \gls*{GFs} of the impurity problem are determined by the Dyson equations 
\begin{subequations}\label{eq:Dyson_holstein_imp}
\begin{align}
&[i\partial_{t}-\mu-\Sigma^{\rm MF}_{\rm imp}(t)]G_{\rm imp}(t,t')-[(\Delta + \Sigma^{\rm corr}_{\rm imp})* G_{\rm imp}](t,t')=\delta_{\mathcal C}(t,t')\label{eq:Dyson_imp},\\
&D_{\rm imp}(t,t^\prime) = D_{0}(t,t^\prime) + [D_{0}*\Pi_{\rm imp}* D_{\rm imp}](t,t^\prime)  \label{eq:D_dyson},
\end{align}
\end{subequations}
and the phonon displacement, $X_{\rm imp}(t)=\langle \hX_{\rm imp}(t)\rangle$, which is described by
\begin{align}
X_{\rm imp}(t) = -\frac{2 g(0)}{\omega_0} n_0(0) + \int^t_0 d\bar{t} D_0^\ret(t,\bar{t})[g({\bar t})n_{\rm imp}(\bar{t})-g(0)n_{\rm imp}(0)]. \label{eq:X}
\end{align}
Here the mean-field contribution ($\Sigma^{\rm MF}_{\rm imp}(t)$) corresponds to 
\begin{align}
\Sigma^{\rm MF}_{\rm imp}(t) &=
g(t)X_{\rm imp}(t)\label{eq:H_mf_Holstein},
\end{align}
$\Sigma^{\rm corr}_{\rm imp}(t,\bar{t})$ is the beyond-mean-field contribution to the self-energy, $D_{0}(t,t')\equiv-i\langle \Delta\hX(t) \Delta\hX(t') \rangle_0$ is for the free phonon system, $\Pi_{\rm imp}$ is the phonon self-energy and $ n_{\rm imp}(t)= \langle \hn_{\rm imp}(t) \rangle$ is the particle number at the impurity site.

After the DMFT loop is converged, 
one can calculate some observables such as different energy contributions.
The expressions for the energies (per site) are given in the following expressions.
The kinetic energy is 
\begin{align}
E_{\rm kin}(t)&=\frac{1}{N}\sum_{\langle i,j\rangle,\sigma}-J(t)\langle \hc_{i,\sigma}^{\dagger}(t)\hc_{j,\sigma}(t)\rangle
=-2i[\Delta \ast G_{\rm loc}]^<(t,t).
\end{align}
The interaction energy is can be expressed as 
\begin{align}
E_{\rm nX}(t)&=\frac{g(t)}{N}\sum_i \langle\hX_i\hn_i\rangle=\Sigma^{\rm MF}_{\rm loc}(t) n(t) - 2i[\Sigma^{\rm corr}_{\rm loc} * G_{\rm loc}]^<(t,t),
\end{align}
The phonon energy is  
\begin{align}\label{eq:ph_energy}
E_{\rm ph}(t)=\frac{\omega_0}{N}\sum_i  \langle \ha^{\dagger}_i \ha_i\rangle=\frac{\omega_{0}}{4} [iD^<(t,t) + X(t)^2]+\frac{\omega_{0}}{4} [iD_{\rm PP}^<(t,t) + P(t)^2].
\end{align}
Here $D_{\rm PP}(t,t')=-i\langle T_\mathcal{C} \Delta\hat{P}_i(t) \Delta\hat{P}_i(t')\rangle$ with $\hat{P}_i=\frac{1}{i}(\ha_i-\ha_i^\dagger)$ and $\Delta\hat{P}_i(t) \equiv \hat{P}_i(t) - \langle \hP_i(t)\rangle$. 
We note that the translational invariance is assumed and  $X(t)=\langle \hX_i(t)\rangle = X_{\rm imp}(t)$, $P(t)=\langle \hP_i(t)\rangle = P_{\rm imp}(t)$, $D=D_{ii}=D_{\rm imp}$, $\Sigma_{\rm loc}=\Sigma_{\rm imp}$.

In this example, we solve the impurity problem using the simplest weak-coupling expansion as an impurity solver, i.e. the unrenormalized Migdal approximation (uMig) \cite{Kemper2014PRB,Sentef2016PRB}, where the phonons act as a glue for the electrons as well as a heat bath. On the web page \href{www.nessi.tuxfamily.org}{www.nessi.tuxfamily.org} we discuss an alternative impurity solver based on the self-consistent Migdal approximation (sMig) \cite{Murakami2015PRB,Murakami2016PRB,Schuler2016PRB}. 
Both solvers are implemented in the C++ module {\tt Holstein\_impurity\_impl.cpp}.

{\it Unrenormalized Migdal approximation as an impurity solver: uMig. ---} 
The impurity self-energy for the electron is approximated as 
\begin{align}
\hat{\Sigma}^{\rm uMig,corr}_{\rm imp}(t,t') & =ig(t)g(t')D_{0}(t,t') G_{\rm imp}(t,t'),\label{eq:uMig_el}
\end{align}
while we do not consider the self-energy of phonons.
In \nessi{}, $\frac{1}{2} D_{0}(t,t')$ is obtained by a cntr routine as 
\begin{lstlisting}[language=C++]
cntr::green_single_pole_XX(D0,Phfreq_w0,beta,h);
\end{lstlisting}

In the sample program, the \gls*{uMig} self-energy is computed  by the routine {\tt Sigma\_uMig}.
We provide two interfaces for $0\leq $ {\tt tstp} $ \leq $ {\tt SolveOrder} (bootstrapping part) and {\tt tstp} $=-1$, {\tt tstp} $>$ {\tt SolveOrder} 
(Matsubara part and the time-stepping part), respectively.
Here, we show the latter as an example:
\begin{lstlisting}[language=C++]
void Sigma_uMig(int tstp, GREEN &G, GREEN &D0, CFUNC &g_el_ph, GREEN &Sigma){
    
        int Norb=G.size1();
        int Ntau=G.ntau();
        
        GREEN_TSTP gGg(tstp,Ntau,Norb,FERMION);
        G.get_timestep(tstp,gGg);//copy time step from G
        gGg.right_multiply(tstp,g_el_ph);
        gGg.left_multiply(tstp,g_el_ph);
        
        //Get Sig(t,t')=ig^2 D_0(t,t') G(t,t') 

        Bubble2(tstp,Sigma,0,0,D0,0,0,gGg,0,0);

 }
\end{lstlisting}
In this routine, evaluating the electron self-energy Eq.~\eqref{eq:uMig_el} is evaluated  using {\tt Bubble2}, see Sec.~\ref{subsec:contourconv}.

\paragraph{Generic structure of the example program}---
The program of DMFT + uMig is implemented in {\tt Holstein\_bethe\_uMig.cpp} for normal states.
As in the case of the Hubbard chain, the program consists of three main steps: 
(i) solving the Matsubara Dyson equation, (ii) bootstrapping ({\tt tstp}$\leq$ {\tt SolveOrder}) and (iii) time propagation for {\tt tstp} $>$ {\tt SolveOrder}.
 Since the generic structure of each part is similar to that of the Hubbard chain, we only show here the time propagation part to illustrate the differences.
\begin{lstlisting}[language=C++]
for(tstp = SolverOrder+1; tstp <= Nt; tstp++){
	// Predictor: extrapolation
	cntr::extrapolate_timestep(tstp-1,G,SolveOrder);
	cntr::extrapolate_timestep(tstp-1,Hyb,SolveOrder);

	// Corrector
	for (int iter=0; iter < CorrectorSteps; iter++){
		//=========================
		// Solve Impurity problem
		// ========================
		cdmatrix rho_M(1,1), Xph_tmp(1,1);
		cdmatrix g_elph_tmp(1,1),h0_imp_MF_tmp(1,1);
	
		//update self-energy
		Hols::Sigma_uMig(tstp, G, D0, g_elph_t, Sigma);

		//update phonon displacement 
		G.density_matrix(tstp,rho_M);
		rho_M *= 2.0;//spin number=2
		n_tot_t.set_value(tstp,rho_M);
		Hols::get_phonon_displace(tstp, Xph_t, n_tot_t, g_elph_t, D0, Phfreq_w0, SolveOrder,h);

		//update mean-field
		Xph_t.get_value(tstp,Xph_tmp);
		g_elph_t.get_value(tstp,g_elph_tmp);
		h0_imp_MF_tmp = h0_imp + Xph_tmp*g_elph_tmp;
		h0_imp_MF_t.set_value(tstp,h0_imp_MF_tmp);
		
		//solve Dyson for impurity
		Hyb_Sig.set_timestep(tstp,Hyb);
		Hyb_Sig.incr_timestep(tstp,Sigma,1.0);
		cntr::dyson_timestep(tstp, G, 0.0, h0_imp_MF_t, Hyb_Sig, beta, h ,SolveOrder);
		
		//===================================
		// DMFT lattice self-consistency (Bethe lattice)
		// ===================================
		//Update hybridization
		Hyb.set_timestep(tstp,G);
		Hyb.right_multiply(tstp,J_hop_t);
		Hyb.left_multiply(tstp,J_hop_t);
	}
}
\end{lstlisting}
At the beginning of each time step, we extrapolate the local 
\gls*{GF} and the hybridization function,
which serves as a predictor.
Next, we iterate the \gls*{dmft} self-consistency loop (corrector) until convergence is reached.
In this loop, we first solve the impurity problem to update the local self-energy and \gls*{GF}.
Then we update the hybridization function by the lattice self-consistency condition, which in the case of the Bethe lattice simplifies to Eq.~\eqref{eq:bethe_condition}.

\paragraph{Running the example programs} ---
The corresponding executable file is named {\tt Holstein\_bethe\_uMig.x}.
In these programs, we use $\mu_{\mathrm{MF}}\equiv \mu - gX(0)$ as an input parameter instead of $\mu$.
($\mu$ is determined in a post processing step.)
Excitations via modulations of the hopping and el-ph coupling are implemented, where we need to provide $dg(t)(\equiv g(t)-g(0))$ and $dJ^*(t)(\equiv J(t)-J(0))$ as inputs.
The driver script {\tt demo\_Holstein.py} located in the {\tt utils/} directory provides a simple interface to the program.
After defining the system parameter, numerical parameters (time step, convergence criterion, etc.) and excitation parameters, the script creates the corresponding input file and starts the simulation.
After the simulation, the number of particles for each site($n(t)=\sum_\sigma n_\sigma(t)$), phonon displacement ($X(t)$), phonon momentum ($P(t)$) and the energies are plotted. In addition, the spectral functions of electron and phonons,
\begin{subequations}\label{eq:spectrum_demo}
\begin{align}
A^\ret(\omega;t_{\rm av}) &= -\frac{1}{\pi} \int dt_{\rm rel} e^{i\omega t_{\rm rel}} F_{\rm gauss}(t_{\rm rel}) G^\ret (t_{\rm rel};t_{\rm av}),\\
B^\ret(\omega;t_{\rm av}) &= -\frac{1}{\pi} \int dt_{\rm rel} e^{i\omega t_{\rm rel}} F_{\rm gauss}(t_{\rm rel}) D^\ret (t_{\rm rel};t_{\rm av}),
\end{align}
\end{subequations}
are plotted using a {\tt python3} script in \nessi{} 
for $t_{\rm av}=\frac{Nt \cdot h}{2}$.
Here $F_{\rm gauss}(t_{\rm rel})$ is a Gaussian window function, which can also be specified in {\tt demo\_Holstein\_impurity.py}.
           
\begin{figure}[t]
  \centering
  \includegraphics[width=1.0\textwidth]{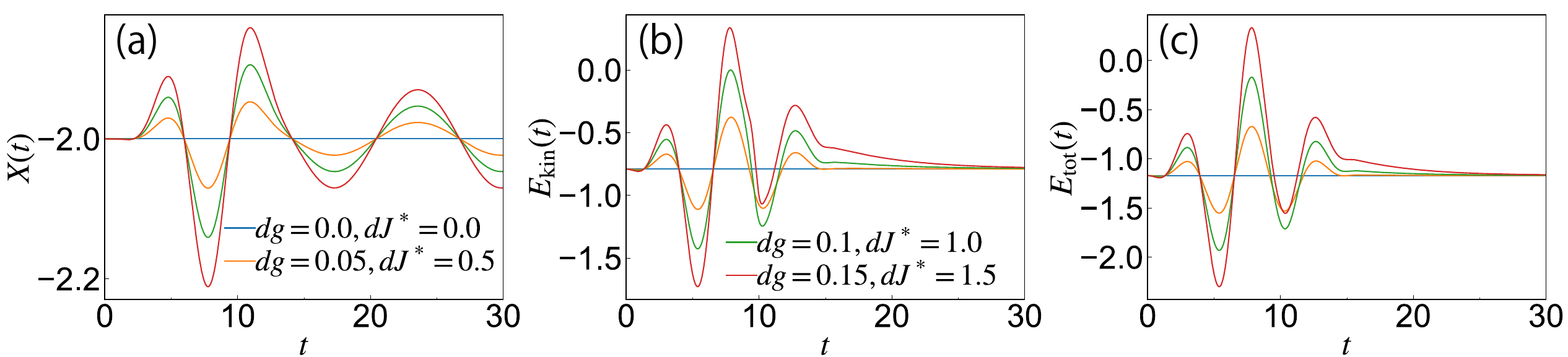}
  \caption{Time evolution of (a) the phonon displacement, (b) the kinetic energy and (c) the total energy after excitation via the simaltaneous modulation of the el-ph coupling and the hopping parameter within (d)\gls*{dmft} + uMig. We use $g=0.5,\omega_0=0.5$ and $\beta=10.0$ and consider the half-filled case.
 Here, we use a $\sin^2$ envelope for the both modulations with excitation frequency $\Omega=1.2$ and pulse duration $T_{\rm end}=15.7$. 
 The size of the excitation is indicated by $dg$ and $dJ^*$.}
  \label{Fig:Holstein_demo}
\end{figure}

\paragraph{Discussion} ---
In Fig.~\ref{Fig:Holstein_demo}(a)--(c), we show the time evolution of the phonon displacement $X(t)$, the kinetic energy $E_{\rm kin}(t)$ and the total energy $E_{\rm tot}(t)$
after excitation by the simultaneous modulation of the el-ph coupling and the hopping parameter. 
Since the energy is gradually absorbed by the phonons after the excitation, both of $E_{\rm kin}(t)$ and $E_{\rm tot}(t)$ are gradually decreased toward the initial value, i.e. the equilibrium value at the phonon temperature. We note that we also provide an example program of the self-consistent Migdal approximation as an impurity solver as well as 
programs of the Nambu formalism for the superconducting states.  Explanations and demonstration of these sample programs are given on the webpage.

\section{MPI parallelization}
\label{sec:mpi}


\subsection{Parallelization}\label{sSec:Parallelization} 

In \libcntr{}, we provide tools for distributed-memory parallelization
via the \gls*{mpi}. In particular, the parallel
layout is tailored to treat vectors of \gls*{GFs}, which is 
relevant for the simulation of 
\emph{extended} systems. In this case, all
quantities are additionally labelled by the reciprocal lattice vector
$\vec{k}$ chosen from the first \gls*{BZ}. For instance, the
Dyson equation for the \gls*{GF} $G_{\vec{k}}(t,t^\prime)$ takes the
form
\begin{align}
  \label{eq:dyson_kspace}
  \left(i \partial_t -\epsilon_{\vec{k}}(t) \right)
  G_{\vec{k}}(t,t^\prime) = \delta_\mathcal{C}(t,t^\prime) +
  \left[\Sigma_{\vec{k}} \ast G_{\vec{k}}\right](t,t^\prime)  \ .
\end{align}
This equation can be solved independently for each $\vec{k}$, which can be performed in
parallel without communication. Constructing the self-energy $\Sigma_{\vec{k}}(t,t^\prime)$ then
typically requires information from different points $\vec{k}'\neq\vec{k}$ in the \gls*{BZ}. However, the computational effort to solve the Dyson equation at a time step $N$ scales like $\mathcal{O}(N^2)$, while the amount of data to be communicated scales only like $\mathcal{O}(N)$, so that the problem can be parallelized using a distributed memory parallelization with moderate communication overhead. 

To handle the all-to-all communication of Green's functions among \gls*{mpi} ranks, we have implemented the  auxiliary class {\tt distributed\_timestep\_array}. (Simple point-to-point communication of time-steps can be done using member functions of the {\tt herm\_matrix\_class}, which is described in the online manual). Below we provide an overview over the {\tt distributed\_timestep\_array} and its parent class {\tt distributed\_array}, and discuss a real-time GW simulation as an advanced example for a parallel application (see Sec.~\ref{subsec:example_gw}).

\begin{figure}[ht]
  \centering
  \includegraphics[width=0.9\textwidth]{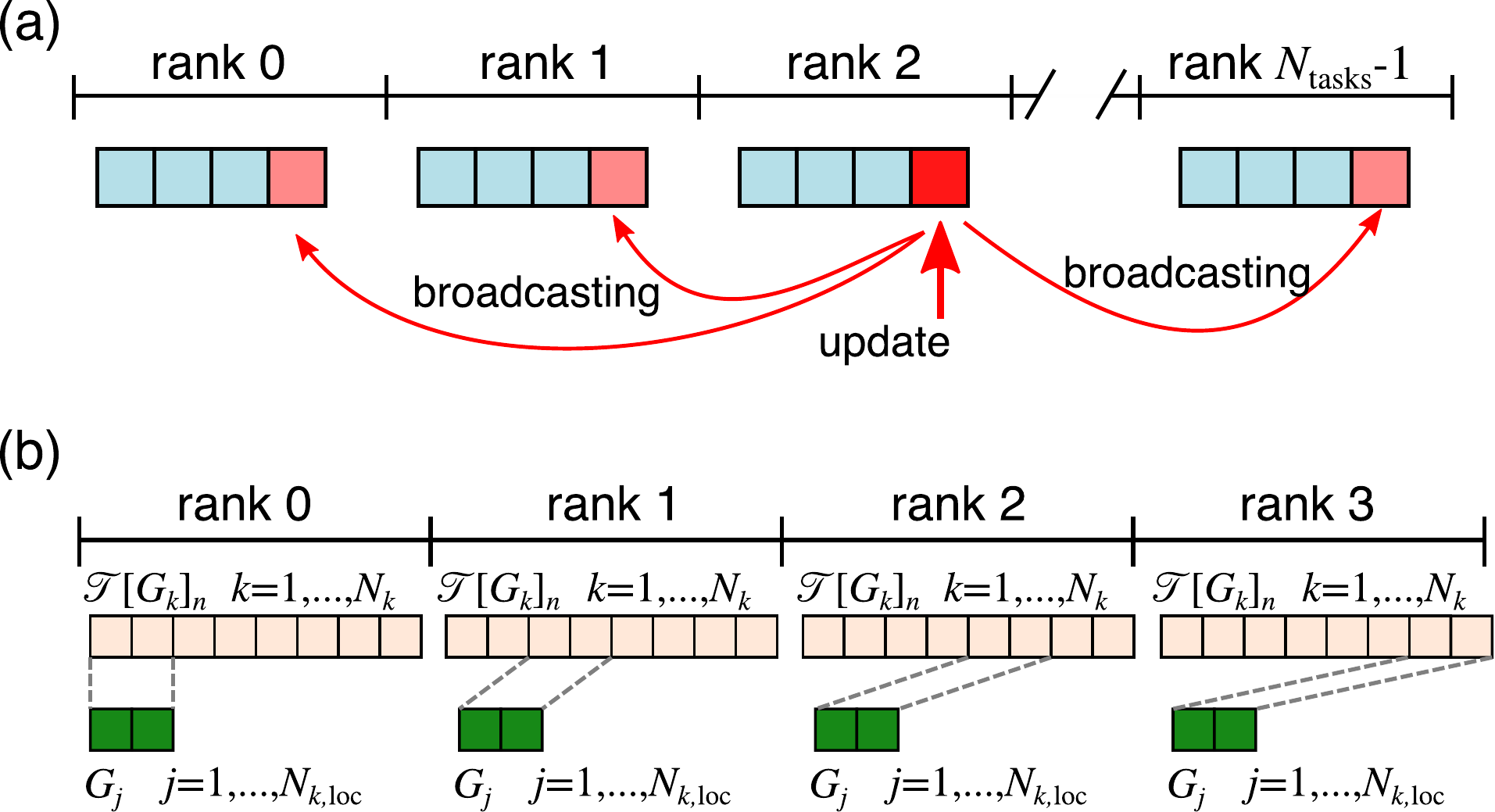}
  \caption{(a) Sketch of the parallization in momentum space using the
    {\tt distributed\_array} class. A local update on one \gls*{mpi}
    rank is broadcasted to all other corresponding elements across
    all ranks. (b) Example of the parallization in momentum space using the
    {\tt distributed\_timestep\_array} class for $N_k=8$, distributed
    over $N_{\mathrm{tasks}}=4$ \gls*{mpi} ranks. The full {\tt herm\_matrix} (represented by the dark green squares)
     is stored for local indices only, while the time slices for a fixed time step 
     are stored on each rank (light pink squares).  
     The gray dashed lines
    indicate the connection of local and global indices.
    \label{fig:mpi_parallel_1}}
\end{figure}

\paragraph{{\tt distributed\_array}}--- The class {\tt
  distributed\_array} provides a generic structure for distributing
and communicating sets of data blocks. Let us assume the total number
of points sampling the \gls*{BZ} is given by $N_k$, and label the points by
$k=1,\dots,N_k$. The {\tt distributed\_array} class is comprised of a
vector of length $N_k$ of any base class (provided by a template
argument) on \emph{every} \gls*{mpi} rank, as illustrated in
Fig.~\ref{fig:mpi_parallel_1}(a). This makes the communication
particularly straightforward. For instance, after updating an element of the
{\tt distributed\_array} on rank 2 (see lower panel in
Fig.~\ref{fig:mpi_parallel_1}(a)), this information is broadcasted to all
other ranks using the \gls*{mpi} collective communication function {\tt
  mpi\_bcast}. Further functionalities include sending and receiving
blocks among different ranks, as well as gathering all data on one
rank (typically the master). The {\tt distributed\_array} thus provides a general framework for \gls*{mpi}
parallelization, which can be adjusted to a particular situation. The
most common usage is distributing instances of the {\tt
  herm\_matrix\_timestep} class.

\paragraph{{\tt distributed\_timestep\_array}}--- The class {\tt
  distributed\_timestep\_array} is a specialization of the class {\tt
  distributed\_array}, for which the distributed base class is \
  {\tt
  herm\_matrix\_timestep}. Let us sketch the typical usage. Due to the
high memory demands for storing two-time \gls*{GFs} as {\tt herm\_matrix}, we
divide the total number of points $N_k$ into a smaller local (with
respect to the \gls*{mpi} rank) number of points $N_{k,\text{loc}}$. Two-time
functions such as the \gls*{GF} are stored as a vector of $N_{k,\text{loc}}$: $G_j$
with $j=1,\dots, N_{k,\text{loc}}$ (see Fig.~\ref{fig:mpi_parallel_1}(b)). For
each rank to have access to the full momentum dependence $G_k$,
$k=1,\dots,N_k$, the  {\tt
  distributed\_timestep\_array} class is used to communicate the time
slice $\mathcal{T}[G_k]_n$ for $k=1,\dots,N_k$. Fig.~\ref{fig:mpi_parallel_1}(b) illustrates this
layout for the example of $N_{\mathrm{tasks}}=4$ \gls*{mpi} ranks, $N_k=8$ and, thus,
$N_{k,\mathrm{loc}}=2$. The precise calls to perform these communications will be 
best apparent from the example below, and are also described in detail in the online manual.

  \subsection{Example: $GW$ for the translationally invariant Hubbard model \label{subsec:example_gw}}

\paragraph{Overview} --- As in Section~\ref{Sec:Chain}, we will consider the Hubbard Hamiltonian, but  we assume a translationally invariant system with periodic boundary conditions. The translational invariance implies that all observables and propagators are diagonal in momentum space. Hence, all quantities can be labelled by the reciprocal lattice vector $\vec k$ within the first \gls*{BZ}.  This reduces the computational and storage complexity from $\mathcal{O}(N_k^2)$ for the real space formalism introduced in Section~\ref{Sec:Chain} to $\mathcal{O}(N_k).$ Moreover, the Dyson equation is diagonal in  momentum space and can be efficiently parallelized using the distributed-memory parallelization based on \gls*{mpi}. 

We will consider a \gls*{1D} chain described by the Hubbard model, see Eq.~\eqref{eq:hubbmodel1}. The single particle part of the Hamiltonian can be diagonalized as
\begin{align}
\hat{H}_0 = \sum_{\vec k,\sigma} \epsilon(\vec k) c_{\vec{k}\sigma}^{\dagger} c_{\vec{k}\sigma},
\end{align}
where we have introduced the free-electron dispersion  $\epsilon(\vec k)=-2 J \cos(\vec k_x).$ We will use a vector notation since the generalization to higher dimensional systems is straightforward. For the \gls*{1D} chain used in the demonstration program, the momentum has only one component $[\vec k]_x=k_x$.

The system is excited via an electromagnetic field, which for a translationally invariant system is conveniently
introduced using the Peierls substitution. The latter involves the vector potential $\vec A(\vec r,t)$ as a phase factor in the
hopping~\cite{peierls1933,luttinger1951}, or, equivalently a time-dependent shift
in the   single-particle dispersion \beq{ \epsilon(\vec k,t)=\epsilon(\vec k- q \vec A(t)/\hbar).}
   The vector potential is obtained from the electric field as $\vec  A(t)=-\int_0^t \vec  E(\bar t ) d\bar t.$ 

In this example, we treat the dynamics within the $GW$ approximation, following an implementation similar to Ref.~\cite{Denis2016_PRB} (the latter has been a simulation for a four-band model).  The numerical evaluation of the respective self-energy expressions is implemented in the C++ module {\tt gw\_selfen\_impl.cpp}, and the main code is found in 
\textcolor{blue}{\tt programs/gw.cpp}. Below we explain the key routines.

\paragraph{Self-energy approximation: $GW$} --- In momentum space, the correlation part of the $GW$ self-energy can be written as 
\begin{align}
  \Sigma_{\vec k}(t,t^\prime)=\frac{\I}{N_k} \sum_{\vec q} G_{\vec k-\vec q}(t,t^\prime) \delta W_{\vec q}(t,t^\prime),
  \label{Eq:Sigmak}
\end{align}
where we have introduced the Fourier transform of the propagator 
$X_{\vec q}(t,t^\prime)=(1/N_k)\sum_{i} \exp(\I (\vec r_i-\vec r_j) \vec q) X_{ij}(t,t^\prime),$ see also  Section~\ref{Sec:Chain} for the definition of the propagators. In line with Sec.~\ref{Sec:Chain}, we have introduced the dynamical part of the effective interaction $\delta W_{\vec q}$ via $W_{\vec q}(t,t^\prime)=U\delta_\mathcal{C}(t,t^\prime)+\delta W_{\vec q}(t,t^\prime)$. 
Due to the translational invariance, the propagators and corresponding Dyson equations are diagonal in momentum space. This leads to a significant speed-up of calculations since the most complex operation, the solutions of the \gls*{vie}, can be performed in parallel.  The retarded interaction is obtained as a solution of the Dyson-like equation 
\beq{
  W_{\vec k}(t,t')=U\delta_\mathcal{C}(t,t^\prime)+ U [ \Pi_{\vec k} \ast W_{\vec
    k} ](t,t^\prime). 
} 
and the Fourier transform of the polarization is given by 
\beq{
\Pi_{\vec k}(t,t')= \frac{-\I}{N_k} \sum_{\vec q} G_{\vec k+\vec q}(t,t') G_{\vec q}(t',t).
\label{Eq:Pol}
}
In the case of a non-local interaction, the polarization is multiplied by a spin factor $s=2.$ 

This structure allows for an easy adaptation of the code to arbitrary lattice geometries. In particular, we provide an implementation of a \gls*{1D} chain geometry in the class {\tt lattice\_1d\_1b} within the C++ module {\tt gw\_lattice\_impl.cpp}. The  routine {\tt add\_kpoints} evaluates the sum or difference of two vectors $\vec k\pm\vec q,$ where slight care has to be taken to map the vector back to the first \gls*{BZ}. For the modification to other lattices and interaction vertices, the user has to define the first \gls*{BZ}, the single particle dispersion $\epsilon(\vec k)$, the interaction vertex $U$ and how vectors sum up. 

The generalization to the long-range interaction
\beq{
  \hat{H}_{\text{int}}=U \sum_i \hat n_{i\uparrow} \hat n_{i\downarrow}+ \frac{1}{2}\sum_{i,j} V(\vec r_i-\vec r_j) \hat n_i \hat n_j
}
is straightforward. For the purpose of demonstration, we have included the nearest-neighbor interaction $V(\vec r_i-\vec r_j)=\delta(|\vec r_i-\vec r_j|=1)V$ into the example program (input parameter {\tt V}). We should comment that for a purely local interaction the Fock term is zero and it only takes a finite value in cases with a non-local interaction.

\paragraph{Distribution of momenta over mpi ranks} --- 
As each momentum point is independent, we have introduced a class {\tt kpoint} in the module {\tt gw\_kpoints\_impl.cpp}. This class includes all propagators at the given momentum  point $\vec k$, as well as corresponding methods, such as the solution of the Dyson equations  for the single particle propagator $G_{\vec k}(t,t^\prime)$, see Eq.~\eqref{eq:dyson_kspace}, and the retarded interaction $W_{\vec k}(t,t^\prime)$. An arbitrary lattice can be represented as a set of   {\tt kpoint} objects. 
In the code, each physical momentum $\vec{k}$ is indexed by an index {\tt k}$\in\{0,...,${\tt Nk-1}$\}$, which we will refer to as the ``global index'' in the following. {\tt lattice} is the variable which stores information on the lattice (in particular the relation between the index {\tt k} and the physical momentum $\vec{k}$, and {\tt lattice.nk\_} returns {\tt Nk}.

Each {\tt kpoint} objects need to be available at only one mpi rank, because the Dyson and vie2 integral equations can be solved independently for each rank. However, the evaluation of the self-energy and polarization diagrams at a given timeslice requires that the timeslice  $\mathcal{T}[G_{\vec{k}}]$ and  $\mathcal{T}[W_{\vec{k}}]$  at {\em all} $\vec{k}$ is made available at {\em all}  mpi ranks, see Eq.~\eqref{Eq:Sigmak}. This is facilitated by introducing a setting with the following variables at a rank which holds {\tt Nkloc}  {\tt kpoint} objects
\begin{itemize}
\item {\tt std::vector <kpoint > corrK\_rank }: A vector of length {\tt Nkloc}, containing the {\tt kpoint} objects stored locally at the rank.
\item
{\tt std::vector <int> kindex\_rank}:
A vector of length {\tt Nkloc}.\\
 {\tt kindex\_rank[j]} returns the global index
 {\tt k}$\in\{0,...,${\tt Nk-1}$\}$ of the {\tt kpoint} $j$.
 \item
{\tt distributed\_timestep\_array  gk\_all\_timesteps}, as described in Section~\ref{sSec:Parallelization}. Can  store $\mathcal{T}_n[G_{\vec{k}}]$ at a given timestep $n$ for all {\tt k}$\,\in\,\{0,...,${\tt Nk-1}$\}$. {\tt gk\_all\_timesteps.G()[k]} returns a reference to the data at $\mathcal{T}[G_{\vec{k}}]$. The class has a copy of {\tt kindex\_rank} and of the inverse map, so that a one can easily launch a communication in which the rank which owns a given {\tt kpoint} would send the corresponding timeslices to all other ranks.
 \item
{\tt distributed\_timestep\_array  wk\_all\_timesteps}: Can store  $\mathcal{T}_n[W_{\vec{k}}]$ at a given timestep $n$ for all {\tt k}$\,\in\,\{0,...,${\tt Nk-1}$\}$. Analogous to {\tt gk\_all\_timesteps}.
\end{itemize}

 The strategy to compute the $GW$
self-energy $\mathcal{T}[\Sigma_{\vec{k}}]_n$ at time step $n$ thus consist of two steps:
\begin{enumerate}
\item At time $t_n$, communicate the latest time slice of the \gls*{GFs} $\mathcal{T}[G_{\vec k}]_n$ and retarded interactions $\mathcal{T}[W_{\vec k}]_n$ for all momentum points among all \gls*{mpi} ranks.
\item Evaluate the self-energy diagram $\mathcal{T}[\Sigma_{\vec k_{\text{rank}}}]_n$ in Eq.~\eqref{Eq:Sigmak}  for a subset of momentum points $\vec k_{\text{rank}}$ present on a given rank using the routine {\tt Bubble2}.
\end{enumerate} 
Step 1 is implemented as
\begin{lstlisting}[language=C++]
void gather_gk_timestep(int tstp,int Nk_rank,DIST_TIMESTEP &gk_all_timesteps,std::vector<kpoint> &corrK_rank,std::vector<int> &kindex_rank){
    gk_all_timesteps.reset_tstp(tstp);
    for(int k=0;k<Nk_rank;k++){
      gk_all_timesteps.G()[kindex_rank[k]].get_data(corrK_rank[k].G_);
    }
    gk_all_timesteps.mpi_bcast_all();
  }
\end{lstlisting}
where the abbreviation {\tt DIST\_TIMESTEP} for  {\tt distributed\_timestep\_array} is used. An analogous routine is used for the bosonic counterpart. We gather the information from all ranks into an object of type {\tt distributed\_timestep\_array} named {\tt gk\_all\_timesteps}. 
{\tt mpi\_bcast\_all()} is a wrapper around the MPI routine {\tt Allgather} adjusted to the type {\tt distributed\_timestep\_array}. 

Step 2 is implemented as
\begin{lstlisting}[language=C++]
void sigma_GW(int tstp,int kk,GREEN &S,DIST_TIMESTEP &gk_all_timesteps,DIST_TIMESTEP &wk_all_timesteps,lattice_1d_1b &lattice,int Ntau,int Norb){
      assert(tstp==gk_all_timesteps.tstp());
      assert(tstp==wk_all_timesteps.tstp());
      GREEN_TSTP stmp(tstp,Ntau,Norb,FERMION);
      S.set_timestep_zero(tstp);
      for(int q=0;q<lattice.nk_;q++){
        double wk=lattice.kweight_[q];
        int kq=lattice.add_kpoints(kk,1,q,-1);
        stmp.clear();
        for(int i1=0;i1<Norb;i1++){
          for(int i2=0;i2<Norb;i2++){
            cntr::Bubble2(tstp,stmp,i1,i2,gk_all_timesteps.G()[kq],gk_all_timesteps.G()[kq],i1,i2,wk_all_timesteps.G()[q],wk_all_timesteps.G()[q],i1,i2);
            }
        }
        S.incr_timestep(tstp,stmp,wk);
      }
  }
\end{lstlisting}

As each rank includes only a subset of momentum points $\vec k_{\text{rank}}$ we only evaluate the self-energy diagrams $\Sigma_{\vec k_{\text{rank}}}$ for this subset of momentum points. After the call to {\tt gather\_gk\_timestep}, all ranks carry information about the latest timestep for all momentum points and the internal sum over momentum $\vec q$ in Eq.~\eqref{Eq:Sigmak} can be performed on each rank. The  evaluation of the self-energy is done using the {\tt bubble2} routines introduced in Section~\ref{subsec:contourconv}.

\paragraph{Generic structure of the example program} ---
As the generic structure is similar to the two previous examples we will focus on the peculiarities connected to the usage of \gls*{mpi}. First, we need to initialize the \gls*{mpi} session

\begin{lstlisting}[language=C++]
    MPI::Init(argc,argv);
    ntasks=MPI::COMM_WORLD.Get_size();
    tid=MPI::COMM_WORLD.Get_rank();
    tid_root=0;
\end{lstlisting}
and the {\tt distributed\_timestep\_array} for the electronic and bosonic propagators
\begin{lstlisting}[language=C++]
    DIST_TIMESTEP gk_all_timesteps(Nk,Nt,Ntau,Norb,FERMION,true); 
    DIST_TIMESTEP wk_all_timesteps(Nk,Nt,Ntau,Norb,BOSON,true);
\end{lstlisting}
The construction of the {\tt DIST\_TIMESTEP} variables generates the map  {\tt kindex\_rank} between the subset of points stored on a given rank and the full \gls*{BZ}. 

The program consists of three main parts, namely Matsubara, Bootstrapping ({\tt tstp} $\leq$ {\tt SolverOrder}) and time propagation for {\tt tstp} $>$ {\tt SolverOrder}. The self-consistency iterations include the communication of all fermionic and bosonic propagators between different ranks using the routine {\tt gather\_gk\_timestep} and the determination of the local propagators. For instance, the Matsubara part~({\tt tstp} $=-1$) looks as follows
\begin{lstlisting}[language=C++]
      for(int iter=0;iter<=MatsMaxIter;iter++){
        // update propagators via MPI
        diag::gather_gk_timestep(tstp,Nk_rank,gk_all_timesteps,corrK_rank,kindex_rank);
        diag::gather_wk_timestep(tstp,Nk_rank,wk_all_timesteps,corrK_rank,kindex_rank);

        diag::set_density_k(tstp,Norb,Nk,gk_all_timesteps,lattice,density_k,kindex_rank,rho_loc);
        diag::get_loc(tstp,Ntau,Norb,Nk,lattice,Gloc,gk_all_timesteps);
        diag::get_loc(tstp,Ntau,Norb,Nk,lattice,Wloc,wk_all_timesteps);
\end{lstlisting}
As on each \gls*{mpi} rank, the momentum-dependent single-particle density matrix $\rho(\vec k)$ is known for the whole \gls*{BZ}, the evaluation of the \gls*{HF} contribution is done as in Section~\ref{Sec:Chain}. The self-energies $\Sigma_{k_{\text{rank}}}$ for the momentum points $k_{\text{rank}}=0,\ldots,N_{\text{rank}}-1$ on each rank are obtained by the routine {\tt sigma\_GW}. 

\begin{lstlisting}[language=C++]
        // update mean field and self-energy
        for(int k=0;k<Nk_rank;k++){
          diag::sigma_Hartree(tstp,Norb,corrK_rank[k].SHartree_,lattice,density_k,vertex,Ut);
          diag::sigma_Fock(tstp,Norb,kindex_rank[k],corrK_rank[k].SFock_,lattice,density_k,vertex,Ut);
          diag::sigma_GW(tstp,kindex_rank[k],corrK_rank[k].Sigma_,gk_all_timesteps,wk_all_timesteps,lattice,Ntau,Norb);
        }
\end{lstlisting}
and the variable {\tt vertex} includes~(possibly time-dependent) values of the interaction $U.$

Similarly, the solution  of the Dyson equation for the fermionic~(bosonic) propagators for each momentum point is  obtained by {\tt step\_dyson\_with\_error} ({\tt step\_W\_with\_error}) which is just a wrapper around the Dyson solver. It returns the error corresponding to the difference between the propagators at the previous and current iterations.  The momentum-dependent error for the fermionic propagators is stored in {\tt err\_ele} and at the end we use {\tt MPI\_Allreduce} to communicate among the ranks. 

\begin{lstlisting}[language=C++]
        // solve Dyson equation
        double err_ele=0.0,err_bos=0.0;
        for(int k=0;k<Nk_rank;k++){
          err_ele += corrK_rank[k].step_dyson_with_error(tstp,iter,SolverOrder,lattice);
          diag::get_Polarization_Bubble(tstp,Norb,Ntau,kindex_rank[k],corrK_rank[k].P_,gk_all_timesteps,lattice);
          err_bos += corrK_rank[k].step_W_with_error(tstp,iter,tstp,SolverOrder,lattice);
        }
        MPI::COMM_WORLD.Allreduce(MPI::IN_PLACE,&err_ele,1,MPI::DOUBLE_PRECISION,MPI::SUM);
        MPI::COMM_WORLD.Allreduce(MPI::IN_PLACE,&err_bos,1,MPI::DOUBLE_PRECISION,MPI::SUM);
\end{lstlisting} 

The structure of the bootstrapping and the real-time propagation are equivalent to the Matsubara solver. The main difference lies in the predictor-corrector scheme as explained in Section~\ref{Sec:Chain}. At the beginning of each time step, we extrapolate the momentum-dependent \gls*{GF} $G_{\vec k}$ and the retarded interactions $W_{\vec k}$, which works as a predictor:
\begin{lstlisting}[language=C++]
        // Predictor: extrapolation
        diag::extrapolate_timestep_G(tstp-1,Nk_rank,SolverOrder,Nt,corrK_rank);
        diag::extrapolate_timestep_W(tstp-1,Nk_rank,SolverOrder,Nt,corrK_rank);
\end{lstlisting}
Then we perform several iterations at a given time step until convergence, which  acts as a corrector.

After the \gls*{NEGFs} are obtained, 
we evaluate the kinetic energy (per spin) $E_{\text{kin}}(t)=\frac{1}{N_k}\sum_{\vec k}\text{Tr}[ \rho_{\vec k}(t) \epsilon_{\vec k}(t)]$.
The interaction energy~(per spin) is obtained from the Galitskii-Migdal
formula 
\begin{align}
 E_{\mathrm{int}}(t)=\frac{1}{2 N_k} \sum_{\vec k} \left ( \mathrm{Tr}\left[\rho_{\vec k}(t) \left(h^{\mathrm{MF}}_{\vec k} - \epsilon_{\vec{k}}\right)\right] + \mathrm{Im}\mathrm{Tr} \left[\Sigma_{\vec k}\ast G_{\vec k}\right]^<(t,t) \right),
\end{align}
using the routine {\tt diag::CorrelationEnergy}. The two operations include an \gls*{mpi} reduction as the momentum sum is performed over the whole \gls*{BZ}.

\paragraph{Running the example program} ---
There is one program for the $GW$ calculation, called {\tt gw.x}. The driver script 
{\tt demo\_gw.py} located in the {\tt utils/} directory provides a simple interface to this program. Similar to the examples in Sec.~\ref{sec:example_programs}, the script creates an input file and launches the program. The user can specify the shape of the electric pulse, but by default, we use a single-frequency pulse with a Gaussian envelope 
\beq{
  E(t)=E_0 \exp(-4.6(t-t_0)^2/t_0^2) \sin(\omega (t-t_0)),
  \label{Eq:pulse}
}
where $t_0=2\pi/\omega N_{p}$ is determined by the number of cycles $N_p.$ After the simulation, the time evolution of the kinetic energy and potential energy are plotted. The output is determined by two optional parameters. If {\tt SaveGreen} is {\tt true} the local fermionic ($G$) and bosonic ($W$) propagators are stored to disk. If {\tt SaveMomentum} is {\tt true} also the  momentum-dependent propagators are stored to disk. As the full momentum and time-dependent propagators would require a large amount of memory, we only save selected time slices and their frequency is determined by the parameter {\tt output}. For example, if {\tt output}=100, every 100th timeslice will be stored to disk.

Running the driver script {\tt demo\_gw.py} produces  the following output files:
\begin{enumerate}
\item By default it produces a file {\tt data\_gw.h5}, which includes information about the time-evolution of observables like kinetic energy, density, etc.
\item Setting the parameter {\tt savegf} to 1 will create two additional groups within the file {\tt data\_gw.h5}, namely {\tt Gloc} and {\tt Wloc}. These groups include the total two-time information about the local single particle propagator $G_{\text{loc}}(t,t')$~({\tt Gloc}) and the local two-particle propagator $W_{\text{loc}}(t,t')$~({\tt Wloc}).
\item Setting the parameter {\tt savegk} to 1 will create a set of files for each momentum point. These files include information about the momentum dependent single-particle propagators $G_{\text{k}}(t,t')$~(group {\tt G}) and the corresponding two-particle propagators $W_{\text{k}}(t,t')$~(group {\tt W}).
\end{enumerate}

\begin{figure}[t]
  \centering
  \includegraphics[width=1.0\textwidth]{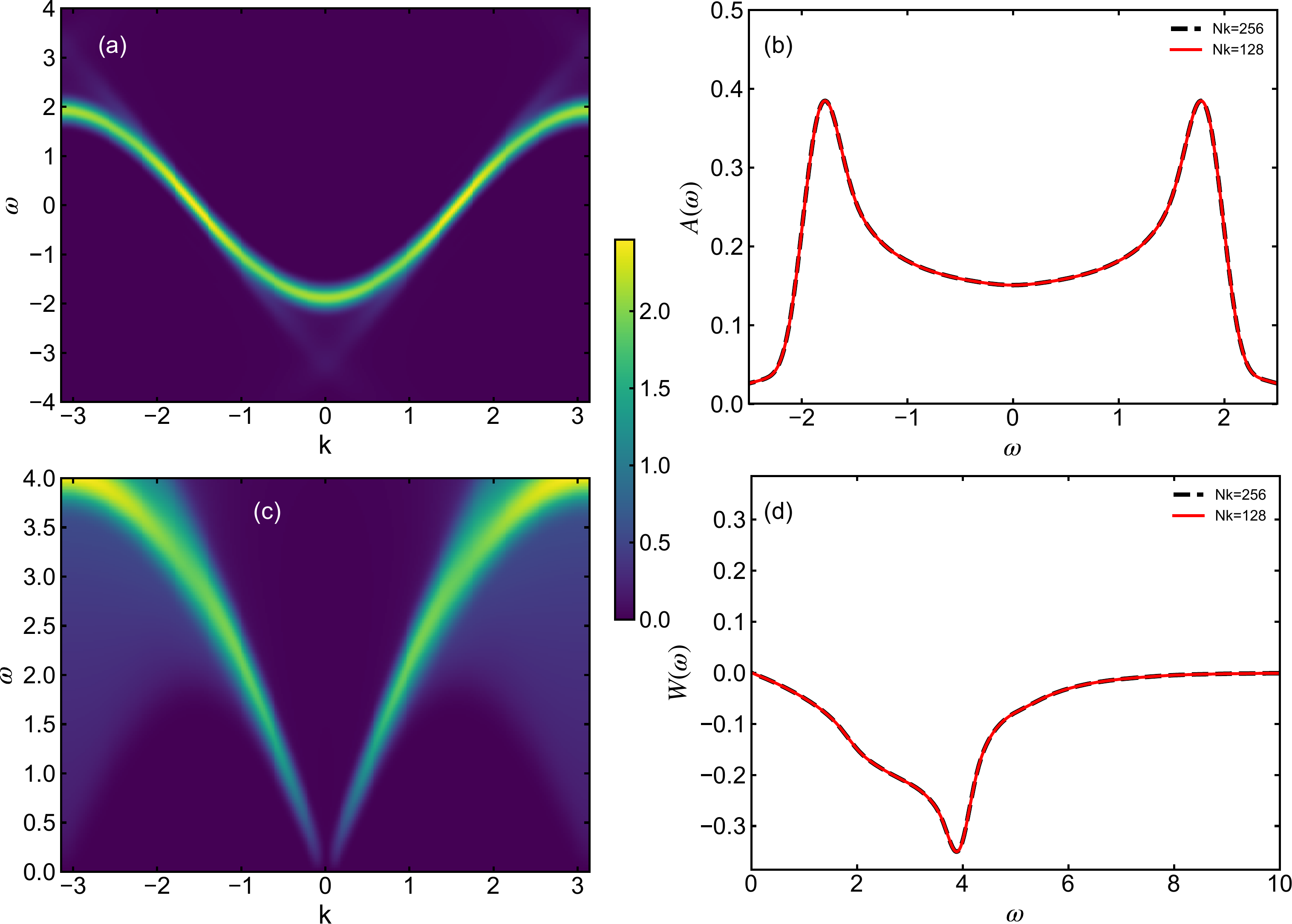}
  \caption{(a)~Momentum-dependent spectral function $A_{\vec k}(\omega)$ of the 1D Hubbard model, obtained within the $GW$ approximation.~(b) Local spectral function $A_{\text{loc}}(\omega)$ for two system sizes $N_k=128$ and $N_k=256$, respectively.
   The second row shows the equivalent pair of panels for the effective interaction:~(c) the momentum-dependent effective interaction $\text{Im}[W_{\vec k}(\omega)]$ and,~(d) its local part $\text{Im}[W_{\text{loc}}(\omega)]$. The parameters for all the plots are $U=2$, the inverse temperature is $\beta=20.0$ and we consider the half-filled case $n=1.$ The momentum-dependent quantities have been obtained with $N_k=256$ momentum points.}
  \label{Fig:gw_spectrum}
\end{figure}

\paragraph{Discussion} ---
The equilibrium momentum-dependent spectral function $A_k(\omega)=-\frac{1}{\pi}\text{Im}\left[ G_{\vec k}(\omega)\right]$ and its local part $A_{\text{loc}}(\omega)=\frac{1}{N_k}\sum_{\vec k}A_k(\omega)$ are presented in Fig.~\ref{Fig:gw_spectrum}. The local spectral function $A_{\text{loc}}(\omega)$ shows the typical van Hove singularities present in \gls*{1D} systems at $\omega \approx \pm 2.$ The comparison between two system sizes, namely $N_k=128$ and $N_k=256$, shows that  the spectrum is converged. The momentum-dependent spectral function $A_{\vec k}(\omega)$ closely follows the single-particle dispersion $\epsilon_{\vec k}$. The broadening due to many-body effects is small close to the Fermi surface points~($\pm \pi/2$), because of the restricted scattering, but it is increasing with increasing energies. Note that the $GW$ approximation cannot capture peculiarities of \gls*{1D} systems, like the absence of the Fermi surface as described by the Tomanaga-Luttinger liquid~\cite{giamarchi2003}. However, this is specific to low-dimensional systems and we consider the 1D case here mainly to avoid heavy calculations in the example program. Another interesting observation is the presence of a shadow band, which is clearly visible for energies away from the chemical potential. The origin of this shadow band is the feedback of the two-particle excitations on the single-particle spectrum. 

\begin{figure}[t] 
  \centering
  \includegraphics[width=0.8\textwidth]{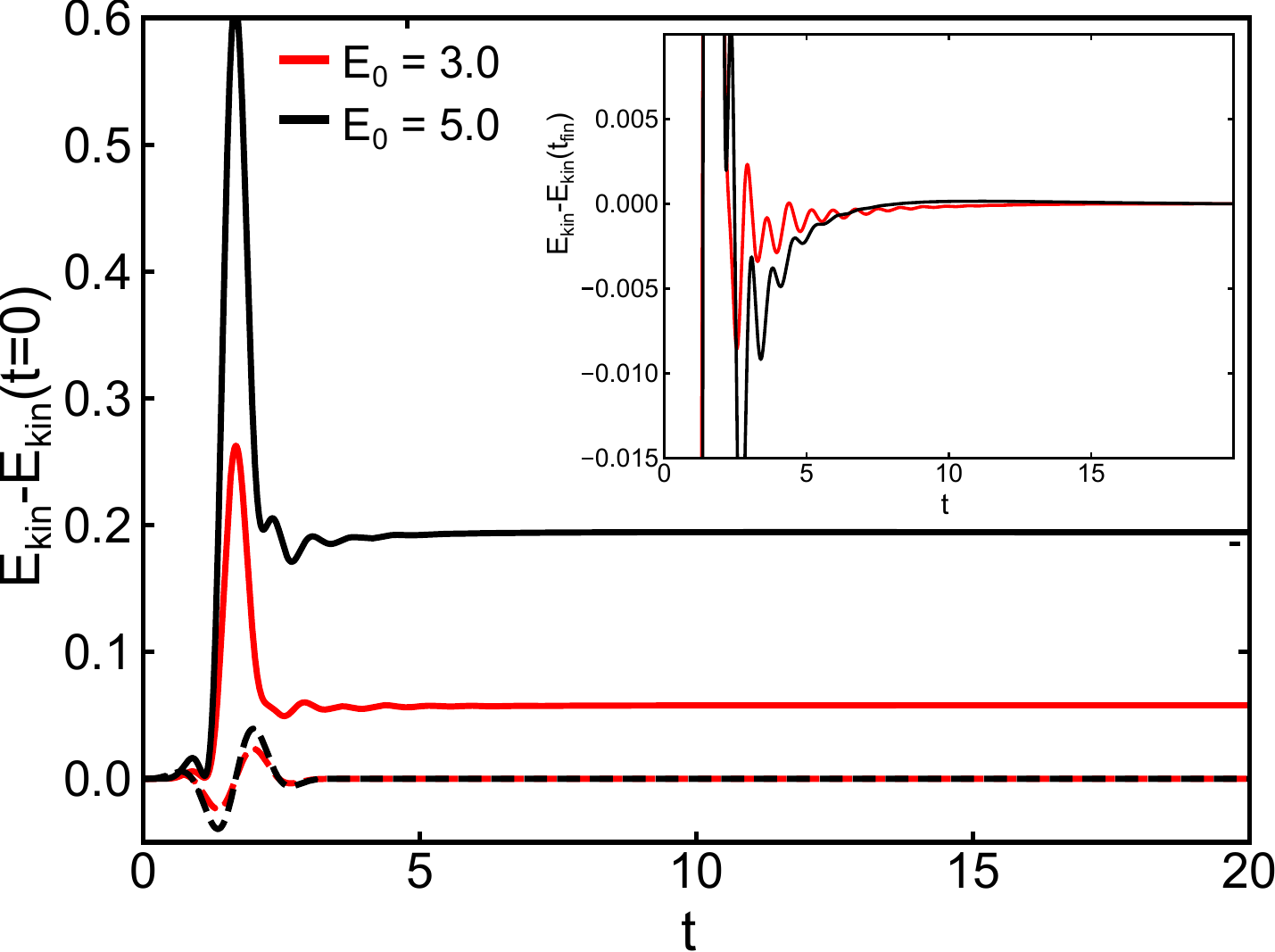}
  \caption{Time evolution of the kinetic energy for the two excitation strengths $E_0=3.0$, $5.0$, respectively. The dashed lines show the shape of the electric field pulse scaled down by 100 to fit on the scale. The inset presents a zoom into the relaxation dynamics  by subtracting the long-time limit $E_{\text{kin}}(t)-E_{\text{kin}}(t_{\text{fin}})$. Both simulations have been performed with $N_k=256$, time step $h=0.01$ and for inverse temperature $\beta=20.$
  }
  \label{Fig:gw_ekin}
\end{figure}

The information about the two-particle excitation spectrum is contained in the bosonic correlator $W$. As the latter is antisymmetric in frequency, $\text{Im}[W(\omega)]=-\text{Im}[W(-\omega)]$, we only present results for positive frequencies, see Fig.~\ref{Fig:gw_spectrum}. The local 
bosonic 
correlator Im[$W_{\text{loc}}(\omega)]$ is presented in Fig.~\ref{Fig:gw_spectrum}(d) for two system sizes $N_k=128$ and $N_k=256$, respectively.  The local component Im[$W_{\text{loc}}(\omega)]$ shows a strong peak around $\omega\approx 4$, which corresponds to particle-hole excitations between the two van-Hove singularities in the single-particle spectrum. 
The effective interaction is rather governed by the particle-hole continuum which for small momenta scales linearly with momentum. The latter is confirmed by the momentum-dependent bosonic correlator Im[$W_{\vec k}(\omega)]$, see Fig.~\ref{Fig:gw_spectrum}(c). At larger momenta, a deviation from the linear dependence is evident, and close to the edge of the \gls*{BZ} the intensity of the bosonic propagator is maximal as it corresponds to the transition between the two van-Hove singularities in the single particle spectrum. 

Now, we turn to the dynamics after the photo-excitation. The system is excited with a short oscillating electric pulse, see Eq.~\eqref{Eq:pulse}, with a single cycle $N_p=1.$  The amplitude of the excitation $E_0$ determines the absorbed energy. In Fig.~\ref{Fig:gw_ekin}, we present the time evolution of the kinetic energy for the two excitation strengths $E_0=3$ and $E_0=5$.
As the energy is increased (during the pulse) and the system heats up, the kinetic energy increases. The observed behavior is consistent with thermalization at a higher temperature, but the transient evolution is complicated by the energy exchange between the electronic and bosonic subsystems (plasmon emission). For the strongest excitations, there is a clear relaxation dynamics to the final state, see inset of Fig.~\ref{Fig:gw_ekin}, accompanied with strongly damped oscillations. 

\begin{figure}[t]
  \centering
  \includegraphics[width=0.8\textwidth]{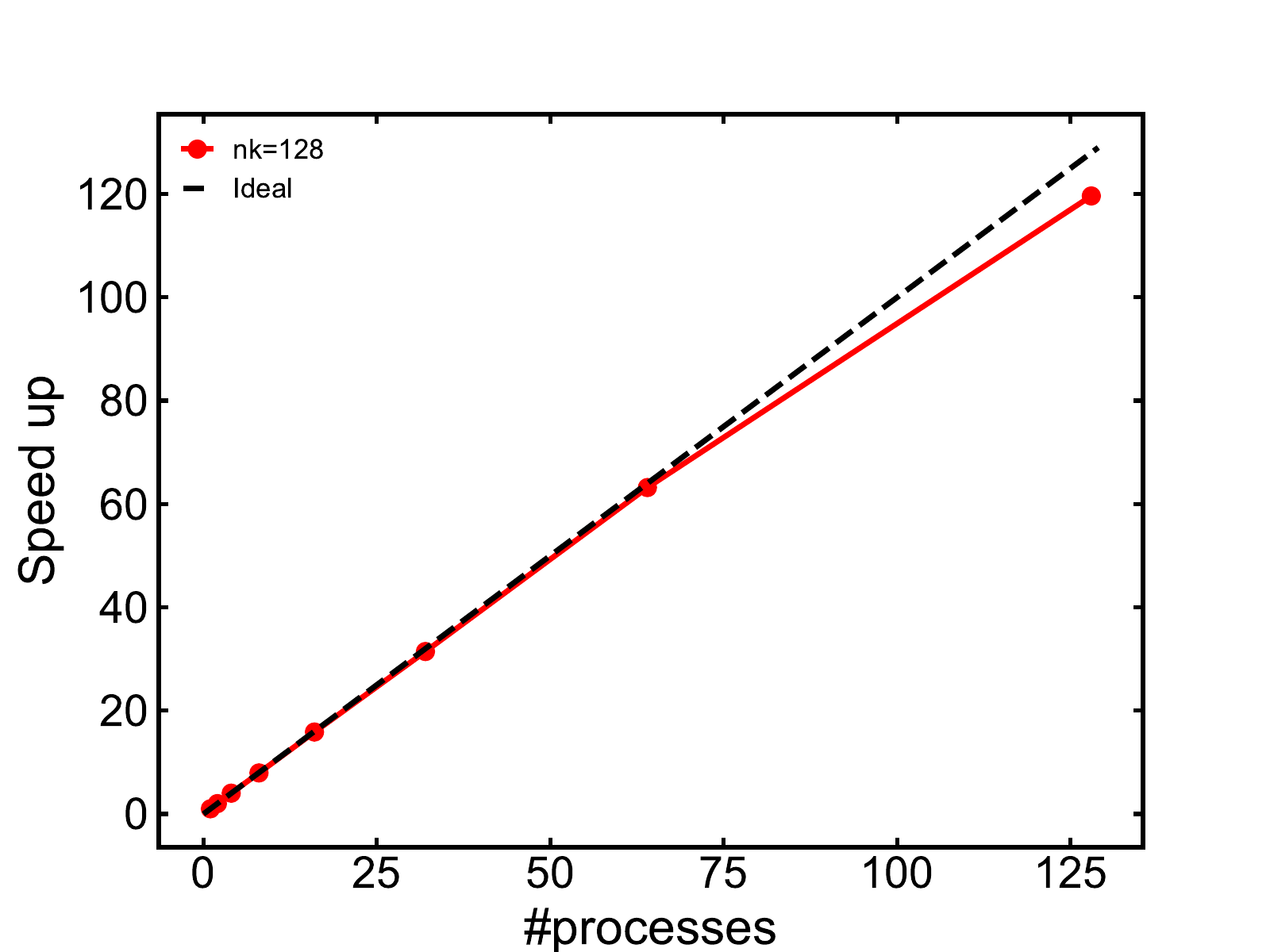}
  \caption{Speed-up  of the total calculation time as a function of the \gls*{mpi} processes for systems with $N_k=128$ momentum points, where we fixed one task per node. The maximum number of time steps used is $N_t=2500.$ These  calculations have been performed on the Popeye cluster at the Flatiron Institute. }
  \label{Fig:gw_scaling}
\end{figure}

In practice, the main bottleneck to reach longer propagation times is the memory restriction imposed by the hardware. The usage of the \gls*{mpi} parallelization scheme over momentum points reduces this issue due to the distribution of memory among different nodes. This is beneficial as long as the number of momentum points  
is an integer multiple of the number of cores.
The usage of the {\tt distributed\_timestep\_array} enables a minimal overlap of the stored information between different nodes, which in all practical cases leads to a linear reduction of the memory requirements per \gls*{mpi} rank.   

Moreover, the \gls*{mpi} parallelization also speeds up the 
execution 
of the program. We have performed a scaling analysis for a system with fixed number of momentum points $N_k=128,$ and parallelization up to 128 processors, see  Fig.~\ref{Fig:gw_scaling}. 
Moreover, for all tests we have fixed the number of tasks per node to one, since in the real-world scenario we want to maximally distribute the memory.
We can see that the scaling is almost perfect up to 128 processors, where a slight deviation from optimal scaling is observed. The main reason for this behavior is the communication overhead, since a substantial amount of data, namely timesteps of propagators for all momentum points, has to be communicated among all nodes. We have tested different communication schemes and the current versions of the library includes the scheme with the best scaling. Of course, we cannot exclude that the scaling for a large number of processors can be improved and this will be an important task for a future update of the library. While the current version can be directly applied to higher dimensional systems~(2D, 3D), future applications to realistic modelling of solids will rely on an efficient parallelization scheme.

\part{Numerical Implementation}
\label{partII}

\section{Basic integration and differentiation rules \label{sec:polynum}}
\label{secqsx01}


In Sections~\ref{secqsx02}--\ref{secqsx05} we describe the
numerics underlying the at least $k^\mathrm{th}$-order accurate solution of the
{\tt dyson}, {\tt vie2}, and {\tt convolution} equations in detail. In this section we define, 
as the first step, the basic notation for polynomial interpolation
as well as approximate relations for evaluating differentials
(backward differentiation) and integrals (Gregory quadrature rules). 

\subsection{Polynomial interpolation\label{subsec:polyinterpolation}}

Consider a function $y(t)$ which takes the values $y_{j}$  at the points $t_j=jh$ of an equidistant mesh $j=0,1,...,k$. We denote the $k^\mathrm{th}$-order polynomial $\tilde{y}(t)$ passing through the points 
$y(j h)=y_j$,
\begin{align}
\label{jgwcik}
\tilde y(jh)=y_{j}, \quad j=0,\dots,k,
\end{align}
by $\mathcal{P}^{(k)}[y_{0},\dots,y_{k}](t)$. The interpolation can be cast into the matrix form,
\begin{align}
\label{dwjvwdjkjxbk}
&\mathcal{P}^{(k)}[y_{0},\dots,y_{k}](t) = \sum_{a,l=0}^k h^{-a}t^a P^{(k)}_{al} y_{l}, 
\\
\label{integrator_P}
&
P^{(k)}_{al}
=
(M^{-1})_{al}\text{~for~}M_{ja}=j^a.
\end{align}
With Eqs.~\eqref{dwjvwdjkjxbk} and \eqref{integrator_P}, Eq.~\eqref{jgwcik} can be verified directly,
\begin{align}
\tilde y(jh)
=
\sum_{a,l=0}^k j^a P^{(k)}_{al} y_{l}
=
\sum_{a,l=0}^k M_{ja}(M^{-1})_{al} y_{l} = y_{j}.
\end{align}
The precomputed weights $P_{al}^{(k)}$ can be obtained from the {\tt
  integrator} class (see Section~\ref{sec:compile_use}).

\subsection{Polynomial differentiation}

An approximation for the derivative $dy/dt$ of a function can be obtained by taking the exact derivative of the polynomial approximant \eqref{dwjvwdjkjxbk},
\begin{align}
\label{dwjvwdjkjxbk01}
\frac{dy}{dt}\Big|_{t=mh}
&\approx
\frac{d}{dt}
\mathcal{P}^{(k)}[y_{0},...,y_{k}](mh)
 = \sum_{a=1}^k\sum_{l=0}^k P^{(k)}_{al}h^{-a} a(mh)^{a-1} y_{l}
 \\
 &
=
h^{-1}\sum_{a=1}^k\sum_{l=0}^k P^{(k)}_{al} a m^{a-1} y_{l}.
\end{align}
We thus arrive at an approximative relation for {\it polynomial differentiation}
\begin{align}
\label{integrator_D01}
&\frac{dy}{dt}\Big|_{t=mh}
\approx
h^{-1} \sum_{l=0}^k D_{ml}^{(k)} y_{l}, \text{~~with}
\\
&
\label{integrator_D}
D_{m,l}^{(k)}=\sum_{a=1}^k P^{(k)}_{al} a m^{a-1}.
\end{align}
The precomputed weights $D_{m,l}^{(k)}$ are stored by the {\tt integrator} class (see Section~\ref{sec:compile_use}).

\subsection{Polynomial integration}

In  some cases below, the polynomial interpolation formula is also used to get the approximation for an integral.
For $0\le m\le n \le k$, 
\begin{align}
\label{dwjvwdjkcw}
\int_{mh}^{nh} dt \,y(t)
&\approx
\int_{mh}^{nh} dt \,
\mathcal{P}^{(k)}[y_0,...,y_k](t)
 = \sum_{a=0}^k\sum_{l=0}^k 
 \int_{mh}^{nh} dt 
P^{(k)}_{al}h^{-a} t^{a} y_{l}
\\
&
 = h\sum_{l=0}^k \left[\sum_{a=0}^k P^{(k)}_{al}
 \int_{m}^{n} dt\,
 t^{a}\right] y_{l}.\
\end{align}
%
We thus use the folowing approximative relation for {\it polynomial integration}
\begin{align}
&\int_{mh}^{nh} dt\, y(t)
\approx
h \sum_{l=0}^k I_{m,n,l}^{(k)} y_{l}, \text{~~with}
\\
\label{integrator_I}
&
I_{m,n,l}^{(k)}=\sum_{a=0}^k P^{(k)}_{al}\frac{n^{a+1}-m^{a+1}}{a+1}.
\end{align}
The precomputed weights $I_{m,n,l}^{(k)}$ are implemented in the {\tt
  integrator} class (see Section~\ref{sec:compile_use}).

\subsection{Backward differentiation} 

Consider a function $y(t)$ which takes the values $y_{j}$ at the points
$t_j=jh$ of an equidistant mesh $j=0,1,...,n$, with $n\ge k$. The \gls*{BDF} 
of order $k$ approximates the
derivative $dy/dt$ at $t=nh$ using the function values
$y_{n}, y_{n-1},...,y_{n-k}$. It is defined via the linear relation
\begin{align}
\frac{dy}{dt}\Big|_{nh}
\approx
h^{-1}
\sum_{j=0}^k
a^{(k)}_j y_{n-j}.
\label{integrator_bd}
\end{align}
Here the coefficients for the $k^\mathrm{th}$ order formula are obtained by the derivative $\tilde y'(t=0)$ of the $k^\mathrm{th}$ order polynomial interpolation $\tilde y(t)$ defined by the values $\tilde y(jh)=y_{n-j}$ (backward differentiation is thus a special case of the polynomial differentiation),
\begin{align}
\frac{dy}{dt}\Big|_{nh}
\approx
-\frac{d}{dt} \mathcal{P}^{(k)}[y_n,y_{n-1},...,y_{n-k}](t=0).
\end{align}
Note that the minus sign is due to the reversed order of the interpolated points.
Therefore, the coefficients of the \gls*{BDF} are directly related to
coefficients for polynomial differentiation: $a^{(k)}_j=-D^{(k)}_{0,j}$. The coefficients for the
first $k$ are tabulated in Table \ref{tab:integrator_bd}. The
precomputed weights $a_{j}^{(k)}$ can be obtained from the {\tt
  integrator} class (see Section~\ref{sec:compile_use}).

\begin{table}[tbp]
\centerline{
\begin{tabular}{|c||c|c|c|c|c|c|c|}
\hline
$k$&$a_0$&$a_1$&$a_2$&$a_3$&$a_4$&$a_5$&$a_6$\\
\hline
\hline
$1$&$1$&$-1$&&&&&\\
\hline
$2$&$\frac{3}{2}$&$-2$&$\frac{1}{2}$&&&&\\
\hline
$3$&$\frac{11}{6}$&$-3$&$\frac{3}{2}$&$-\frac{1}{3}$&&&\\
\hline
$4$&$\frac{25}{12}$&$-4$&$\frac{6}{2}$&$-\frac{4}{3}$&$\frac{1}{4}$&&\\
\hline
$5$&$\frac{137}{60}$&$-5$&$\frac{10}{2}$&$-\frac{10}{3}$&$\frac{5}{4}$&$-\frac{1}{5}$&\\
\hline
$6$&$\frac{49}{20}$&$-6$&$\frac{15}{2}$&$-\frac{20}{3}$&$\frac{15}{4}$&$-\frac{6}{5}$&$\frac{1}{6}$\\
\hline
\end{tabular}
}
\caption{Weights of the backward differentiation formula~\eqref{integrator_bd} up to $k=6$.}
\label{tab:integrator_bd}
\end{table}

\subsection{Gregory Integration}

The solution of \gls*{vies} 
discussed below is based on a combination of backward-differentiation formulae with so-called Gregory quadrature rules for the integration. The $k^\mathrm{th}$ Gregory quadrature rule on a linear mesh is defined by the equation
\begin{align}
\mathcal{I}_n\equiv \int_{0}^{nh}
\!\!dt\, y(t)
\approx
h
\sum_{j=0}^{m(n,k)}
w^{(k)}_{n,j}
y_j,
\,\,\,\,
m(n,k)=\begin{cases}
n& n>k
\\
k & n\le k
\end{cases}.
\label{integrator_greg}
\end{align}
The weights $w^{(k)}_{n,j}$ are explained below. In general,  the approximation for the integral is obtained from function values $\{y_j : 0\le j\le n\}$ 
{\em within} the integration interval $[0,nh]$, i.e. $m(n,k)=n$. However, this is not possible for $n<k$, because a $k^\mathrm{th}$ order accurate quadrature rule cannot be constructed  from less than $k+1$ function values. In the Gregory quadrature for $n<k$, we assume that the function $y(t)$ exists outside the interval $[0,nh]$, and construct an approximation for the integral  from values $\{y_j : 0\le j\le k\}$, i.e., $m(n,k)=k$. 

The simplest example of a Gregory quadrature rule is the trapezoidal approximation,
\begin{align}
\label{trazi}
\int_{0}^{nh}
\!\!dt\, y(t)
\approx h
\Big(
\tfrac12 y_0 + y_1 + \cdots + y_{n-1}+\tfrac12 y_n
\Big),
\end{align}
which corresponds to $k=0$, $m(n,k)=n$, and the weights $w^{(k)}_{n,j}=\tfrac12$ for $j\in\{0,n\}$ and $w^{(k)}_{n,j}=1$ for $0<j<n$. The weights $w^{(k)}_{n,j}$ for a general $k^\mathrm{th}$ order accurate rule are implicitly defined by the following procedure:
\begin{itemize}
\item
$n\le k$: 
$I_n$ is approximated by the exact integral over the polynomial interpolation $\mathcal{P}^{(k)}[y_0,...y_{k}](t)$,
\begin{align}
\mathcal{I}_n
\approx
\int_0^{nh}
dt \,\mathcal{P}^{(k)}[y_0,...y_{k}](t)
\equiv
h
\sum_{j=0}^k
s^{(k)}_{n,j}
y_j.
\label{ldnljwcn}
\end{align}
Hence the weights for $n\le k$, which are denoted as {\em starting weights} $w^{(k)}_{n,j}=s^{(k)}_{n,j}$, 
are equivalent to the polynomial integration weights \eqref{integrator_I}, 
\begin{align}
w^{(k)}_{n,j} = I^{(k)}_{0,n,j}\equiv s^{(k)}_{n,j} \ ,\quad 0\le n \le k.
\end{align}
\item
$n>k$:
To generate an approximation for the integral 
$\mathcal{I}(t)=\int_0^t d\bar t \,y(\bar t)$ at $t=nh$ and $n>k$, we consider the differential equation
\begin{align}
\frac{d}{dt}\mathcal{I}(t) = y(t),\,\,\,\mathcal{I}(0)=0.
\end{align}
This equation is solved by taking the values  $\mathcal{I}_{j}$ for $0\le j\le k$ from the approximation \eqref{ldnljwcn}, and solving for $\mathcal{I}_n$ at $n>k$ by applying the \gls*{BDF}
. The resulting set of linear equations for $\mathcal{I}_n$ with $n>k$,
\begin{align}
h^{-1}
\sum_{l=0}^k
a^{(k)}_l
\mathcal{I}_{m-l}
=
y_m,
\,\,\,
m=k+1,...,n,
\end{align}
implicitly determines the values of the integral.
\end{itemize}
This  procedure defines a weight matrix with the following structure:
\begin{align}
w^{(k)}_{n,j}
=
\begin{pmatrix}
s^{(k)}_{00} & \cdots & s^{(k)}_{0k} & 0 & 0
\\
\vdots &&  \vdots  & 
\\
s^{(k)}_{k0} & \cdots & s^{(k)}_{kk} & 0 & 0
\\
\Sigma^{(k)}_{00} & \cdots & \Sigma^{(k)}_{0k} & \omega^{(k)}_{0} & 0  
\\
\vdots &&  \vdots  & &\ddots & \ddots
\\
\Sigma^{(k)}_{k0} & \cdots & \Sigma^{(k)}_{kk} & \omega^{(k)}_{k} & \cdots &  \omega^{(k)}_{0} & 0 
\\
\omega^{(k)}_{0} & \cdots & \omega^{(k)}_{k} & 1 & \omega^{(k)}_{k} & \cdots &  \omega^{(k)}_{0} & 0 
\\
\omega^{(k)}_{0} & \cdots & \omega^{(k)}_{k} & 1 & 1 & \omega^{(k)}_{k} & \cdots &  \omega^{(k)}_{0} & 0 
\\
\omega^{(k)}_{0} & \cdots & \omega^{(k)}_{k} & 1 & \cdots & 1 & \omega^{(k)}_{k} & \cdots &  \omega^{(k)}_{0} & 0
\end{pmatrix}.
\end{align}
Here the upper block are the weights obtained from the polynomial
approximation. For $n>k$ we have $m(n,k)=n$. For $n\ge 2k+1$ the
weights are symmetric $w^{k}_{n,j}=w^{k}_{n,n-j}\equiv \omega^{(k)}_j$
for $j\le k$.
For $n\ge 2k+1$ the
weights satisfy $w^{(k)}_{n,n-j-1}\equiv \omega^{(k)}_j$
for $j\le k$, 
and furthermore 
$w^{(k)}_{n,j}=1$ for $k<j<n-k-1$.
The
latter property makes this quadrature rule different from, e.\,g., the Simpson
rule, where the weights alternate between $\tfrac23$ and $\tfrac43$ but
never become one. Gregory quadrature rules for $n>2k+1$ can thus be
understood as a simple Riemann sum
$\mathcal{I}_n\approx h\sum_{j=0}^n y_j$ with a {\em boundary
  correction} obtained from the function values
$\{y_j, y_{n-j}: 0\le j \le k\}$, thus generalizing the structure of
the trapezoidal rule \eqref{trazi}. For completeness, the integration
rules for some of the lowest $k$ are presented in Table
\ref{tab:gregory}. The weights for $k=1,\dots,5$ can be obtained from the {\tt
  integrator} class (see Section~\ref{sec:compile_use}).

\begin{table}
{ $k=0$:}
\begin{tabular}{|c||c|}
\hline
$0$ & $\frac{1}{2}$ 
\\
\hline
\end{tabular}
\\[1mm]
{$k=1$:}
\begin{tabular}{|c|c||c|c|}
\hline
$0$ & $0 $& $\frac{5}{12}$ &  $\frac{7}{6}$
\\
\hline
$\frac{1}{2}$ & $\frac{1}{2}$ &  $\frac{5}{12}$ & $\frac{13}{12}$
\\
\hline
\end{tabular}
\\[1mm]
{$k=2$:}
\begin{tabular}{|c|c|c||c|c|c|}
\hline
 $0$ &$0$ &$0$ &$\frac{3}{8}$ &$\frac{9}{8}$ &$\frac{9}{8}$
 \\
 \hline
 $\frac{5}{12}$ &$\frac{2}{3}$ &$-\frac{1}{12}$ &$\frac{3}{8}$ &$\frac{7}{6}$ &$\frac{11}{12}$
 \\
 \hline
 $\frac{1}{3}$ &$\frac{4}{3}$ &$\frac{1}{3}$ &$\frac{3}{8}$ &$\frac{7}{6}$ &$\frac{23}{24}$
\\
\hline
\end{tabular}
\\[1mm]
\begin{tabular}{|c|c|c||c|c|c|}
\hline
$s^{(k)}_{0,0}$ & $\cdots$  & $s^{(k)}_{0,k}$ & $\Sigma^{(k)}_{0,0}$ & $\cdots$  & $\Sigma^{(k)}_{0,k}$
\\
$\vdots$ & & $\vdots $&$\vdots$ & & $\vdots $
\\
$s^{(k)}_{k,0} $& $\cdots$  & $s^{(k)}_{k,k}$ &$\Sigma^{(k)}_{k,0}=\omega^{(k)}_0$ & $\cdots$  & $\Sigma^{(k)}_{k,k}=\omega^{(k)}_k$ 
\\
\hline
\end{tabular}\caption{Weights of the first few Gregory integration rules, Eq.~\eqref{integrator_greg}. In the table for each $k$, the numbers in the left (right) $(k+1)\times(k+1)$ block  define the weights $s^{(k)}_{l,j}$ ($\Sigma^{(k)}_{l,j}$), respectively, as shown in the last table. The $\omega$ weights can be read off from the last row of the $\Sigma$ weights, $\omega^{(k)}_j=\Sigma^{(k)}_{k,j}$, $j=0,...,k$.}
\label{tab:gregory}
\end{table}

\begin{figure}[t]
  \centering
  \includegraphics[width=\textwidth]{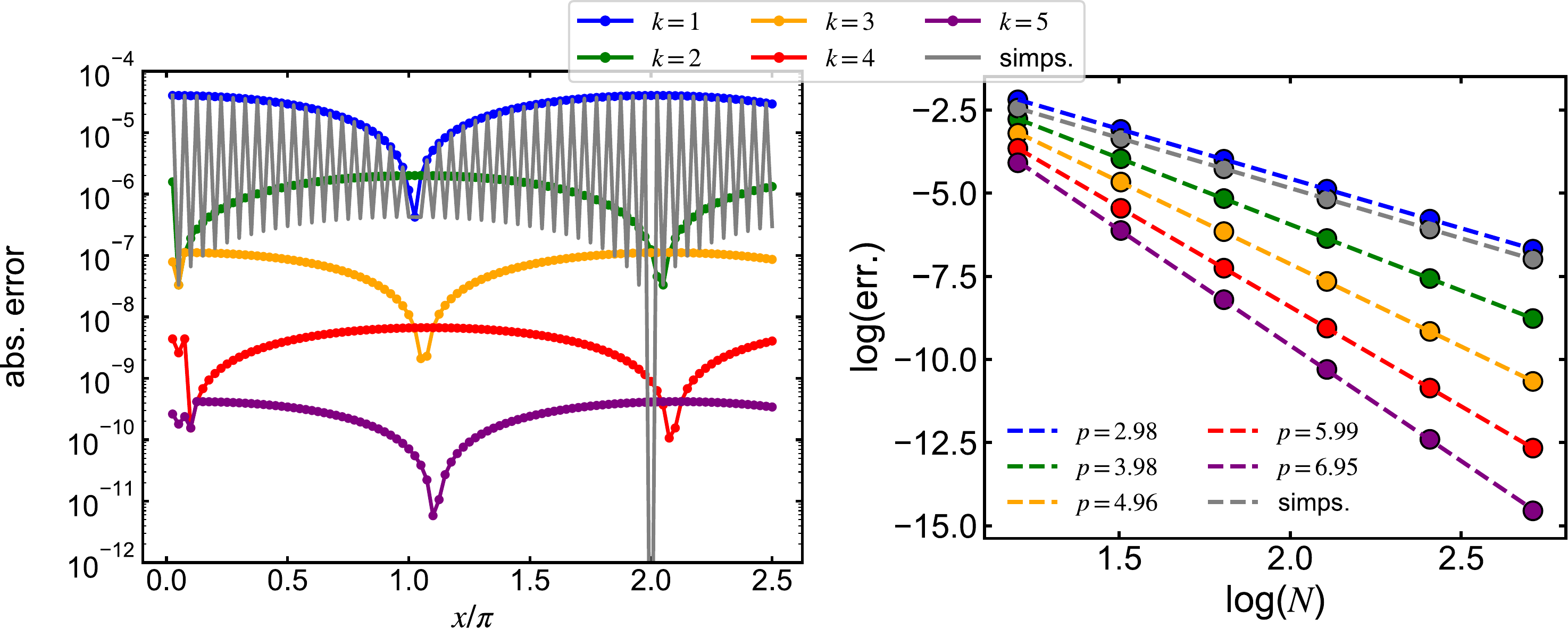}
  \caption{Left 
  panel: absolute error $|\mathcal{I}_n-\mathcal{I}^\mathrm{ex}_n|$
    of the Gregory integration for different orders $k$. Right panel:
    mean absolute error for integrating up to $x_\mathrm{max}=5\pi/2$,
    discretizing the interval into $N$ points with $h=0.025\pi$. The dashed lines are
    linear fits confirming the order of the Gregory quadrature.
    \label{fig:gregoryintegration}}
\end{figure}

The advantage of the Gregory integration is a uniform
approximation of $\mathcal{I}_n$: for any $n$, the error scales as
$\mathcal{O}(h^{k+2})$~\cite{steinberg_numerical_1972}.
This is different from Newton-Cotes rules of the same order $k$, 
which are only $k^\mathrm{th}$ order accurate
for a certain number of grid points. For instance, the Simpson rule
requires an odd number of grid points. 
The accuracy of the Gregory
integration is illustrated in Fig.~\ref{fig:gregoryintegration} for the integral
$y(x) = \exp(i x)$ with exact integral
$\int_0^{x} dx' y(x')=-i(\exp(i x)-1)$. In particular, the panel on
the right-hand side of Fig.~\ref{fig:gregoryintegration} confirms
that the average absolute error $(1/N) \sum^N_{n=0} |\mathcal{I}_n-\mathcal{I}^\mathrm{ex}_n|$ as a function of
the number of points $N$ scales as $\mathcal{O}(N^{-p})$ with
$p\approx k+2$. 
\footnote{In the solution of the Volterra integral equations, discussed in Sec.~\ref{secqsx02}, the overall error depends on the accuracy of this quadrature rule, but also of the starting procedure and the differential operator.}
  We also compare to the Simpson's rule, employing the
  trapezoidal rule for integrating over $[nh, (n-1)h]$ if $n$ is odd
  (the total number of points $n+1$ is even). As
  Fig.~\ref{fig:gregoryintegration} shows, this primitive extension of
  Simpson's rule induces oscillatory behavior of the error, as the
  accuracy is hampered by the trapezoidal rule. Thus, the scaling of
  the averaged error is effectively reduced to first order ($k=1$).

  Furthermore, the construction of the Gregory weights makes
  computing $\mathcal{I}(t)$ by Gregory quadrature and solving the corresponding
  differential equation by the \gls*{BDF} method 
  numerically equivalent, ensuring consistency of the integral or
  differential formulation.
  

\subsection{Boundary convolution}
\label{hjqxkbjaxLXNS}

In this paragraph we introduce a $k^\mathrm{th}$-order accurate approximation for a special kind of convolution integral, which appears in the context of imaginary time convolutions. Consider the convolution integral 
\begin{align}
c(t) = \int_0^t dt^\prime F(t-t^\prime)G(t^\prime)
\end{align}
between two functions $F$ and $G$ which are only defined for $t>0$, and cannot be continued into a differentiable function on the domain $t<0$. On an equidistant mesh with $t=mh$ and $m<k$, the integration range includes less than $k$ points, and the functions must be continued outside the integration range in order to obtain an $k^\mathrm{th}$-order accurate approximation. Because of the structure of the convolution integral, $F(t-t^\prime)$ can only be continued to the domain $t^\prime<0$, while $G(t^\prime)$ should be continued to the domain $t^\prime>t$.

We use the approximation
\begin{align}
c(mh) = \int_0^{mh} dt^\prime  \mathcal{P}^{(k)}[F_{0},...,F_k](mh-t^\prime)\mathcal{P}^{(k)}[G_0,...,G_k](t^\prime).
\end{align}
Using Eq.~\eqref{integrator_P}, this  can be transformed into
\begin{align}
c(mh) &= 
\sum_{r,s,a,b=0}^k
\int_0^{mh} dt^\prime  P^{(k)}_{ar} h^{-a}(mh-t^\prime)^a F_{r}P^{(k)}_{bs} h^{-b}(t^\prime)^b G_s 
\\
&=
\sum_{r,s=0}^k F_{r} G_s \Big[ \sum_{a,b=0}^k \int_0^{mh} dt^\prime  P^{(k)}_{ar} h^{-a}(mh-t^\prime)^a P^{(k)}_{bs} h^{-b}(t^\prime)^b \Big].
\end{align}
The terms in brackets are coefficients which can be precomputed, so that finally 
\begin{align}
\label{integrator_R}
&\int_0^{mh} dt^\prime F(t-t^\prime)G(t^\prime) = h \sum_{r,s=0}^k F_{r} G_s R^{(k)}_{m;r,s},
\\
&R^{(k)}_{m;r,s} 
=
\sum_{a,b=0}^k
P^{(k)}_{a,r}
P^{(k)}_{b,s} 
\int_0^{m} dx
 (m-x)^a 
 x^b.
\end{align}
The precomputed weights $R^{(k)}_{m;r,s}$ can be obtained from the
{\tt integrator} class.

\section{Numerical details: Volterra integral equations\label{sec:num_vie}}
\label{secqsx02}

The numerical solution of the {\tt vie2} and {\tt dyson} integral
equations is based on a mapping of these equations onto a set of
coupled \gls*{vies} or \gls*{vides}. In this section we first explain
$k^\mathrm{th}$-order accurate algorithms for the solution of Volterra
equations. These algorithms are discussed in detail in the book by
Brunner and van Houven~\cite{brunner_numerical_1986}.

\subsection{Volterra integro-differential equation}

We consider a \gls*{vide} of the form
\begin{align}
\label{MBCVSAMx}
\frac{dy}{dt} + p(t)y(t) + \int_{0}^t ds \,k(t,s)y(s) = q(t).
\end{align}
For given $k(t,s)$, $p(t)$, and $q(t)$ and an {\em initial condition}
specifying $y(0)$, this equation must be solved to determine $y(t)$
in the domain $t>0$.  In the numerical solution all functions are
known or determined on an equidistant mesh $t_j=jh$, $j=0,1,2,...$,
and we use the notation $k_{jl}=k(jh,lh)$, $p_l=p(lh)$, etc.  Here and
in the following, the values of the functions $k,p,q,y$ can be complex
matrices of 
size
$d>1$.

The at least $k^\mathrm{th}$-order accurate solution of this equation is obtained by combining the $k^\mathrm{th}$-order Gregory quadrature  
in two steps:
\begin{itemize}
\item[1)]
{\it Start-up:}
A procedure which is used to obtain a solution $y_j$ for $j=1,...k$. 
\item[2)] 
{\it Time-stepping:} A procedure to obtain $y_{n}$ for $n>k$ from $\{y_{j} : j< n\}$.
\end{itemize}
In the algorithm explained below, the time-stepping is causal,
i.e., the solution $y_n$ does not depend on the input $k_{lj}$, $p_j$,
$q_j$ at $l>n$ or $j>n$. For the start-up, the numerical error at
$y_{n}$ for $n<k$ can depend on the input $k_{lj}$, $p_j$, $q_j$ at
$0\le l,j \le k$. Furthermore, in the numerical implementation we
assume that the kernel $k(t,t^\prime)$ can be defined as a
differentiable function on the whole domain $0\le t,t^\prime$,
although only the values at $0\le t^\prime\le t$ enter the exact
integral.

\subsubsection{Start-up procedure}

We express the derivative in \eqref{MBCVSAMx} in terms of the
polynomial differentiation Eqs.~\eqref{integrator_D01} and
\eqref{integrator_D}, and use the Gregory integration
\eqref{integrator_greg} for the convolution
\begin{align}
h^{-1}
\sum_{l=0}^k D^{(k)}_{n,l}y_l
 + p_n y_n+ 
h \sum_{l=0}^k w^{(k)}_{n,l} \, k_{nl}y_l 
= q_n \text{~for~}n=1,...,k.
\end{align}
This defines a linear equation
\begin{align}
\label{vdiff_start}
\begin{pmatrix}
M_{1,1} & \cdots & M_{1,k}
\\
\vdots&&\vdots
\\
M_{k,1} & \cdots & M_{k,k}
\end{pmatrix}
\begin{pmatrix}
y_1
\\
\vdots
\\
y_k
\end{pmatrix}
=
\begin{pmatrix}
q_1 - M_{1,0} y_0
\\
\vdots
\\
q_n - M_{k,0} y_0
\end{pmatrix},
\end{align}
where the Matrix $M$ is given by
\begin{align}
M_{n,l}
=
h^{-1}
D^{(k)}_{n,l}+\delta_{nl} p_n + hw^{(k)}_{nl} \, k_{n,l}
\end{align}
This $k\times k$ dimensional linear equation is solved directly. Note that when $y,k,p$, and $q$ are $d$-dimensional matrices, the solution of the linear system amounts to inverting 
a matrix of size $(kd)\times (kd)$.

\subsubsection{Time-stepping}

The time-stepping is done using a combination of backward
differentiation \eqref{integrator_bd} and Gregory integration
\eqref{integrator_greg}
\begin{align}
h^{-1}
\sum_{l=0}^k
a^{(k)}_{l}
y_{n-l}
+
p_n y_n
+
h
\sum_{l=0}^n
w^{(k)}_{n,l}
k_{n,l}y_l
=q_n.
\end{align}
If the $y_j$ are known for $j<n$ we obtain a linear equation  for $y_n$,
\begin{align}
\label{MBCVSAMx_t}
&\Big[h^{-1}a^{(k)}_0 + p_n + hw^{(k)}_{nn} k_{n,n}\Big] y_n
=\Big[
q_n -h^{-1}\sum_{l=1}^k
a^{(k)}_{l}
y_{n-l}
-
h
\sum_{l=0}^{n-1}
w^{(k)}_{n,l}
k_{n,l}y_l
\Big],
\end{align}
which is solved for $y_n$.

\subsubsection{Conjugate equation}

For later convenience we also define the start-up and time-stepping
relations to solve an equivalent conjugate equation
\begin{align}
\label{MBCVSAMx_cc}
\frac{dy}{dt} + y(t)p(t) + \int_{0}^t ds \,y(s)k(s,t) = q(t).
\end{align}
The start-up determines the values $y_1,...,y_k$ by solving the $k\times k$ linear equation
\begin{align}
\label{MBCVSAMx_cc_start}
\sum_{l=1}^k y_l M_{l,n} = q_n - y_0M_{0,n},\,\,\,\,1\le n \le k,
\end{align}
where the Matrix $M$ is given by
\begin{align}
M_{l,n}
=
h^{-1}
D^{(k)}_{n,l}+\delta_{nl} p_n + hw^{(k)}_{n,l} \, k_{l,n}.
\end{align}
In the time-stepping, $y_n$ for $n>k$ is determined by solving the linear equation
\begin{align}
\label{MBCVSAMx_cc_t}
&y_n \Big[h^{-1}a^{(k)}_0 + p_n + hw^{(k)}_{nn} k_{n,n}\Big] = \Big[
q_n -h^{-1}\sum_{l=1}^k
a^{(k)}_{l}
y_{n-l}
-
h
\sum_{l=0}^{n-1}
w^{(k)}_{n,l}
y_lk_{l,n}
\Big].
\end{align}

\subsection{Volterra Integral equation of the second kind\label{subsec:standardvie2}}

Small modifications of Eqs.~\eqref{MBCVSAMx} and \eqref{MBCVSAMx_cc} lead to the 
\gls*{vies} of the second kind
\begin{align}
\label{MBCVSAMx01}
&y(t) + \int_{0}^t ds \,k(t,s)y(s) = q(t),
\\
\label{MBCVSAMx01_cc}
&y(t) + \int_{0}^t ds \,y(s)k(s,t) = q(t),
\end{align}
with the same assumptions on the domain and the kernel. These equations are solved with the initial condition $y_0=q_0$.

Start-up and time-stepping procedures are 
obtained from the integro-differen-tial equation by setting $p(t)=1$ and omitting the differential: The start-up procedure for Eq.~\eqref{MBCVSAMx01}
determines the values $y_1,...,y_k$ by solving the $k\times k$ linear equation
\begin{align}
\label{MBCVSAMx01_start}
\sum_{l=1}^k  M_{n,l} y_l = q_n - M_{n,0}y_0,\,\,\,\,1\le n \le k,
\end{align}
where the Matrix $M$ is given by
\begin{align}
M_{n,l}
=
\delta_{nl} + hw^{(k)}_{n,l} \, k_{n,l}.
\end{align}
In the time-stepping, $y_n$ for $n>k$ is determined by solving the linear equation
\begin{align}
\label{MBCVSAMx01_t}
&\Big[1 + hw^{(k)}_{nn} k_{n,n}\Big] y_n = \Big[
q_n - h
\sum_{l=0}^{n-1}
w^{(k)}_{n,l}
k_{n,l}y_l
\Big].
\end{align}
The start-up procedure for the conjugate Eq.~\eqref{MBCVSAMx01_cc}
determines the values $y_1,...,y_k$ by solving the $k\times k$ linear equation
\begin{align}
\label{MBCVSAMx01_cc_start}
\sum_{l=1}^k  y_l M_{l,n}  = q_n - y_0M_{0,n},\,\,\,\,1\le n \le k,
\end{align}
where the Matrix $M$ is given by
\begin{align}
M_{l,n}
=
\delta_{nl} + hw^{(k)}_{n,l} \, k_{l,n}.
\end{align}
In the time-stepping, $y_n$ for $n>k$ is determined by solving the linear equation
\begin{align}
\label{MBCVSAMx01_cc)t}
&y_n 
\Big[1 + hw^{(k)}_{nn} k_{n,n}\Big] = \Big[
q_n - h
\sum_{l=0}^{n-1}
w^{(k)}_{n,l}
y_l k_{l,n}
\Big].
\end{align}

\section{Numerical details: convolution integrals on
  $\mathcal{C}$ \label{sec:num_convolution}}
\label{secqsx03}

In this section we present $k^\mathrm{th}$-order 
discrete
approximations for various contour convolution integrals which appear
in the solution of the {\tt dyson}, {\tt vie2}, and {\tt convolution}
problems. The integrals constitute different contributions to the
convolution \eqref{convolution}, 
which we separate 
into
the Matsubara, retarded, mixed or lesser components
of a contour function $C$. All equations are obtained in a straightforward way from the 
Gregory integration \eqref{integrator_greg} if the integration
interval includes more than $k+1$ function values, and from the polynomial
integration \eqref{integrator_I} or the boundary convolution
\eqref{integrator_R} otherwise.

Below, 
we indicate by the label
$\stackrel{c}{=}$ those equations which are {\em exactly} causal, i.\,e., the result $C$
at real time arguments $\le nh$ does not depend on the input functions
$A$, $f$, $B$ with (one or both) real time arguments larger than
$n$. For the other equations, causality is satisfied only up to the
numerical accuracy. Furthermore, by adding  a tilde $\tilde A$, $\tilde B$
over the input functions in an equation we indicate that some of the input
values of $A$ and $B$ lie outside the domain of the {\tt
  herm\_matrix} type and must be 
  obtained 
  from the hermitian
conjugates $A^\ddagger$ and $B^\ddagger$, respectively
(see Section~\ref{sec:basic}).

\paragraph{Matsubara}
\begin{align}
\label{cm1}
C^\mat_1&[A,f,B]
(m) 
= 
\int_0^{mh_\tau}d\tau' A^\mat(mh_\tau-\tau^\prime) f(0^-)B^\mat(\tau^\prime)
=
\\
&
\begin{cases} 
\,\,\, \stackrel{c}{=} h_\tau
\sum_{j,l=0}^kR^{(k)}_{m;j,l}A^\mat_{j} f_{-1} B^\mat_l
& m\le k
\\
\,\,\, \stackrel{c}{=} h_\tau
\sum_{l=0}^m w^{(k)}_{m,l} A^\mat_{m-l}f_{-1} B^\mat_l
&
m>k
\end{cases}.
\\
\label{cm2}
C^\mat_2&[A,f,B]
(m) = 
\int_{mh_\tau}^\beta d\tau'  A^\mat(mh_\tau-\tau^\prime) f(0^-) B^\mat(\tau^\prime)
=
\\
&
\begin{cases} 
\,\,\, \stackrel{c}{=} h_\tau
\sum_{j,l=0}^k
R^{(k)}_{N_\tau-m;j,l} \xi A^\mat_{N_\tau- j}f_{-1}B^\mat_{N_\tau-l}
& m \ge N_\tau -k
\\
\,\,\, \stackrel{c}{=} h_\tau
\sum_{l=0}^{N_\tau -m} w^{(k)}_{N_\tau-m,l} \xi
  A^\mat_{N_\tau- l}f_{-1} B^\mat_{m+l}
&
m<N_\tau-k
\end{cases}.
\end{align}
In the second equation $A^\mat(\tau)$ at the values
  $\tau\in[-\beta,0]$ 
  is obtained by using the 
  periodicity
property $A^\mat(\tau+\beta) = \xi A^\mat(\tau)$.
\paragraph{Retarded}
\begin{align}
\label{cr1}
C^{\mathrm{R}}&[A,f,B](n,m) 
= \int_{mh}^{nh} d\bar t A^{\mathrm{R}}(nh,\bar t ) f(\bar t) B^{\mathrm{R}}(\bar t,mh)
=
\\
&\begin{cases}
\,\,\,\stackrel{c}{=} h\sum_{j=m}^{n} w^{(k)}_{n-m,j-m}
 A^{\mathrm{R}}_{n,j} f_j B^{\mathrm{R}}_{j,m} 
 &n>k, n-m>k
 \\
\,\,\,\stackrel{c}{=}  h \sum_{j=0}^{k} w^{(k)}_{n-m,j}
 A^{\mathrm{R}}_{n,n-j} f_{n-j} \tilde B^{\mathrm{R}}_{n-j,m} 
 &n>k, n-m\le k
\\
\,\,\,= h\sum_{j=0}^k
I^{(k)}_{m,n;j} 
\tilde A^{\mathrm{R}}_{n,j} f_j \tilde B^{\mathrm{R}}_{j,m}
&n\le k 
\end{cases}. 
\end{align}
As mentioned above, the tilde $\tilde B^{\mathrm{R}}_{j,m}$ in the third
equation indicates that $B^{\mathrm{R}}_{j,m}$ is also evaluated 
outside the domain $j\ge m$ of the {\tt herm\_matrix} type, and thus
needs to be reconstructed from $B^\ddagger$, i.e.,
$\tilde B^{\mathrm{R}}_{j,m}=B^{\mathrm{R}}_{j,m}= -
(B^\ddagger)^{\mathrm{R}}_{m,j}$. Analogous definitions hold for
$\tilde A^{\mathrm{R}}_{n,j}$ in the third equation, and
$\tilde B^{\mathrm{R}}_{n-j,m}$ in the second equation.

\paragraph{Left-mixing Components}
\begin{align}
\label{ctv1}
C^{\rceil}_1&[A,f,B](n,m)
=\int_{0}^{nh} d\bar t A^{\mathrm{R}}(nh,\bar t ) f(\bar t) B^{\rceil}(\bar t,mh_\tau) 
\\
&
\begin{cases}
\,\,\,
\stackrel{c}{=}
h
\sum_{j=0}^{n} w^{(k)}_{n,j}
 A^{\mathrm{R}}_{n,j} f_j B^{\rceil}_{j,m} 
 &n>k,
 \\
 \,\,\,
=h
\sum_{j=0}^{k} w^{(k)}_{n,j}
 \tilde A^{\mathrm{R}}_{n,j} f_j B^{\rceil}_{j,m} 
 &n\le k
\end{cases},
\\
\label{ctv2}
C^{\rceil}_2&[A,f,B](n,m)
=\int_0^{mh_\tau} d\tau A^{\rceil}(nh,\tau^\prime) f(0^-) B^\mat(\tau^\prime-mh_\tau)
\\
&
\begin{cases} 
\,\,\,
\stackrel{c}{=}h_\tau
\sum_{j,l=0}^kR^{(k)}_{m;j,l}A^{\rceil}_{l} f_{-1} \xi B^\mat_{N_\tau-j}
& m\le k
\\
\,\,\,
\stackrel{c}{=}h_\tau
\sum_{l=0}^m w^{(k)}_{m,l} A^{\rceil}_{m-l}f_{-1}\xi B^\mat_{N_\tau-l}
&
m>k
\end{cases},
\\
\label{ctv3}
C^{\rceil}_3&[A,f,B](n,m)
=\int_{mh_\tau}^\beta d\tau A^{\rceil}(nh,\tau') f(0^-)B^\mat(\tau'-mh_\tau)
\\
&
\begin{cases} 
\,\,\,
\stackrel{c}{=}h_\tau
\sum_{j,l=0}^k
R^{(k)}_{N_\tau-m;j,l}A^{\rceil}_{N_\tau-l}f_{-1}B^\mat_{j}
& m \ge N_\tau -k
\\
\,\,\,
\stackrel{c}{=}h_\tau
\sum_{l=0}^{N_\tau -m} w^{(k)}_{N_\tau-m,l} A^{\rceil}_{m+l}f_{-1}B^\mat_{l}
&
m<N_\tau-k.
\end{cases}
\end{align}

\paragraph{Lesser Components $n\le m$}

\begin{align}
\label{cles1}
C^{<}_1&[A,f,B](n,m) 
=
\int_{0}^{nh} d\bar t A^{\mathrm{R}}(nh,\bar t ) f(\bar t)B^{<}(\bar t,mh) 
\\
&\begin{cases}
\,\,\,\stackrel{c}{=} 
h\sum_{j=0}^{n} w^{(k)}_{n,j}
 A^{\mathrm{R}}_{n,j} f_j B^<_{j,m} 
 &n>k,
 \\
\,\,\,=  h \sum_{j=0}^{k} w^{(k)}_{n,j}
 \tilde{A}^{\mathrm{R}}_{n,j} f_{j} B^<_{j,m} 
 &n\le k
\end{cases}.
\\
\label{cles2}
C^{<}_2&[A,f,B](n,m) 
=
\int_{0}^{mh} d\bar t A^<(nh,\bar t ) f(\bar t) B^{\mathrm{A}}(\bar t,mh) 
=
\\
&\begin{cases}
\,\,\,\stackrel{c}{=} 
h\sum_{j=0}^{m} w^{(k)}_{m,j}
A^<_{n,j} f_j B^A_{j,m} 
 &m>k,
 \\
\,\,\,=  
h\sum_{j=0}^{k} w^{(k)}_{m,j}
 A^<_{n,j} f_j \tilde{B}^\mathrm{A}_{j,m} 
 &m\le k
\end{cases}.
\\
\label{cles3}
C^{<}_3&[A,f,B](n,m) 
=
-i \int_0^\beta d\tau A^{\rceil}(nh,\tau') f(0^-)B^\lceil(\tau,mh)
=
\\
&=-ih_\tau \sum_{j=0}^{N_\tau} w^{(k)}_{N_\tau,j}
A^{\rceil}_{n,j} f_{-1} B^\lceil_{j,m}.
\end{align}
Because the advanced and right-mixing components are not stored by the {\tt
  herm\_matrix} type, these quantities must be reconstructed from the
hermitian conjugate. For example,
$B^\lceil_{j,m} = -\xi [B^\ddagger]^{\rceil}_{m,j}$ in the third
equation, and $B^{\mathrm{A}}_{j,m} = [B^\ddagger]^{\mathrm{R}}_{j,m}$ in
the second equation.

\section{Implementation: \tt convolution \label{sec:impl_convolution}}

\subsection{Langreth rules}

In this section we  present the implementation of the {\tt
  convolution} routine which solves Eq.~\eqref{convolution}.  Using
the Langreth rules, the convolution integral \eqref{convolution} is split
into contributions from the Matsubara, retarded, left-mixing, and lesser
components
\begin{align}
  \label{eq:conv_langreth}
C^\mat(\tau) 
&= \int_0^\beta d\tau ' A^\mat(\tau-\tau')f(0^-) B^\mat(\tau'),
\\
C^{\mathrm{R}}(t,t^\prime) 
&
= \int_{t^\prime}^t d\bar t A^{\mathrm{R}}(t,\bar t ) f(\bar t) B^{\mathrm{R}}(\bar t,t^\prime),
\\
C^{\rceil}(t,\tau)
&
=
\int_{0}^t d\bar t A^{\mathrm{R}}(t,\bar t ) f(\bar t) B^{\rceil}(\bar t,\tau) 
\nonumber
\\
&\,\,\,\,
+
\int_0^\beta d\tau A^{\rceil}(t,\tau') f(0^-)B^\mat(\tau'-\tau),
\\
C^{<}(t,t^\prime) 
&
=
\int_{0}^t d\bar t A^{\mathrm{R}}(t,\bar t ) f(\bar t)B^{<}(\bar t,t^\prime)
+
\int_{0}^{t^\prime} d\bar t A^<(t,\bar t ) f(\bar t) B^{\mathrm{A}}(\bar t,t^\prime) 
\nonumber
\\
&\,\,\,\,
-i \int_0^\beta d\tau A^{\rceil}(t,\tau') f(0^-)B^\lceil(\tau,t^\prime).\label{conv_les}
\end{align}
$k^\mathrm{th}$-order approximations to these individual components have been presented in Section~\ref{secqsx03}. 

It is also convenient to introduce the convolution of a two-time
contour object with a function as
\begin{align}
  \label{eq:resp_conv}
  c(t) = \int_\mathcal{C}\! d\bar{t} A(t,\bar{t}) f(\bar{t}) \ .
\end{align}
Eqs.~\eqref{eq:conv_langreth}--\eqref{conv_les} can be adapted to this case by replacing $B(t,t^\prime)$ by the
identity function.

\subsection{Matsubara}

The evaluation of $C^\mat(\tau)$, i.e., $C$ at timeslice $\mathcal{T}[C]_{-1}$
is implemented as
(c.f. Eqs.~\eqref{cm1} and \eqref{cm2})
\begin{align}
C^\mat(mh_\tau)
=
C^\mat_1[A,f,B](m)
+
C^\mat_2[A,f,B](m)
\text{~for~}m=0,...,N_\tau.
\end{align}

\subsection{Time steps}

The evaluation of $C$ at timeslice $\mathcal{T}[C]_{n}$ for $n\ge 0$
is implemented as follows:
\begin{itemize}
\item
For $m=0,...,n$ [c.f. Eq. ~\eqref{cr1}]:
\begin{align}
C^{\mathrm{R}}(nh,mh)
=
C^{\mathrm{R}}_1[A,f,B](n,m) .
\end{align}
\item For $m=0,...,N_\tau$  [c.f. Eqs.~\eqref{ctv1}, \eqref{ctv2}, \eqref{ctv3}]:
\begin{align}
C^{\rceil}(nh,mh_\tau)
=&\,
C^{\rceil}_1[A,f,B](n,m)
+
C^{\rceil}_2[A,f,B](n,m)
\nonumber
\\
&
+
C^{\rceil}_3[A,f,B](n,m).
\end{align}
\item For $m=0,...,n$  [c.f. Eqs.~\eqref{cles1}, \eqref{cles2}, \eqref{cles3}]:
\begin{align}
C^{<}(mh,nh)
=&\,
C^{<}_1[A,f,B](m,n)
+
C^{<}_2[A,f,B](m,n)
\nonumber
\\
&
+
C^{<}_3[A,f,B](m,n).
\end{align}
\end{itemize}
Comparison with the causal properties of Eqs.~\eqref{cm1} to
\eqref{cles3} shows that the causal time-dependence indicated in Table
\ref{tab:convolution} is satisfied.

Response convolutions of the type of Eq.~\eqref{eq:resp_conv} are
obtained by replacing $B \rightarrow 1$ and simplifying the integration formalae in Section~\ref{sec:num_convolution} accordingly.

\section{Implementation: {\tt dyson} \label{sec:impl_dyson}}
\label{secqsx04}

\subsection{Langreth rules}

In this section we  present the implementation of the {\tt
  dyson} routine which solves  Eq.~\eqref{dyson}. To solve
  Eq.~\eqref{dyson}, we again invoke the Langreth rules to split the
equation of motion on the \gls*{KB} contour into the respective equations
for the Matsubara, lesser, and left-mixing components,
\begin{align}
&-\partial_\tau
G^\mat(\tau) 
-
\epsilon(0^-) 
G^\mat(\tau) 
-
\int_0^\beta d\tau ' \,\Sigma^\mat(\tau-\tau')G(\tau')
=
\delta(\tau),
\label{dyson_langreth_m}
\\
&i\partial_t
G^{\mathrm{R}}(t,t^\prime) 
-
\epsilon(t) 
G^{\mathrm{R}}(t,t^\prime)
-
 \int_{t^\prime}^t d\bar t \,\Sigma^{\mathrm{R}}(t,\bar t ) G^{\mathrm{R}}(\bar t,t^\prime)
 =0
 \label{dyson_langreth_r}
\\
&i\partial_t G^{\rceil}(t,\tau)
-
\epsilon(t) G^{\rceil}(t,\tau)
-
\int_{0}^t d\bar t \,\Sigma^{\mathrm{R}}(t,\bar t ) G^{\rceil}(\bar t,\tau) 
\nonumber
\\
&\hspace{10mm} =
\int_0^\beta d\tau \Sigma^{\rceil}(t,\tau') G^\mat(\tau'-\tau),
\label{dyson_langreth_tv}
\\
&i\partial_t
G^{<}(t,t^\prime) 
-
\epsilon(t) 
G^{<}(t,t^\prime) 
-
\int_{0}^t d\bar t \,\Sigma^{\mathrm{R}}(t,\bar t ) G^{<}(\bar t,t^\prime)
\nonumber
\\
&\hspace{10mm} =\int_{0}^{t^\prime} d\bar t \,\Sigma^<(t,\bar t ) G^{A}(\bar t,t^\prime) 
-i \int_0^\beta d\tau \,\Sigma^{\rceil}(t,\tau') G^\lceil(\tau,t^\prime).
\label{dyson_langreth_les}
\end{align}
Here   Eq.~\eqref{dyson_langreth_m} must be solved with the  boundary condition
\begin{align}
\label{dyson_ini_m}
G^\mat(-\tau)=\xi G^\mat(\beta-\tau),
\end{align}
and the remaining equations are solved with initial conditions
\begin{align}
\label{dyson_ini_r}
G^{\mathrm{R}}(t,t)
&=-i,
\\
\label{dyson_ini_tv}
G^{\rceil}(0,\tau)
&=iG^\mat(-\tau)=i\xi G^\mat(\beta-\tau),
\\
\label{dyson_ini_les}
G^{<}(0,t^\prime)
&=-[G^\rceil(t^\prime,0)]^\dagger.
\end{align}
In the solution of the {\tt dyson} problem, we will use in part the
conjugate equation \eqref{dyson_cc} for the retarded and lesser
component. These equations translate into
\begin{align}
&-i\partial_{t^\prime}
G^{\mathrm{R}}(t,t^\prime) 
-
G^{\mathrm{R}}(t,t^\prime)
\epsilon(t^\prime) 
-
 \int_{t^\prime}^t d\bar t \,G^{\mathrm{R}}(t,\bar t ) \Sigma^{\mathrm{R}}(\bar t,t^\prime)
 =0,
 \label{dyson_langreth_r_cc}
\\
&-i\partial_{t^\prime}
G^{<}(t,t^\prime) 
-
G^{<}(t,t^\prime) 
\epsilon(t^\prime) 
-
\int_{0}^t d\bar t\, G^{\mathrm{R}}(t,\bar t ) \Sigma^{<}(\bar t,t^\prime)
\nonumber
\\
&\hspace{10mm}=\int_{0}^{t^\prime} d\bar t \,G^<(t,\bar t ) \Sigma^{\mathrm{A}}(\bar t,t^\prime) 
-i \int_0^\beta d\tau \,G^{\rceil}(t,\tau') \Sigma^\lceil(\tau,t^\prime).
\label{dyson_langreth_les_cc}
\end{align}
which are solved with the initial conditions \eqref{dyson_ini_r} and
\begin{align}
\label{dyson_ini_les_cc}
G^{<}(t,0)
&=G^{\rceil}(t,0).
\end{align}

\subsection{Matsubara\label{subsec:impl_dyson_matsubara}}

The Matsubara \gls*{GF} is obtained by solving
  Eq.~\eqref{dyson_langreth_m}.
Unlike the Dyson equations for the
real-time and mixed components,   Eq.~\eqref{dyson_langreth_m}
constitutes a boundary-value integro-differential equation.

\paragraph{Fourier series representation} --- 
The 
(anti-) periodicity $G^\mat(\tau+\beta)=\xi G^\mat(\tau)$ allows
to express the Matsubara \gls*{GF} by the Fourier series
\begin{align}
  \label{eq:Gmat_fourier}
  G^\mat(\tau) = \frac{1}{\beta} \sum^{N_\omega}_{m=-N_\omega} e^{-i
  \omega_m \tau}
  G^\mat(i\omega_m)
\end{align}
with $N_\omega\rightarrow \infty$, where
\begin{align}
  \omega_m = \begin{cases} \frac{2m \pi}{\beta} & : \mathrm{bosons} \\
    \frac{2(m+1) \pi}{\beta} & : \mathrm{fermions}
  \end{cases}
\end{align}
denote the Matsubara frequencies. The Fourier coefficients
$G^\mat(i\omega_m)$ are, in turn, determined by
\begin{align}
  \label{eq:Gmat_fourier_coeff}
  G^\mat(i\omega_m) = \int^\beta_0\! d\tau\, G^\mat(\tau)e^{i \omega_m
  \tau} \ .
\end{align}
Defining the imaginary frequency representation of the self-energy
$\Sigma^\mat(i \omega_m)$ in an analogous fashion, the Dyson
equation~\eqref{dyson_langreth_m} is transformed into the algebraic
equation
\begin{align}
  \left(i \omega_m - \epsilon(0^-) \right) G^\mat(i \omega_m) =
  G^\mat(i \omega_m) \Sigma^\mat(i \omega_m) \ , 
\end{align}
which is readily solved for $G^\mat(i \omega_m)$. Evaluating the
Fourier sum~\eqref{eq:Gmat_fourier} then yields $G^\mat(\tau)$.

Due to the discontinuity of $G^\mat(\tau)$ at $\tau=0$ and $\tau=\beta$, \gls*{GFs} 
show the asymptotic behavior $G(i\omega_n)\sim (i\omega_n)^{-1}$. These tails must be treated exactly in order to assure convergence of the Fourier sum~\eqref{eq:Gmat_fourier}. Modifying the Matsubara \gls{GF} in $0\le \tau\le \beta$
  according to
  \begin{align}
    \widetilde{G}^\mat(\tau) = \begin{cases}
      G^\mat(\tau) + \frac12 & : \text{fermions} \\
      G^\mat(\tau) + \frac{\tau}{\beta}- \frac12 & : \text{bosons}
    \end{cases},
    \label{kcwbljDKHB1}
  \end{align}
  and in an (anti-) periodic fashion outside this interval, removes the discontinuity at $\tau=0$ and $\tau=\beta$, so that
  $\widetilde{G}^\mat(\tau)$ becomes a continous function. The Fourier
  coefficients are thus obtained by 
  \begin{align}
    \label{eq:Gmat_fourier_coeff_mod}
    G^\mat(i \omega_m) = -\frac{\xi}{i \omega_m} +
    \widetilde{G}^\mat(i\omega_m) \ ,
  \end{align}
  where $\widetilde{G}^\mat(i\omega_m)$ is analogous to
   Eq.~\eqref{eq:Gmat_fourier_coeff}. We numerically perform the back-transformation \eqref{eq:Gmat_fourier} on 
  $\widetilde{G}^\mat(i\omega_m)$, and then obtain  $G^\mat(\tau)$ from \eqref{kcwbljDKHB1}.
  
  For the Fourier transform, we use a piecewise cubic interpolation, yielding a cubically corrected
  discrete Fourier transformation as described in chapter 13.9 of
   Ref.~\cite{press_numerical_2007}.  The convergence of this method is determined by the number of frequency points $N_\omega$. We chose $N_\omega= p N_\tau$ in the  Fourier sum~\eqref{eq:Gmat_fourier}, where $p$ is an \emph{oversampling} factor (typically $p=10$). 
  
In practice, the convergence of this method is limited by the tail
correction and thus the average error scales as
$\mathcal{O}(h^2_\tau)$ (see Section~\ref{sec:example-scaling} for an
illustrative example). The accuracy can be improved to
$\mathcal{O}(h^{k+2}_\tau)$ 
\footnote{The accuracy of solution of an integral equation  $G+F\ast G=Q$ is identical the accuracy of the quadrature rule if the convolution integral is bounded such that 
$||F\ast \delta G|| < const. ||\delta G||$.}
by solving the integral
equation~\eqref{dyson_langreth_m}. For convenience, we reformulate the
Dyson equation in terms of the integral equation
\begin{align}
  \label{eq:Gmat_dyson_integ}
  G^\mat(\tau) = g^\mat(\tau) + [K\ast G]^\mat(\tau) \ , \ K^\mat(\tau) = [g\ast
  \Sigma]^\mat(\tau) \ ,
\end{align}
where $g^\mat(\tau)$ solves   Eq.~\eqref{dyson_langreth_m} with $\Sigma
^\mat = 0$. The exact solution reads
\begin{align}
  g^\mat(\tau) = -\bar{f}_{\xi}(\epsilon(0^-)-\mu)\exp(-\epsilon(0^-) \tau)
  \ ,
\end{align}
where $\bar{f}_{\xi}(\omega) = 1 + \xi f_{\xi}(\omega)$ and
$f_{\xi}(\omega)$ denote the Fermi ($\xi=-1$) or Bose ($\xi=1$)
distribution, respectively.
Equation~\eqref{eq:Gmat_dyson_integ} constitutes a linear
equation for $G^\mat(m h_\tau)$. 

We have implemented a variation of
Newton's method for solving this equation iteratively:
\paragraph{Newton iteration} ---
After solving for $G^\mat(m h_\tau)$ via the Fourier method, the
residual
\begin{align}
  \label{eq:Gmat_res}
  R(m h_\tau) = G^\mat(m h_\tau) - [K\ast G]^\mat(m h_\tau) -g^\mat(m h_\tau) 
\end{align}
is generally not zero, as the accuracy of the Fourier method is
different from the $k^\mathrm{th}$-order accurate convolution. 
We can regard $R$ defined in Eq.~\eqref{eq:Gmat_res} as a functional $R[G]$. 
Finding the root $R[G]=0$ of the functional is equivalent to solving the Dyson
equation in integral form. To find the root, we set up an iteration in the form
\begin{align}
  G^{\mat,(i+1)}(m h_\tau) = G^{\mat,(i)}(m h_\tau) - \Delta
  G^{\mat,(i)}(m h_\tau) \ ,
\end{align}
where the update to the $i^\mathrm{th}$ iteration, $\Delta
G^{\mat,(i)}(m h_\tau)$, obeys the equation
\begin{align*}
   \Delta G^{\mat,(i)}(m h_\tau)- [K\ast \Delta G^{(i)}]^\mat(m h_\tau) = R^{(i)}(m h_\tau) \ .
\end{align*}
To estimate the update, the above equation is solved using the Fourier
method. This procedure provides a rapidly converging\footnote{If the error of solving the auxiliary equation for $\Delta
  G^\mat$ can be neglected, exactly one iteration is required to
  reach convergence.}
iteration to
minimize the magnitude of the resolvent~\eqref{eq:Gmat_res}. As an
initial guess $G^{\mat,(0)}(m h_\tau)$, we again employ the Fourier
method. This procedure can be considered as the Newton iteration for finding the root of the functional $R[G]$ with an approximation for the derivative $\delta R/\delta G$.


The routine {\tt dyson\_mat}
provides a general interface for both methods. The optional
argument {\tt method} can be set to {\tt CNTR\_MAT\_FOURIER} if the 
Fourier method is to be used, or to {\tt CNTR\_MAT\_FIXPOINT} for the Newton iteration.

\subsection{Start\label{subsec:impl_dyson_start}}

The {\tt dyson\_start} routine evaluates $G$ on the time-slices
$\mathcal{T}[G]_n$ for $0\le n \le k$ (c.f. Table \ref{tab:dyson}).

\begin{itemize}
\item To determine $G^{\mathrm{R}}(nh,mh)$ for $0\le n\le k$ and
  $n\le m \le k$ we consider   Eq.~\eqref{dyson_langreth_r} with initial
  condition \eqref{dyson_ini_r}. The solution is similar to the
  start-up procedure for a Volterra equation \eqref{MBCVSAMx}: At each
  fixed $m$, we use a polynomial approximation for
  $y(t)=G^{\mathrm{R}}(t,mh)$ with $G^{\mathrm{R}}_{n,m}= y_n$, 
\begin{align}
y_n=
\begin{cases}
G^{\mathrm{R}}_{n,m} & m < n \le k
\\
-i & m=n
\\
-[G^{\mathrm{R}}_{m,n}]^\dagger & 0 \le n <m
\end{cases}.
\end{align}
Here the values $y_{n}$ for $n<m$ amount to a continuous extrapolation
of $G^{\mathrm{R}}(t,t^\prime)$ to the domain $t<t^\prime$. When
  Eq.~\eqref{dyson_langreth_r} is solved successively for $m=0,1,...,k$,
the values $y_n$ are already known for $n\le m$. Inserting the
polynomial ansatz for $y(t)$ into \eqref{dyson_langreth_r} yields
\begin{align}
ih^{-1}\sum_{l=0}^k  D^{(k)}_{n,l}y_l + \epsilon_n y_n - h\sum_{l=0}^k I^{(k)}_{m,n;l}\tilde{\Sigma}^\ret_{n,l}y_l=0.
\end{align}
This is transformed into a $(k-m)\times(k-m)$ linear problem,
\begin{align}
&\sum_{l=m+1}^k M_{n,l}y_l = - \sum_{l=0}^{m} M_{n,l}y_l\equiv Q_n,\,\,\,n=m+1,...,k,
\\
&M_{n,l} =ih^{-1}D^{(k)}_{n,l} + \delta_{n,l}\epsilon_n- h I^{(k)}_{m,n;l}\tilde{\Sigma}^\ret_{n,l}.
\end{align}
Because the input $y_{l\le m}$ for $Q_n$ has been computed previously, this equation can be solved for $y_{l>m}$.
\item To determine $G^{\rceil}(nh,mh_\tau)$ for $0\le n\le k$ and
  $0\le m \le N_\tau$ we consider   Eq.~\eqref{dyson_langreth_tv} with
  initial condition \eqref{dyson_ini_tv}.  For each given $m$, this
  equation provides a Volterra equation of standard type~\eqref{MBCVSAMx}, with the replacement
\begin{align}
&y(t)=G^{\rceil}(t,\tau) \ , \ p(t)=i\epsilon(t) \ , \ k(t,s)=i\Sigma^{\mathrm{R}}(t,s), 
\\
&q(t)= -i\int_{0}^\beta d\bar \tau \Sigma^{\rceil}(t,\bar \tau) G^\mat(\bar \tau-\tau).
\end{align}
For $0\le n\le k$, the Volterra equation is solved using the start-up algorithm \eqref{vdiff_start}, where   
the convolution routines Eqs.~\eqref{ctv2} and \eqref{ctv3} are used to evaluate $q(nh)$,
\begin{align}
q(t)
=
-iC^{\rceil}_2[\Sigma,1,G](n,m)-iC^{\rceil}_3[\Sigma,1,G](n,m).
\end{align}
\item To determine $G^{<}(mh,nh)$ for $0\le n\le k$ and $0\le m \le n$
  we consider   Eq.~\eqref{dyson_langreth_les} with the initial condition
  \eqref{dyson_ini_les}.  For each given $n$, this equation
  corresponds to a
  Volterra equation of standard type~\eqref{MBCVSAMx}, with the
  replacement
\begin{align}
&y(t)=G^{<}(t,nh) \ , \ p(t)=i\epsilon(t) \ ,
                \ k(t,s)=i\Sigma^{\mathrm{R}}(t,s) \ , 
\end{align}
and a source term $q(t)$ which is obtained from the convolution
routines Eqs.~\eqref{cles2} and \eqref{cles3},
\begin{align} 
q(t)
=
-iC^{<}_2[\Sigma,1,G](m,n)-iC^{<}_3[\Sigma,1,G](m,n).
\end{align}
Note that $G^<$ must be calculated {\em after } $G^{\rceil}$ and $G^{\mathrm{R}}$ have been evaluated at  time-slices $\mathcal{T}[G]_{0\le n \le k}$, so that the input for  the latter convolution is already known at this stage of the algorithm. For $0\le m\le k$, the Volterra equation is solved using the start-up algorithm \eqref{vdiff_start}. 
\end{itemize}

\begin{figure}[ht]
  \centering
  \includegraphics[width=0.9\textwidth]{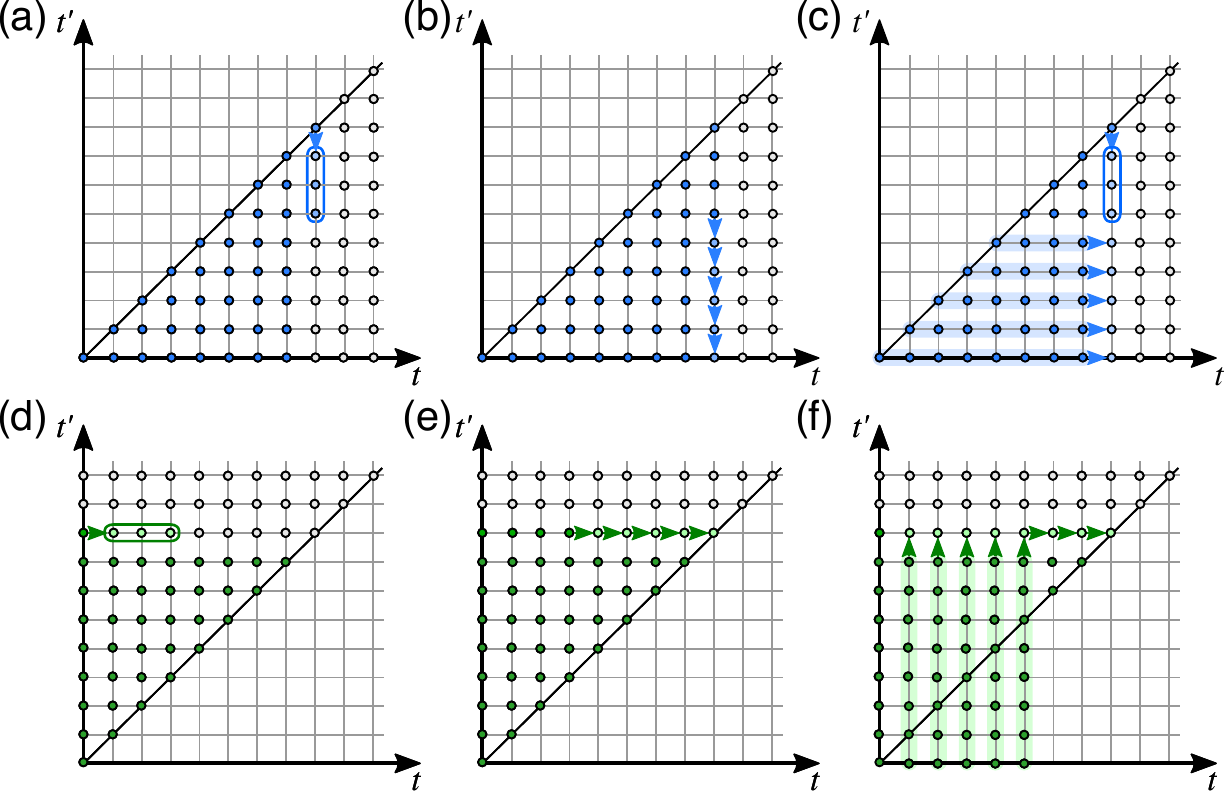}
  \caption{Propagation scheme of {\tt dyson} with $k=3$. (a) Starting at the
    diagonal $G^\ret_{n,n}$ with the initial condition~\eqref{dyson_ini_r}, the start-up algorithm determines
    $G^\ret_{n,m}$ for $m=n-1,\dots,n-k$. (b) After the start-up
    procedure, the remaining values of $G^\ret_{n,m}$,
    $m=n-k=1,\dots,0$ can be computed. (c) Parallel version of the
    {\tt dyson} solver for the retarded component: the values
    $G^\ret_{m,n}$ can be computed in parallel for $m=0,\dots,n-k$,
    while the boundary values are obtained as in (a). (d) Start-up
    procedure for $G^<_{m,n}$ for $m=0,\dots,k$ and subsequent
    time stepping (e). (f) Parallel algorithm for calculating
    $G^<_{m,n}$ for $m=1,\dots,n-k$.
    \label{fig:propscheme_dyson} }
\end{figure}

\subsection{Time stepping\label{subsec:impl_dyson_step}}

The {\tt dyson\_timestep} routines evaluate $G$ from   Eq.~\eqref{dyson}
on time-slice $\mathcal{T}[G]_n$ for $n>k$, provided that $G$ is
already known at time-slices $\mathcal{T}[G]_j$ for $j<n$ (c.f. Table
\ref{tab:dyson}). $\mathcal{T}[G]_n$ is calculated successively for
the retarded, left-mixing, and lesser components:

\begin{itemize}
\item
To determine $G^{\mathrm{R}}(nh,mh)$ for fixed $n$ and $0\le m \le n$ there are two alternatives:
\begin{itemize}
\item[(A)] We can consider   Eq.~\eqref{dyson_langreth_r_cc} with
  initial condition \eqref{dyson_ini_r}. The equation reduces to a
  standard Volterra equation~\eqref{MBCVSAMx_cc}, with the replacement
\begin{align}
&y(\bar t)=G^{\mathrm{R}}(nh,nh-\bar t) \ , \ p(\bar t)=i\epsilon(nh-\bar t), 
\nonumber
\\
&k(\bar t,s)=i\Sigma^{\mathrm{R}}(nh-s,nh-\bar t) \ , \ y(0)=-i.
\end{align}
The equation is solved using the start-up algorithm~\eqref{MBCVSAMx_cc_start}
for $\bar t=lh$, $0\le l\le k$ (i.\,e., to compute
$G^{\mathrm{R}}_{n,m}$ for $n-k \le m \le n$), while the time stepping algorithm~\eqref{MBCVSAMx_cc_t} in $\bar
t$ is applied for $\bar t=lh$, $l>k$ (i.\,e.,
to compute $G^{\mathrm{R}}_{n,m}$ for $0\le m < n-k$). The
time-stepping scheme is sketched in Fig.~\ref{fig:propscheme_dyson}(a)
and (b).
\item[(B)] We can consider Eq.~\eqref{dyson_langreth_r} with the initial
  condition~\eqref{dyson_ini_r}. The equation reduces to a standard
  Volterra equation~\eqref{MBCVSAMx} with the replacement
\begin{align}
&y(\bar t)=G^{\mathrm{R}}(mh+\bar t,mh) \ , \ p(\bar t)=i\epsilon(mh+\bar t), 
\nonumber
\\
&k(\bar t,s)=i\Sigma^{\mathrm{R}}(mh+\bar t,mh+s) \ ,\ y(0)=-i.
\end{align}
For each $0\le m < n-k$ this equation is solved for the single time
$\bar  t = (n-m)h$ (i.\,e. $t=nh$), using the time-stepping method
\eqref{MBCVSAMx_t}. Implementation (B) seems to have, in some cases, a
slightly larger numerical error than the alternative $(A)$. However, 
the Volterra time-steps for $0\le m < n-k$ can be
carried out in parallel, while the implementation (A) is inherently
serial. Hence we use alternative (B) for the {\tt openMP} parallel
implementations {\tt dyson\_timestep\_omp}, while (A) is used for the
serial implementation  {\tt dyson\_timestep}. For simplicity and better stability, the
values $G^{\mathrm{R}}_{n,m}$ for $n-k\le m \le n$ are always
determined from the implementation (A).
Figure~\ref{fig:propscheme_dyson}(c) illustrates the parallel
propagation scheme.
\end{itemize}
\item To determine $G^{\rceil}(nh,mh_\tau)$ for fixed $n>k$, we
  consider   Eq.~\eqref{dyson_langreth_tv} with initial condition
  \eqref{dyson_ini_tv}.  For each given $m$, this equation provides a
  Volterra equation of standard type \eqref{MBCVSAMx_cc}, with the
  replacement
\begin{align}
&y(t)=G^{\rceil}(t,\tau) \ , \ p(t)=i\epsilon(t) \ , \ k(t,s)=i\Sigma^{\mathrm{R}}(t,s)
\end{align}
with a source term $q(t)$ that is evaluated using the convolution
routines Eqs.~\eqref{ctv2} and \eqref{ctv3},
\begin{align}
q(nh)
=
-iC^{\rceil}_2[\Sigma,1,G](n,m)-iC^{\rceil}_3[\Sigma,1,G](n,m).
\end{align}
The Volterra equation is solved using the time stepping
\eqref{MBCVSAMx_t} at the single step $n$.
\item 
To determine $G^{<}(mh,nh)$ for given $n$ and $0\le m \le n$ we again have two alternatives: 
\begin{itemize}
\item[(A)] We consider   Eq.~\eqref{dyson_langreth_les} with the initial
  condition \eqref{dyson_ini_les}.  For each given $n$, this equation
  becomes a Volterra equation of standard type \eqref{MBCVSAMx}, with
  the replacement
\begin{align}
&y(t)=G^{<}(t,nh) \ , \ p(t)=i\epsilon(t) \ , \ k(t,s)=i\Sigma^{\mathrm{R}}(t,s), 
\end{align}
and a source term $q(t)$ which is obtained from the convolution
routines Eqs.~\eqref{cles2} and \eqref{cles3}
\begin{align} 
q(t)
=
-iC^{<}_2[\Sigma,1,G](m,n)-iC^{<}_3[\Sigma,1,G](m,n).
\end{align}
The equation is solved using the start-up algorithm
\eqref{vdiff_start} for $0\le m \le k$ (see Fig.~\ref{fig:propscheme_dyson}(d)) and the successive
time stepping according to   Eq.~\eqref{MBCVSAMx_t} for $k< m \le n$ (Fig.~\ref{fig:propscheme_dyson}(e)). 
\item[(B)] Alternatively, we consider
    Eq.~\eqref{dyson_langreth_les_cc} with the initial condition
  \eqref{dyson_ini_les_cc}.  For each given $m$, this equation
  provides a Volterra equation of standard type~\eqref{MBCVSAMx_cc}, with
  the replacement
\begin{align}
&y(\bar t)=G^{<}(mh,\bar t) \ , \ p(\bar t)=-i\epsilon(\bar t) \ , \
                k(s,\bar t)=-i\Sigma^{\mathrm{R}}(s,\bar t), 
\end{align}
and a source term $q(t)$ which is obtained from the convolution routines Eqs.~\eqref{cles2} and \eqref{cles3},
\begin{align} 
q(nh)
=
iC^{<}_2[G,1,\Sigma](m,n)+iC^{<}_3[G,1,\Sigma](m,n).
\end{align}
The equation is  solved using a single time step \eqref{MBCVSAMx_cc_t}
$\bar t=nh$ for each $0\le m < n-k$.
Since all these steps are independent, they can be performed in
parallel. Hence, we have implemented a parallelized version {\tt
  dyson\_timestep\_omp} based on {\tt
  openMP} threads.
The boundary values $m=n-k,...,n$ are
obtained using the serial implementation (A), {\em after}
$G^{<}_{m,n}$ has ben obtained from implementation (B) at $0\le m\le
n-k$. The scheme is sketched in Fig.~\ref{fig:propscheme_dyson}(f).
\end{itemize}
\end{itemize}

\section{Implementation: {\tt vie2} \label{sec:impl_vie2}}
\label{secqsx05}

\subsection{Langreth rules}

In this section we present the implementation of the {\tt vie2}
routine which solves Eq.~\eqref{vie2}. The solution is largely
equivalent to {\tt dyson}, but it reduces to a 
\gls*{vie} instead of a \gls*{vide}. To
solve Eq.~\eqref{vie2}, we again employ the Langreth rules to obtain
the individual equations for the Matsubara, lesser, and left-mixing
components,
\begin{align}
&
G^\mat(\tau) 
+
\int_0^\beta d\tau ' \,F^\mat(\tau-\tau')G(\tau')
=
Q^\mat(\tau),
\label{vie2_langreth_m}
\\
&
G^{\mathrm{R}}(t,t^\prime) 
+
 \int_{t^\prime}^t d\bar t \,F^{\mathrm{R}}(t,\bar t ) G^{\mathrm{R}}(\bar t,t^\prime)
 =Q^{\mathrm{R}}(t,t^\prime)
 \label{vie2_langreth_r}
\\
&
G^{\rceil}(t,\tau)
+
\int_{0}^t d\bar t \,F^{\mathrm{R}}(t,\bar t ) G^{\rceil}(\bar t,\tau)
\nonumber
\\
&\hspace{10mm}=
Q^{\rceil}(t,\tau)-\int_0^\beta d\tau F^{\rceil}(t,\tau') G^\mat(\tau'-\tau),
\label{vie2_langreth_tv}
\\
&
G^{<}(t,t^\prime) 
+
\int_{0}^t d\bar t F^{\mathrm{R}}(t,\bar t ) G^{<}(\bar t,t^\prime)
\nonumber
\\
&\hspace{10mm}=Q^<(t,t^\prime)
-
\int_{0}^{t^\prime} d\bar t \,F^<(t,\bar t ) G^{A}(\bar t,t^\prime) 
+i \int_0^\beta d\tau \,F^{\rceil}(t,\tau) G^\lceil(\tau,t^\prime).
\label{vie2_langreth_les}
\end{align}
Here Eq.~\eqref{vie2_langreth_m} must be solved with the boundary condition
\begin{align}
\label{vie2_ini_m}
G^\mat(-\tau)=\xi G^\mat(\beta-\tau) \ ,
\end{align}
while the remaining equations are solved with initial conditions
\begin{align}
\label{vie2_ini_r}
G^{\mathrm{R}}(t,t)
&=Q^{\mathrm{R}}(t,t)
\\
\label{vie2_ini_tv}
G^{\rceil}(0,\tau)
&=iG^\mat(-\tau)=i\xi G^\mat(\beta-\tau),
\\
\label{vie2_ini_les}
G^{<}(0,t^\prime)
&=-[G^\rceil(t^\prime,0)]^\dagger
\end{align}

\subsection{Matsubara\label{subsec:impl_vie_mat}}

The solution of the \gls*{vie} for the Matsubara component
(Eq.~\eqref{vie2_langreth_m}) is analogous to {\tt dyson\_mat}. After
transforming to the imaginary frequency representation
(cf. Eq.~\eqref{eq:Gmat_fourier_coeff_mod}),
Eq.~\eqref{vie2_langreth_m} is transformed to the algebraic equation
\begin{align}
  G^\mat(i \omega_m ) + F^\mat(i \omega_m) G^\mat(i \omega_m ) =
  Q^\mat(i \omega_m) \ . 
\end{align}
Solving this linear system and calculating the Fourier
sum~\eqref{eq:Gmat_fourier} then yields $G^\mat(\tau)$.

The accuracy of solving Eq.~\eqref{vie2_langreth_m} can again be
elevated to $\mathcal{O}(h^{k+2}_\tau)$ 
order by the Newton iteration. The algorithm
is analogous to the one discussed in
Section~\ref{subsec:impl_dyson_matsubara}, upon replacing
$g^\mat\rightarrow Q^\mat$, $K^\mat \rightarrow -F^\mat$.

The interface {\tt vie2\_mat} allows to choose either method by
specifying the argument {\tt method = CNTR\_MAT\_FOURIER} for the
Fourier method, and {\tt method = CNTR\_MAT\_FIXPOINT} for the Newton
iteration, respectively. By default, 
Newton's method is employed.

\subsection{Start\label{subsec:impl_vie_start}}

The {\tt vie2\_start} routine evaluates $G$ on the time-slices
$\mathcal{T}[G]_n$, $0\le n \le k$ (c.f. Table~\ref{tab:vie2}).

\begin{itemize}
\item To determine $G^{\mathrm{R}}(nh,mh)$ for $0\le n\le k$ and
  $n\le m \le k$ we consider Eq.~\eqref{vie2_langreth_r} with initial
  condition \eqref{vie2_ini_r}. The solution is similar to the
  start-up procedure for a Volterra equation \eqref{MBCVSAMx}: At each
  fixed $m$, we use a polynomial approximation for
  $y(t)=G^{\mathrm{R}}(t,mh)$ with $G^{\mathrm{R}}_{n,m}= y_m$
  \begin{align}
    \label{eq:y_to_GR}
y_n=
\begin{cases}
G^{\mathrm{R}}_{n,m} & m < n \le k
\\
Q^{\mathrm{R}}_{n,n} & m=n
\\
-[G^{\mathrm{R}}_{m,n}]^\dagger & 0 \le n <m
\end{cases}.
\end{align}
Here the values $y_{n}$ for $n<m$ amount to a continuous extrapolation
of $G^{\mathrm{R}}(t,t^\prime)$ to the domain $t<t^\prime$. When
Eq.~\eqref{vie2_langreth_r} is solved successively for $m=0,1,...,k$,
the values $y_n$ are already known for $n\le m$. Inserting the
polynomial ansatz for $y(t)$ into \eqref{vie2_langreth_r} yields
\begin{align}
y_n + h\sum_{l=0}^k I^{(k)}_{m,n;l}\tilde{F}^\ret_{n,l}y_l=Q^{\mathrm{R}}_{n,m}.
\end{align}
This is transformed into an $(k-m)\times(k-m)$ linear problem,
\begin{align}
&\sum_{l=m+1}^k M_{n,l}y_l = - \sum_{l=0}^{m} M_{n,l}y_l,\,\,\,n=m+1,...,k,
\\
&M_{n,l} =\epsilon_n+ h I^{(k)}_{m,n;l}\tilde{F}^\ret_{n,l}.
\end{align}
Because the input $y_{l\le m}$ for the right-hand side has been
computed previously, this equation can be solved for $y_{l>m}$.
\item To determine $G^{\rceil}(nh,mh_\tau)$ for $0\le n\le k$ and
  $0\le m \le N_\tau$ we consider Eq.~\eqref{vie2_langreth_tv} with the
  initial condition \eqref{vie2_ini_tv}.  For each given $m$, this
  equation provides a Volterra equation of standard type
  \eqref{MBCVSAMx01}, with the replacement
\begin{align}
&y(t)=G^{\rceil}(t,\tau), \  k(t,s)=F^{\mathrm{R}}(t,s),
\end{align}
where the source $q(t)$ is evaluated using the convolution routines Eqs.~\eqref{ctv2} and \eqref{ctv3},
\begin{align}
  \label{eq:map_vie_tv}
q(n h)
=
-C^{\rceil}_2[F,1,G](n,m)-C^{\rceil}_3[F,1,G](n,m)+Q^{\rceil}_{n,m}.
\end{align}
For $0\le n\le k$, the Volterra equation is solved using the start-up algorithm \eqref{MBCVSAMx01_start}.
\item 
To determine $G^{<}(mh,nh)$ for $0\le n\le k$ and $0\le m \le n$ we consider Eq.~\eqref{vie2_langreth_les} with the initial condition \eqref{vie2_ini_les}.  For each given $n$, this equation provides a  Volterra equation of standard type \eqref{MBCVSAMx01}, with the replacement
\begin{align}
 \label{eq:map_les_vie1}
y(t)=G^{<}(t,nh), \ k(t,s)=F^{\mathrm{R}}(t,s), 
\end{align}
and a source term $q(t)$ which is obtained from the convolution routines Eqs.~\eqref{cles2} and \eqref{cles3},
\begin{align}
  \label{eq:map_les_vie2}
q(t)
=
-C^{<}_2[F,1,G](m,n)-C^{<}_3[F,1,G](m,n) + Q^<_{m,n}.
\end{align}
Note that $G^<$ must be calculated {\em after } $G^{\rceil}$ and $G^{\mathrm{R}}$ have been evaluated at the time-slices $\mathcal{T}[G]_{0\le n \le k}$, so that the input for  the latter convolution is already known at this stage of the algorithm. For $0\le m\le k$, the Volterra equation is solved using the start-up algorithm \eqref{MBCVSAMx01_start}. 
\end{itemize}

\subsection{Time stepping \label{subsec:impl_vie_step}}

Once the start-up problem has been solved and $\mathcal{T}[G]_n$ is
known for $n=0,\dots,k$, time-stepping can be
employed (see Table~\ref{tab:vie2}). Mapping the
\gls*{vies}~\eqref{vie2_langreth_r}--\eqref{vie2_langreth_les} to the
standard \gls*{vie}~\eqref{MBCVSAMx01} allows to directly adopt the algorithm
from Section~\ref{subsec:standardvie2}. Suppose $G^\ret(j h, m h)$,
$G^\rceil(j h, l h_\tau)$ and $G^<(m h, j h)$ are known for
$j=0,\dots,n-1$, $m=0,\dots,j$ and $l=0,\dots, N_\tau$. Then the next
time step $\mathcal{T}[G]_n$ is obtained as follows:
\begin{itemize}
  \item In order to compute $G^\ret(n h, m h)$, for $m=0,\dots, n-1$
    (since $G^\ret(n h, n h)$ = $Q^\ret(n h, n h)$),  we approximate
    the convolution by Eq.~\eqref{cr1}. Hence, setting $y(t) = G^\ret(t, m
    h)$ for fixed $m$ maps Eq.~\eqref{vie2_langreth_r} to the standard
    \gls*{vie}~\eqref{MBCVSAMx} for $n-m > k$. One obtains
    \begin{align*}
      y_n = q_n + h \sum^n_{j=m} w^{(k)}_{n-m,j-m} F^\ret_{n,j} y_j \ ,
    \end{align*}
    where the continuous extension~\eqref{eq:y_to_GR} is implied. The
    above equation is then solved for $y_n$. For $n-m \le k$, the
    procedure is similar: via the approximation~\eqref{cr1}, the \gls*{vie}
    translates to
    \begin{align*}
      y_n = q_n + h \sum^k_{j=0} w^{(k)}_{n-m,j} F^\ret_{n,j} y_{n-j}
      \ ,
    \end{align*}
    which is readily solved for $y_n$. Note that only $G^\ret(n h, m
    h)$ needs to be extrapolated to the upper triangle, while the
    kernel $F^\ret_{n,m} = F^\ret(n h, m h)$ is strictly
    causal. Except for the case $n-m \le k$, the time step
    $n-1\rightarrow n$ can be carried out independently for every
    $m=0$. Therefore, these time steps can be performed in parallel,
    as implemented in the {\tt openMP}-based function {\tt
      vie2\_timestep\_omp}. 
   \item The \gls*{vie}~\eqref{vie2_langreth_tv} maps to the standard
     \gls*{vie}~\eqref{MBCVSAMx01} upon identifying $y(t)=G^\rceil(t, m
     h_\tau)$, $k(t,s) = F^\ret(t,s)$, while the source term $q(t)$ is
     obtained by the identification~\eqref{eq:map_vie_tv}. The
     time-stepping algorithm~\eqref{MBCVSAMx01_t} can be used
     directly. All steps depend only parametrically on $m$, so
     parallel propagation is straightforward.
     
    \item Once $G^\ret(n h, m h)$ and $G^\rceil(n h, l h_\tau)$
      ($m=0,\dots,n$, $l=0,\dots,N_\tau$) have been obtained, the
      lesser component $G^<(m h, n h)$ can be computed. As for {\tt
        dyson\_timestep}, there are two options for proceeding:
      \begin{itemize}
        \item [(A)] The substitutions~\eqref{eq:map_les_vie1}
        and \eqref{eq:map_les_vie2} map the lesser
        \gls*{vie}~\eqref{vie2_langreth_les} to the standard
        \gls*{vie}~\eqref{MBCVSAMx01}. For $m=0,\dots,k$, the resulting
        equation is solved by the start-up
        method~\eqref{MBCVSAMx01_start}, using the initial
        condition~\eqref{vie2_ini_les}. For $m=k+1,\dots n$, the time
        propagation proceeds by solving Eq.~\eqref{MBCVSAMx01_t}. This
        scheme of time stepping is sequential by construction.
        \item[(B)] Instead of starting from
          Eq.~\eqref{vie2_langreth_les}, the conjugate equation
          \begin{align*}
            G^<(t,t^\prime) + [G\ast F^\ddagger]^<(t,t^\prime) = Q^<(t,t^\prime)
          \end{align*}
          can serve as a starting point. The substitution
          \begin{align*}
            y(t) &= G^<(m h, t) , \ k(s, t) =
                   [F^\ddagger]^\mathrm{A}(s,t) \\
            q(m h) &= -C^<_1[G,1,F^\ddagger](m,n) -
            C^<_3[G,1,F^\ddagger](m,n) +  Q^<(m h, n h)
          \end{align*}
          leads the to the conjugate \gls*{vie}~\eqref{MBCVSAMx01_cc}, which
          can then be propagated by invoking
          Eq.~\eqref{MBCVSAMx01_cc)t}. This time-stepping scheme can
          be performed for all $m=0,\dots,n-1$ in parallel, as
          implemented in {\tt vie2\_timestep\_omp}. The last point
          $G^<(n h, n h)$ can be computed once $G^<(n h, (n-1) h)
          =-[G^<((n-1) h, n h)]^\dagger$ is known.
        \end{itemize}
      \end{itemize}

\section{Implementation: Free Green's
    functions\label{sec:impl_green}}

Free \gls*{GFs} $G_0(t,t^\prime)$ are determined from the  equation of motion [c.f.~Eq.~\eqref{freefreefree}]
\begin{align}
  \left[i \partial_t - \epsilon(t)\right] G_0(t,t^\prime) = \delta_\mathcal{C}(t,t^\prime)
\end{align}
as follows. Let us denote the eigenvalues of the Hamiltonian matrix
$\epsilon(0^-)$ by $\varepsilon_\alpha$ and the corresponding 
basis transformation 
matrix by $R$, such that $\epsilon(0^-) = R\,
\mathrm{diag}\{\epsilon_\alpha\} R^\dagger$
($\mathrm{diag}\{\varepsilon_\alpha\}$ stands for the diagonal matrix containing
the energies $\varepsilon_\alpha$).
The Matsubara component is then given by
\begin{align}
  G^\mat_0(\tau) = R\, \mathrm{diag}\{f_\xi(\mu-\varepsilon_\alpha)
  e^{-(\epsilon_\alpha-\mu)\tau} \} R^\dagger
\end{align}
for $\tau\in (0,\beta)$.

All other Keldysh components of $G_0(t,t^\prime)$ are governed by the 
unitary time evolution (defined in Eq.~\eqref{eq:timevol}) with
respect to the single-particle Hamiltonian $\epsilon(t)$. 

\subsection{Commutator-free matrix exponentials}
On the equidistant grid
$t_n=n h$, we approximate the propagator $U_{n,j}\equiv
U(n h, j h)$ by
the commutator-free matrix exponential approximation described in 
Ref.~\cite{ALVERMANN20115930}. In particular, we have implemented the fourth-order
approximation
\begin{align}
  \label{eq:cf4}
  U_{n+1,n} &= \exp\left[-i(a_1 \epsilon((n+c_1)h) + a_2
  \epsilon((n+c_2)h))\right] \\
  &\quad \times \exp\left[-i(a_2 \epsilon((n+c_1)h) + a_1
  \epsilon((n+c_2)h))\right] + \mathcal{O}(h^5) \nonumber  \ ,
\end{align}
where $a_1=(3-2\sqrt{3})/12$, $a_2=(3+2\sqrt{3})/12$,
$c_1=(1-1/\sqrt{3})/2$ and $c_2=(1+1/\sqrt{3})/2$. Using the semi-group
property
$U_{n,j}=U_{n,n-1} U_{n-1,n-2} \dots U_{j+1,j}$, we can
thus express the propagator up to
$\mathcal{O}(h^4)$.  The Hamiltonian at the intermediate points
$(n+c_{1,2}) $ entering Eq.~\eqref{eq:cf4} is approximated by
polynomial interpolation, using the points $n-k+1,\dots,n,n+1$ (see
Section~\ref{sec:polynum}). If $\epsilon(t)$ represents a mean-field
Hamiltonian, which is self-consistently determined in the course of
the time step $n\rightarrow n+1$, $\epsilon_{n+1}$ is typically not
known before the \gls*{GF} at time step $n+1$ has been computed. Hence, we
employ polynomial extrapolation to provide a guess for
$\epsilon_{n+1}$ before interpolating. 

\subsection{Real-time and mixed components}

Based on the commutator-free matrix exponential approximation, the
remaining Keldysh components are determined by
\begin{subequations}
  \label{eq:G0_free}
  \begin{equation}
    G_0^\rceil(n h,\tau) = -i\xi U_{n,0}(n h,0) R\,
    \mathrm{diag}\{f_\xi(\varepsilon_\alpha-\mu)e^{(\epsilon_\alpha-\mu)\tau}\}
    R^\dagger \ , 
  \end{equation}
  \begin{equation}
    G_0^\ret(nh,j h) = -i U_{n,j} = U_{n,0}[U_{j,0}]^\dagger \ ,
  \end{equation}
  \begin{equation}
    G_0^<(j h, n h) = i U_{j,0} R\, \mathrm{diag}\{f_\xi(\varepsilon_\alpha-\mu)\} R^\dagger [U_{n,0}]^\dagger \ .
  \end{equation}
\end{subequations}
Note that for a time-independent Hamiltonian, Eq.~\eqref{eq:G0_free}
is numerically exact up to round-off errors. Furthermore, the
structure of Eq.~\eqref{eq:G0_free} allows to compute the time slice
$\mathcal{T}[G_0]_n$ directly.

\section{Conclusions\label{sec:conclusions}}
We have presented the \nessi{} library, a \textbf{N}on-\textbf{E}quilibrium \textbf{S}ystems \textbf{Si}mulation package. 
This open-source computational physics library provides a simple and efficient framework for simulations of quantum
many-body systems out of equilibrium, based on the Greens function formalism. The 
numerical routines employed in the solution of the Kadanoff-Baym equations and the evaluation of Feynman diagrams have been described in detail. We have  exemplified the usage of the library by several applications ranging from
simple two-level problems to the state-of-the-art simulations of interacting lattice systems. This information should enable users of the library to implement and run custom applications.

\nessi{} is an open source library and we encourage contributions and feedback 
from the user community. 
We will continue to work on extensions of the library. 
Planned near-term improvements include the publication of a software package for nonequilibrium impurity and dynamical mean-field theory calculations based on strong-coupling perturbative solvers, 
non-equilibrium steady state solvers, and  truncation schemes for the memory integrals in the integral equations.
The latest updates will posted on the web page {\tt www.nessi.tuxfamily.org}, which also contains a link to the repository, installation instructions, a detailed manual of all relevant classes and routines, and additional example programs.
Contributions to the future extensions are welcome, although we recommend to coordinate with the main \nessi{} developers before embarking on any major coding effort. Any issues encountered in the use of the library should be exclusively reported via the contact address specified on the web site {\tt www.nessi.tuxfamily.org}.


\appendix

\section{Contour function utilities}
\label{app:utilities}

In this appendix, we describe how contour functions can be extrapolated by polynomial extrapolation. Furthermore, we define a Euclidean distance norm for contour functions.

\subsection{Extrapolation of contour functions \label{subsec:extrapolation}}

For the time-stepping algorithm, a guess for the \gls*{GF} $\mathcal{T}[G]_{n}$ or the self-energy $\mathcal{T}[\Sigma]_{n}$ is usually required for starting the self-consistency cycle at time step $n$. In many cases it is useful to employ a polynomial extrapolation as a predictor $\mathcal{T}[G]_{n-1} \rightarrow \mathcal{T}[G]_{n}$, as explained in the following. 

Based on polynomial interpolation (see Section~\ref{subsec:polyinterpolation}), we define the polynomial extrapolation by
\begin{align}
  \label{eq:polyextrap}
  y_{n+1} = \sum^k_{l=0} C^{(k)}_l y_{n-l} \ ,
\end{align}
where $y_l = y(l h)$. The coefficients $C^{(k)}_l$ are obtained by inserting $t=(n+1)h$ into Eq.~\eqref{dwjvwdjkjxbk}. For extrapolations in the two-time plane, we have implemented the following algorithm:
\begin{itemize}
  \item To approximate $G^\rceil((n+1)h,\tau)$ we set $y(t) = G^\rceil(t,\tau)$ for fixed $\tau$ and apply Eq.~\eqref{eq:polyextrap}. 
  \item For extrapolating the retarded component, we set $y(t) = \widetilde{G}^\ret(t,j h)$ for $j=0,\dots,k$ and apply Eq.~\eqref{eq:polyextrap}. For the remaining points, we extrapolate along lines parallel to the time diagonal by identifying $y(t)=G^\ret(t,t-j h)$ for $j=0,\dots,n-k$. Using Eq.~\eqref{eq:polyextrap} then yields the extrapolation to $G^\ret((n+1)h,(n+1-j)h)\approx y_{n+1}$.
  \item Similarly, the lesser component can be extrapolated by identifying
  \begin{align*}
  y(t) = \begin{cases} G^<(j h, t) & : t \ge j h \\ -[G^<(t,j h)]^\dagger & : t < j h \end{cases} \ .
  \end{align*}
  Polynomial extrapolation~\eqref{eq:polyextrap} then yields $G^<(jh, (n+1)h)$ for $j=0,\dots,k$.
  Analogous to the retarded component, the remaining points in the two-time plain are obtained by
  applying Eq.~\eqref{eq:polyextrap} to $y(t) = G^<((j-n-1)h + t,t)$ for $j=k+1,\dots,n+1$. Note that this includes the diagonal $G^<((n+1)h,(n+1)h)$.
\end{itemize}
Equation~\eqref{eq:polyextrap} can also be applied to single-time contour functions $f(t)$. The above algorithm is implemented in the function {\tt extrapolate\_timestep}.

\subsection{Euclidean norm\label{subsec:distance}}

For assessing the convergence of self-consistent algorithms, we introduce an Euclidean norm for contour functions. Consider two time slices $\mathcal{T}[A]_n$, $\mathcal{T}[B]_n$ at time step $n$. We define the distance for the individual components as
\begin{subequations}
\begin{equation}
  \left\Vert A - B \right\Vert^\mat = \sum^{N_\tau}_{m=0}\sum_{a,b} \left| A^\mat_{a,b}(m h_\tau) - B^\mat_{a,b}(m h_\tau)\right| \ ,
\end{equation}
\begin{equation}
  \left\Vert A - B \right\Vert^\rceil_n = \sum^{N_\tau}_{m=0}\sum_{a,b} \left| A^\rceil_{a,b}(n h,m h_\tau) - B^\rceil_{a,b}(n h,m h_\tau)\right| \ ,
\end{equation}
\begin{equation}
  \left\Vert A - B \right\Vert^\ret_n = \sum^{n}_{j=0}\sum_{a,b} \left| A^\ret_{a,b}(n h,j h) - B^\ret_{a,b}(n h,j h)\right| \ ,
\end{equation}
\begin{equation}
  \left\Vert A - B \right\Vert^<_n = \sum^{n}_{j=0}\sum_{a,b} \left| A^<_{a,b}(j h, n h) - B^<_{a,b}(j h, n h)\right| \ . 
\end{equation}
\end{subequations}
The total distance at time step $n$ is then defined by
\begin{align}
  \label{eq:distancenorm}
  \left\Vert A - B \right\Vert_n = \begin{cases}\left\Vert A - B \right\Vert^\mat & : n=-1 \\
   \left\Vert A - B \right\Vert^\rceil_n + \left\Vert A - B \right\Vert^\ret_n + \left\Vert A - B \right\Vert^<_n & : n \ge 0 \end{cases} \ .
\end{align}
The Euclidean norm Eq.~\eqref{eq:distancenorm} is implemented in the function {\tt distance\_norm2}.

\section{Installation instructions: nessi\_demo
example programs}
\label{app:install_nessidemo}

We assume that the \libcntr{} library has been compiled successfully and installed under the prefix {\tt /home/opt}. Hence, {\tt /home/opt/lib} contains the shared library {\tt libcntr.so} (or {\tt libcntr.dylib} under MacOSX), while {\tt /home/opt/include} contains the directory {\tt cntr} with all required headers. After downloading or cloning the repository {\tt nessi\_demo}, navigate into it and create a build directory (for instance, {\tt cbuild}). The installation procedure is similar to the compilation of \libcntr{} (see Section~\ref{subsec:installation_libcntr}). We recommend creating a configuration script similar to
\begin{lstlisting}[language=bash]
CC=[C compiler] CXX=[C++ compiler] \
cmake \
    -DCMAKE_BUILD_TYPE=[Debug|Release] \
    -Domp=[ON|OFF] \
    -Dhdf5=[ON|OFF] \
    -Dmpi=[ON|OFF] \
    -DCMAKE_INCLUDE_PATH=[include directory] \
    -DCMAKE_LIBRARY_PATH=[library directory] \
    -DCMAKE_CXX_FLAGS="[compiling flags]" \
    ..
\end{lstlisting}
For compiling all examples including the translationally invariant Hubbard model (Section~\ref{subsec:example_gw}), \gls*{mpi} compilers need to be provided for the C and the C++ compiler. Furthermore, set {\tt mpi=ON}. 

{\tt CMAKE\_INCLUDE\_PATH} needs to include the path used to compile \libcntr{} (containing the {\tt eigen3} and {\tt hdf5} headers) and, additionally, {\tt /home/opt/include}. The paths provided to {\tt CMAKE\_LIBRARY\_PATH} should include all the library paths used to compile \libcntr{}, extended by {\tt /home/opt/lib}. We recommend using the same compiler flags as for the compilation of \libcntr{}, including
\begin{lstlisting}[language=bash]
-std=c++11
\end{lstlisting}
After creating the above configure script (for instance, {\tt configure.sh}), navigate to the build directory and run
\begin{lstlisting}[language=bash]
sh ../configure.sh
make
\end{lstlisting}
to compile the example programs. The executables are placed under {\tt nessi\_demo/exe}. 




\mbox{}\\
{\it Acknowledgements}

We thank 
Marcus Kollar,
Naoto Tsuji,
Jiajun Li,
and Nagamalleswararao Dasari,
for important feedback while using the library, and for collaborations on early stages of the library.
The development of this library has been supported by the Swiss National Science Foundation through SNF Professorship PP0022-118866 (ME,PW), Grants 200021-140648 and 200021-165539 (DG), and NCCR MARVEL (MS,YM), as well as the European Research Council through ERC Starting Grants No. 278023 (AH,HS,PW) and No. 716648 (ME), and ERC Consolidator Grant No. 724103 (MS,NB,PW,YM). The Flatiron institute as a division of the Simons Foundation.

\nocite{*}







\end{document}